\documentclass[twocolumn]{aastex62}
\usepackage{graphicx}
\usepackage{natbib}
\usepackage{amsmath,amsthm,amssymb}
\usepackage{color}
\usepackage{ulem}
\usepackage{epstopdf}
\newcommand{\gaia}{{\it Gaia}}
\newcommand{\gaiao}{{\it Gaia}~DR1}
\newcommand{\gaiat}{{\it Gaia}~DR2}
\newcommand{\PS}{\protect \hbox {Pan-STARRS1}}
\newcommand{\spitzer}{{\it Spitzer}}
\newcommand{\WISE}{{\it WISE}}

\newcommand{\ktwo}{\ensuremath{K_{S,{\rm 2MASS}}}}

\newcommand{\jmko}{\ensuremath{J_{\rm MKO}}}
\newcommand{\kmko}{\ensuremath{K_{\rm MKO}}}

\newcommand{\fldg}{\mbox{\textsc{fld-g}}}
\newcommand{\intg}{\mbox{\textsc{int-g}}}
\newcommand{\vlg}{\mbox{\textsc{vl-g}}}

\newcommand{\rchi}{\ensuremath{\chi^2_\nu}}

\newcommand{\radec}{\ensuremath{(\alpha,\delta)}}                                
\newcommand{\radeco}{\ensuremath{(\alpha_0,\delta_0)}}                                
\newcommand{\mua}{\ensuremath{\mu_\alpha{\rm cos}\,\delta}}     
\newcommand{\mud}{\ensuremath{\mu_\delta}}                                
\newcommand{\teff}{\ensuremath{T_{\rm eff}}}
\newcommand{\dphot}{\ensuremath{d_{\rm phot}}}

\newcommand{\plx}{\ensuremath{\varpi}}
\newcommand{\uas}{$\mu$as}
\newcommand{\my}{\protect \hbox {mas yr$^{-1}$}}
\newcommand{\uy}{\protect \hbox {\uas\ yr$^{-1}$}}
\newcommand{\kms}{\protect \hbox {km s$^{-1}$}}
\newcommand{\mjup}{\ensuremath{M_{\mathrm{Jup}}}}           
\newcommand{\vmax}{\ensuremath{V/V_{\rm max}}}

\newcommand{\banyans}{\protect \hbox{BANYAN $\Sigma$}}
\newcommand{\banyanii}{\protect \hbox{BANYAN II}}

\newcommand{\varntargets}{356}
\newcommand{\varngoodplx}{348}
\newcommand{\varnnoplx}{8}
\newcommand{\varngaia}{104}
\newcommand{\varngaiadof}{103}
\newcommand{\varngaiagood}{97}
\newcommand{\varngaiagooddof}{96}
\newcommand{\varngaiadirect}{104}
\newcommand{\varngaiacomp}{10}
\newcommand{\varpgaia}{30}
\newcommand{\varpgaiaearly}{86}
\newcommand{\varnlitstart}{24}
\newcommand{\varnlitnew}{7}
\newcommand{\varmedrchi}{0.96}
\newcommand{\varmedrchibin}{1.00}
\newcommand{\varminplxerr}{1.5}
\newcommand{\varmedplxerr}{3.5}

\newcommand{\varplxerrltsix}{97}
\newcommand{\varnfirst}{165}
\newcommand{\varntfirst}{121}
\newcommand{\varnimprove}{53}

\shorttitle{\varngoodplx\ L and T Dwarf Parallaxes}
\shortauthors{Best, W. M. J. et al}

\received{April 25, 2019}
\revised{February 19, 2020}
\accepted{February 28, 2020}

\begin{document}

{\large Accepted by {\it The Astronomical Journal}, 2020 February 28}

\title{The Hawaii Infrared Parallax Program. \\
IV. A Comprehensive Parallax Survey of L0--T8 Dwarfs with UKIRT}

\correspondingauthor{William M. J. Best}
\email{wbest@utexas.edu}

\author[0000-0003-0562-1511]{William M. J. Best}
\affil{Institute for Astronomy, University of Hawaii, 2680 Woodlawn Drive, Honolulu, HI 96822}
\affil{University of Texas at Austin, Department of Astronomy, 2515 Speedway C1400, Austin, TX 78712, USA}

\author[0000-0003-2232-7664]{Michael C. Liu}
\affil{Institute for Astronomy, University of Hawaii, 2680 Woodlawn Drive, Honolulu, HI 96822, USA}

\author[0000-0002-7965-2815]{Eugene A. Magnier}
\affil{Institute for Astronomy, University of Hawaii, 2680 Woodlawn Drive, Honolulu, HI 96822, USA}

\author[0000-0001-9823-1445]{Trent J. Dupuy}
\affil{Gemini Observatory, 670 N. A`ohoku Place, Hilo, HI 96720, USA}

\begin{abstract}

  We present parallaxes, proper motions, and $J$-band photometry for
  \varngoodplx\ L and T dwarfs measured using the wide-field near-infrared
  camera WFCAM on the United Kingdom Infrared Telescope.
  This is the largest single batch of infrared parallaxes for brown dwarfs to date.
  Our parallaxes have a median uncertainty of \varmedplxerr~mas, similar to most
  previous ground-based infrared parallax surveys.
  Our target list was designed to complete a volume-limited parallax sample of L0--T8
  dwarfs out to 25~pc spanning declinations $-30\degr$ to $+60\degr$ (68\% of
  the sky).
  We report the first parallaxes for \varnfirst~objects,
  and we improve on previous measurements for another \varnimprove~objects.
  Our targets include \varngaia\ objects (mostly early-L~dwarfs) having \gaiat\
  parallax measurements, with which our parallaxes are consistent.
  We include an extensive comparison of previous literature parallaxes for L and
  T~dwarfs with both our results and \gaiat\ measurements, identifying
  systematic offsets for some previous surveys.
  Our parallaxes confirm that 14 objects previously identified as wide
  common proper motion companions to main-sequence stars have distances
  consistent with companionship.
  We also report new \jmko\ photometry for our targets, including the first
  measurements for 193 of our targets and improvements over previously published
  \jmko\ photometry for another 60~targets.
  Altogether, our parallaxes will enable the first population studies using a
  volume-limited sample spanning spectral types L0--T8 defined entirely by
  parallaxes.

\end{abstract}

\keywords{Late-type dwarf stars (906); Parallax (1197); Proper motions (1295);
Infrared photometry (792); Trigonometric parallax (1713); L dwarfs (894); T dwarfs (1679)}

\section{Introduction}
\label{intro}

Brown dwarfs are objects more massive than giant planets
\citep[$\gtrsim13$~\mjup;][]{Spiegel:2011ip} but not massive enough to sustain
hydrogen fusion in their cores and become stars
\citep[$\lesssim70$~\mjup;][]{Dupuy:2017ke}. Brown dwarfs therefore cool as they
age, from late-M ($\teff\approx3000$~K) spectral types through the L, T, and Y
($\approx300$~K) types. All along their cooling sequence, brown dwarfs show a
surprising diversity of near- and mid-infrared colors, thought to be caused by
variations in surface gravity and condensate clouds
\citep[e.g.,][]{Burrows:2006ia} or thermo-chemical instabilities \citep[but see
\citealt{Leconte:2018ft}]{Tremblin:2016hi}. Young ($\lesssim$100~Myr) low-mass
brown dwarfs serve as analogs for directly imaged exoplanets given their
overlapping temperature and age ranges. Since exoplanets are far more difficult
to observe in the glare of their host stars, young brown dwarfs are vital
templates for understanding giant exoplanets.

As the lowest-mass products of star formation, brown dwarfs also hold unique
clues about the history of star formation in our galaxy. However, because brown
dwarfs cool continuously after formation, their luminosities and temperatures
depend on both age and mass, with the result that a younger, less massive brown
dwarf can have the same luminosity and temperature (and thus spectral type) as
an older, more massive brown dwarf. This observational degeneracy makes
evolutionary trends in brown dwarf populations difficult to identify. A
well-defined sample of brown dwarfs with precise luminosities would considerably
improve our ability to test models of formation, evolution, and atmospheres.

Trigonometric parallaxes provide the most accurate and direct measures of
distance, and they are therefore vital for measuring luminosities. Parallaxes
are also the best means to determine membership in volume-limited samples, long
regarded as the gold standard for population studies as they minimize the
selection biases inherent in magnitude-limited samples. However, parallax
measurements are observationally expensive and, therefore, lacking for many
brown dwarfs. Large digital sky surveys such as the Sloan Digital Sky Survey
\citep[SDSS;][]{York:2000gn}, Two Micron All Sky Survey
\citep[2MASS;][]{Skrutskie:2006hl}, UKIRT Infrared Deep Sky Survey
\citep[UKIDSS;][]{Lawrence:2007hu}, Wide-Field Infrared Survey Explorer
\citep[\WISE;][]{Wright:2010in}, and the Panoramic Survey Telescope And Rapid
Response System (\PS)~3$\pi$ Survey (PS1; K. C. Chambers et al. 2020, in
preparation) have discovered more than 2000 L, T, and Y~dwarfs, but precise
parallaxes have been measured for fewer than 20\% of them, and no complete
volume-limited sample spanning all LTY spectral types has been established.

The earliest ultracool parallax programs targeted M, L, and T dwarfs as they
were newly discovered, collectively observing $\approx$50 L and T~dwarfs
\citep{Dahn:2002fu,Tinney:2003eg,Vrba:2004ee}. As the rate of discoveries
accelerated, however, more recent ultracool parallax programs have frequently
prioritized extreme or unusual objects. Our Hawaii Infrared Parallax Program
(HIPP) has previously observed mostly L/T transition dwarfs \citep[hereinafter
DL12]{Dupuy:2012bp}, young ultracool dwarfs \citep[hereinafter
LDA16]{Liu:2016co}, and ultracool binaries \citep[DL12;][hereinafter
DL17]{Dupuy:2017ke} with WIRCam on the Canada France Hawaii Telescope (CFHT),
while others have focused on ultracool subdwarfs \citep{Schilbach:2009ja} and
young L~dwarfs \citep{ZapateroOsorio:2014cu}. Several programs have targeted
late-T and Y dwarfs, the coldest substellar objects in the solar neighborhood
\citep{Dupuy:2013ks,Marsh:2013hk,Beichman:2014jr,Tinney:2014bl,Martin:2018hc,Kirkpatrick:2019kt}.
Very high-precision parallax measurements have also constrained the orbital
motion of binaries and planetary-mass companions
\citep[DL17]{Sahlmann:2014hu,Sahlmann:2015ku,Sahlmann:2015ip,Dupuy:2015gl,Sahlmann:2015jt}.

Two large parallax programs targeting nearby stars, the Research Consortium On
Nearby Stars \citep[RECONS; e.g.,][]{Winters:2017er,Henry:2018gh} and the
Carnegie program at Las Campanas \citep{Weinberger:2013cc,Weinberger:2016gy}
have also observed a handful of ultracool dwarfs, but the optical detectors used
for those programs limited them to observing mostly late-M and early-L dwarfs.
Most recently, \gaiat\ \citep{GaiaCollaboration:2016cu,GaiaCollaboration:2018io}
has measured parallaxes for hundreds of L~dwarfs, but {\gaia}'s optical
detectors face the same limitation and have obtained only a handful of
parallaxes for L6 and later objects. (We discuss {\gaia}'s ultracool parallaxes
in Section~\ref{results}.)

A comprehensive understanding of brown dwarfs requires a large volume-limited
sample representing the full population.  Parallax programs that have targeted
broad swaths of L and T~dwarfs include 
the United States Naval Observatory CCD and near-infrared (NIR) programs
\citep[USNO;][]{Dahn:2002fu,Vrba:2004ee,Dahn:2017gu}, which so far have
published parallaxes for 91 L and T dwarfs;
the Brown Dwarf Kinematics Project
\citep[BDKP;][]{Faherty:2012cy,Faherty:2016fx}, which obtained 88 parallaxes of
late-M, L, and T dwarfs;
and PARallaxes of Southern Extremely Cool Objects
\citep[PARSEC;][]{Andrei:2011jm,Marocco:2013kv,Smart:2018en} and New Technology
Telescope (NTT) PARSEC \citep[NPARSEC;][]{Smart:2013km}, which have measured 127
L and T~dwarf parallaxes.
While these programs have contributed substantially to the ultracool parallax
census, collectively, their results form a patchwork of varying sky and spectral
type coverage. Previously, the most complete volume-limited sample assembled was
the 2MASS-based sample of \citet{Reid:2008fz}, comprising 196 late-M, L, and T
dwarfs out to 20~pc over 65\% of the sky, but complete only for spectral types
M9-L6 and using photometric distances for two-thirds of the members.  The
full-sky 8~pc volume-limited sample of \citet{Kirkpatrick:2012ha} includes all
spectral types but contains only 33 L, T, and Y dwarfs.  \citet[hereinafter
K19]{Kirkpatrick:2019kt} have recently constructed a 20~pc volume-limited sample
using their and literature parallaxes for T6~and later-type dwarfs along with
\gaiat\ and literature parallaxes for L0--L5 dwarfs, containing 286 objects but
lacking L6--T5 dwarfs.  The new Ultracool SpeXtroscopic sample of
\citet{BardalezGagliuffi:2019gn} contains 410~M7--L5~dwarfs of which 93\% have
parallaxes, and is $83^{+10}_{-9}\%$ complete for L0--L5~dwarfs, but with its
mid-L cutoff, it includes only a handful of the warmest brown dwarfs.

We have obtained infrared parallaxes for \varngoodplx\ L and T~dwarfs using the
United Kingdom Infrared Telescope (UKIRT), with the goal of completing a
volume-limited L and T~dwarf sample that we will present in an upcoming paper
(W. Best et al., AAS Journals, submitted). The present work describes the
measurement and validation of our UKIRT parallaxes. We describe our target
selection in Section~\ref{sample}, and our observations in Section~\ref{obs}. In
Section~\ref{reduc} we describe our data reduction pipeline, and we explain
details of our parallax measurements in Section~\ref{plx}. We present our
results in Section~\ref{results}.

\section{Target Selection}
\label{sample}
\begin{figure}
    \includegraphics[width=1\columnwidth]{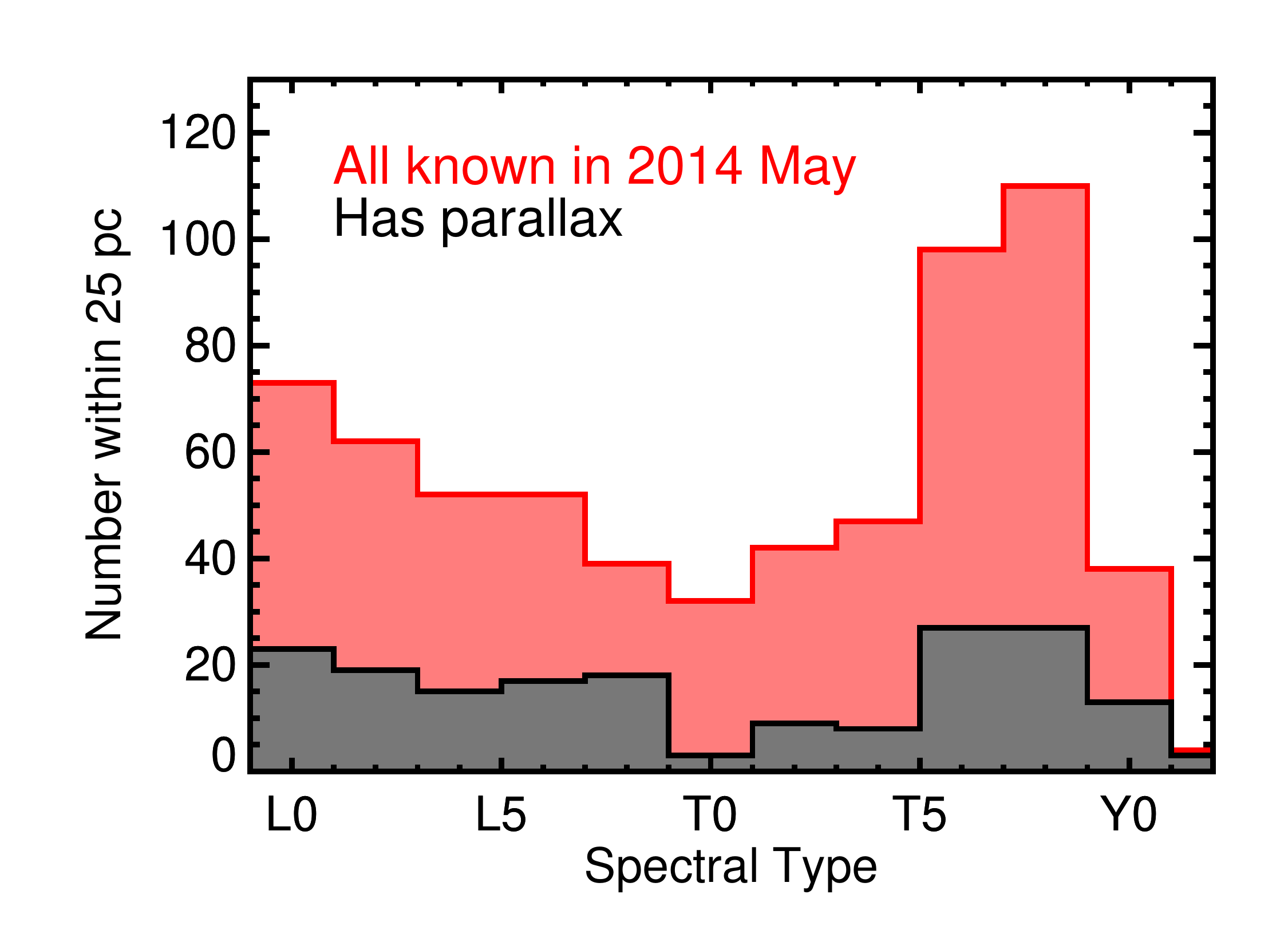}
    \caption{Spectral types of known M9--Y1 dwarfs out to 25~pc, as of 2014 May
      when we began our parallax observations. The gray histogram indicates the
      objects having published parallax measurements at the time. For other
      objects lacking parallax measurements at the time (red), we used
      photometric distances based on \WISE\ $W2$ photometry to select objects
      within 25~pc.  The parallax census in 2014 May had uneven spectral type
      coverage and included only $\approx$1/3 of the known 25~pc sample.}
  \label{fig.spt.dist}
\end{figure}

Our goal is to establish a complete volume-limited sample of L and T~dwarfs
large enough for robust statistical analysis. In order to do this, we require a
sample defined by parallaxes to avoid the pitfalls that come with photometric
distances (e.g., incorrect distances for binaries or spectrally peculiar
objects).  Historically, 25~pc has often been used as the boundary of stellar
volume-limited samples, including the landmark all-sky Gliese catalog
\citep{Gliese:1991uy}, the Palomar-Michigan State University spectroscopic
survey of M dwarfs \citep{Reid:1995kw}, the Nearby Stars project
\citep[NStars;][]{Gray:2003fz}, the RECONS parallax program
\citep[e.g.,][]{Henry:2006jp}, and, most recently, the Ultracool SpeXtroscopic
sample of \citet{BardalezGagliuffi:2019gn}.  Prior to our observations, space
density estimates from \citet{Cruz:2007kb}, \citet{Metchev:2008gx}, and
\citet{DayJones:2013hm} indicated that 25~pc would enclose $\sim$400 L and
T~dwarfs over half the sky, a significant improvement in number and spectral
type coverage over previous volume-limited samples, so we likewise chose 25~pc
as our limiting distance.  Figure~\ref{fig.spt.dist} shows the spectral type
distribution of known M9--Y1~dwarfs within 25~pc at the beginning of our
parallax observations in 2014~May, indicating that the subset of objects having
parallax measurements at the time comprised only $\approx$1/3 of the known 25~pc
sample.

The {decl.} boundaries of our target list come from the observing limits of
UKIRT and PS1. UKIRT cannot observe north of $\delta=+60\degr$. PS1 provided the
reference catalog we used to calibrate our astrometric measurements
(Section~\ref{reduc.measurement}), and it contributes photometry and proper
motions \citep{Best:2018kw}, so we did not observe farther south than the
$\delta=-30\degr$ limit of PS1. These boundaries encompass 68\% of the sky. We
placed no restrictions in {R.A.} on our target list.

We identified all spectroscopically confirmed L and T~dwarfs in the literature
with $-30\degr\le\delta\le60\degr$ that lacked a parallax measurement with
${\rm error}\le10\%$. We selected those objects having photometric distances
based on \WISE\ $W2$ photometry and the spectral type-absolute magnitude
relations of DL12, or low-quality (${\rm errors}>10\%$) trigonometric distances
where available, placing them within 25~pc or at most $1\sigma$ beyond 25~pc
using the measurement errors from the literature. Many of these objects have
spectral types from both optical and NIR spectra; if both were available, we
used the the optical spectral types for L~dwarfs and the NIR spectral types for
T~dwarfs.\footnote{One of our targets, 2MASS~J00464841+0715177, had an optical
  spectral type L0:: \citep{Reid:2008fz} that was updated to M9~$\beta$
  \citep{Faherty:2016fx} during the course of our observations.  We continued to
  observe this object, and we report its parallax in this
  paper.}$^,$\footnote{We included one target, 2MASSW~J0320284$-$044636, which
  has an optical spectral type M8 \citep{Cruz:2003fi} and an NIR spectral type
  L0.5 \citep{Wilson:2003tk}.  This object has been identified as a
  spectroscopic binary \citep{Blake:2008co} with component spectral types
  M8.5+T5: \citep{Burgasser:2008cj}, making it a target of particular interest.}

We used the \vmax\ statistic \citep{Schmidt:1968jc} to estimate that our targets
combined with objects having literature parallaxes (${\rm errors}\le10\%$)
formed a sample $\approx$85\% volume-complete for L0--T6 dwarfs out to 25~pc,
with most of the missing objects lying at low galactic latitudes where few
searches for ultracool dwarfs have ventured. At spectral types later than T6,
the 25~pc sample of known objects (with and without parallaxes) is expected to
be more incomplete due to the limiting magnitude of \WISE\ and the limited sky
coverage ($\approx$4000~deg$^2$) of UKIDSS, which have discovered most of the
known T6+ dwarfs. Most targets become prohibitively faint for NIR astrometry
from the ground beyond spectral type T8. We decided to retain T6.5--T8 targets
brighter than $J=19.2$~mag (Section~\ref{obs.strategy}), and we discarded
fainter and later-type targets. Fainter targets with spectral types $\ge$T7 are
being observed to greater depth for parallaxes with CFHT
\citep[e.g.,][]{Liu:2011hb}, Magellan \citep[e.g.,][]{Tinney:2014bl}, and
\spitzer\ \citep[e.g.,][K19]{Dupuy:2013ks,Beichman:2014jr}, and those results
can be combined with ours to form a complete volume-limited sample out
$\approx$20~pc for the latest-type objects.

Our target volume included 97 L0--T8 objects with previously published
parallaxes (${\rm errors}\le10\%$) placing them within $25+1\sigma$~pc. For
comparison purposes, we retained \varnlitstart\ of those parallax-confirmed
objects in our target list, including seven companions to stars with Hipparcos
parallaxes. We eliminated most targets already being observed as part of HIPP at
CFHT, but we retained 84 targets known to be on the target lists for other
ultracool parallax programs, including PARSEC, NPARSEC, USNO, and four from
HIPP/CFHT.

Our final target list contained \varntargets\ L0--T8~dwarfs.  Parallaxes with
${\rm errors}\le10\%$ were published by others for \varnlitnew\ of our targets
after our program began, and we continued observations for those targets.

\section{Observations}
\label{obs}

\subsection{UKIRT/WFCAM}
\label{obs.wfcam}
We used the near-infrared Wide Field Camera \citep[WFCAM;][]{Casali:2007ep} on
UKIRT, the same instrument and telescope that produced the UKIRT Infrared Deep
Sky Survey \citep[UKIDSS;][]{Lawrence:2007hu} and the UKIRT Hemisphere Survey
\citep[UHS;][]{Dye:2018jo}. \citet{Marocco:2010cj},
\citet{Smart:2010gd,Smart:2017hr}, and K19 have previously used WFCAM to measure
33 parallaxes of mid-T through Y~dwarfs with precisions of $\approx$2--5~mas,
demonstrating the potential of WFCAM for parallaxes.

WFCAM consists of four $2048\times2048$ Rockwell Hawaii-II (HgCdTe) infrared
arrays, each with a field of view of $13.\!'65 \times 13.\!'65$, arranged in a
$2\times2$ grid and spaced by $12.\!'83$. WFCAM has a pixel scale of
$0.\!''4$~pixel$^{-1}$. Based on advice from UKIRT staff and challenges with
obtaining precise astrometric solutions across wide fields on separated arrays,
we only used data from WFCAM's Camera 3 (northeast array).

All UKIRT observations were performed in service (queue) mode, which allowed us
to obtain good parallax phase coverage by observing every object at different
times of the year (when visible) while minimizing disruptions from poor weather.
Service mode also facilitated repeating observations at similar airmasses, which
enabled us to reduce the impact of differential chromatic refraction (DCR) to an
insignificant level (Sections \ref{obs.strategy} and~\ref{plx.dcr}). In
practice, irregular scheduling due to instrument changes on UKIRT and a
nonintegrated observing queue (programs from specific institutions had priority
on specific nights) meant that some {R.A.} ranges received better coverage in
different years.

As part of the instrument changes, WFCAM was typically removed from the
telescope a few times a year. \citet{Smart:2010gd} investigated the impact on
astrometry of these remove-replace cycles and found no systematic effects larger
than $\approx$10~mas, which is consistent with the astrometric precision per
epoch that we achieve with our measurements (Section~\ref{reduc.errors}).

\begin{figure}
    \includegraphics[width=1\columnwidth]{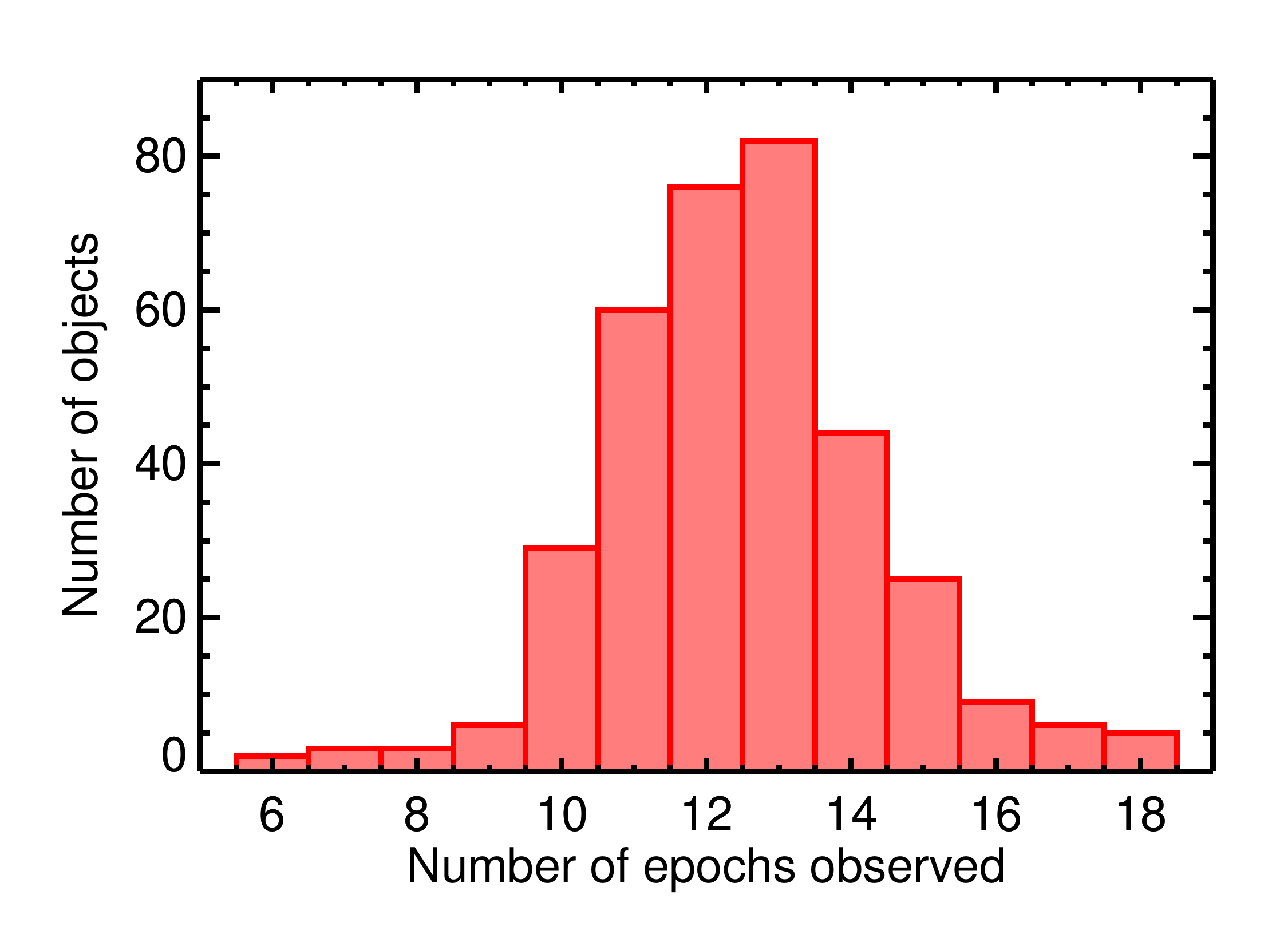}
    \includegraphics[width=1\columnwidth]{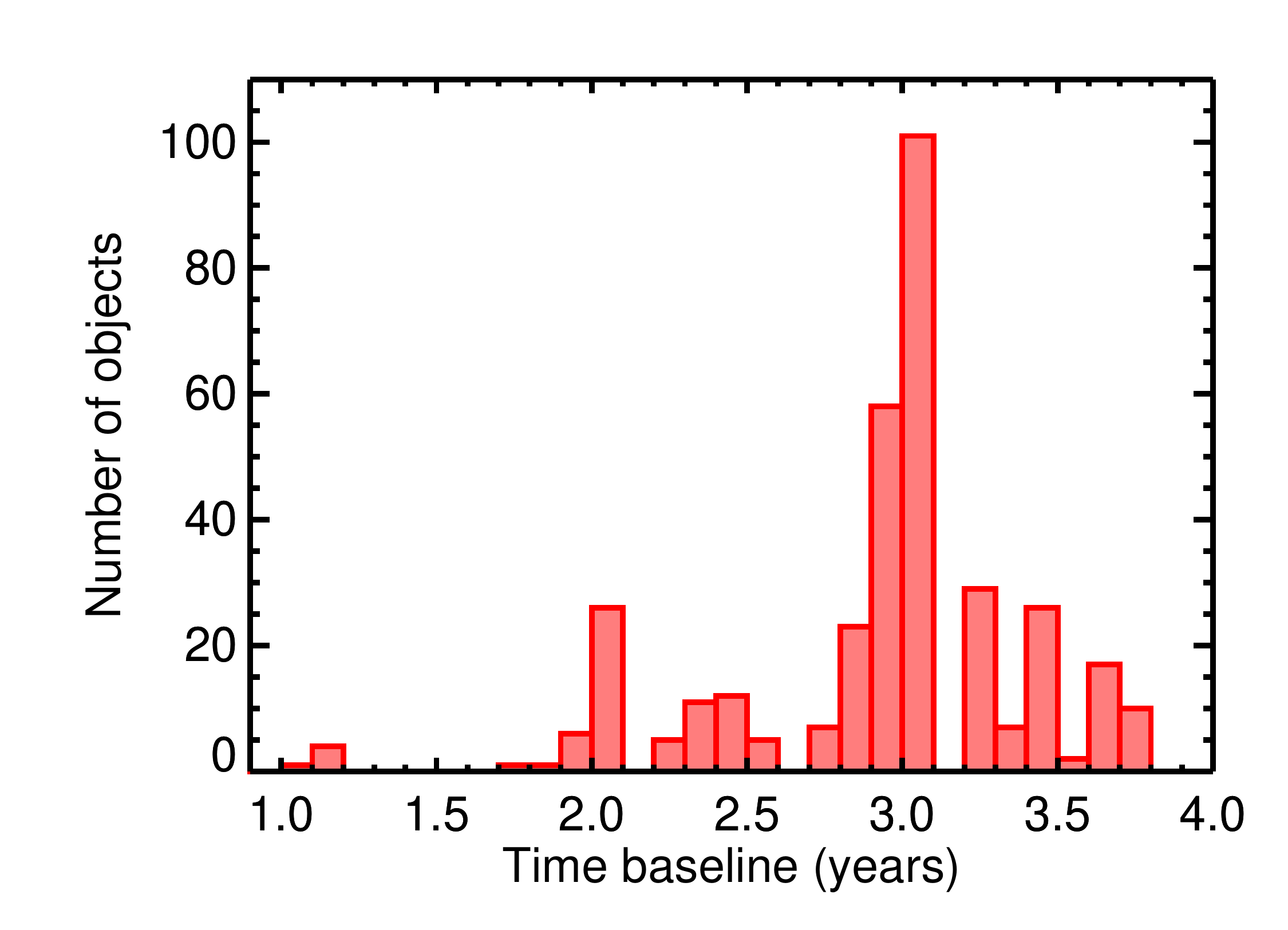}
    \caption{Distributions of the number of epochs ({\it top panel}) and time
      baseline ({\it bottom panel}) for observations of each target. 97\% of
      targets have at least eight observations spanning at least 1.7~yr,
      enabling robust parallax measurements.}
  \label{fig.epochs.dyr}
\end{figure}

The observations presented in this paper span \hbox{2014 May 5 UT} through
\hbox{2018 January 16 UT}. Details of our observations are given in
Table~\ref{tbl.obs}. Figure~\ref{fig.epochs.dyr} shows the number of observed
epochs and time baselines for our targets.

\begin{longrotatetable}
\begin{deluxetable*}{lcCCccccccccCc}
\centering
\tablecaption{Observations \label{tbl.obs}}
\tabletypesize{\scriptsize}
\tablewidth{0pt}
\rotate
\tablehead{   
  \colhead{Object} &
  \colhead{SpT} &
  \colhead{\jmko} &
  \colhead{FWHM} &
  \colhead{$\Delta$airmass} &
  \colhead{$t_{\rm int}\tablenotemark{a}$} &
  \colhead{$N_{\rm fr}$} &
  \colhead{$N_{\rm ep}$} &
  \colhead{$\Delta t$} &
  \colhead{$N_{\rm ref}$} &
  \colhead{$N_{\rm cal}$} &
  \colhead{$N_{\rm abs}$} &
  \colhead{$\plx_{\rm abs} - \plx_{\rm rel}$} &
  \colhead{References} \\
  \colhead{} &
  \colhead{Optical/NIR} &
  \colhead{(mag)} &
  \colhead{($''$)} &
  \colhead{} &
  \colhead{(s)} &
  \colhead{} &
  \colhead{} &
  \colhead{(yr)} &
  \colhead{} &
  \colhead{} &
  \colhead{} &
  \colhead{(mas)} &
  \colhead{(Disc; SpT)}
}
\startdata
SDSS J000112.18+153535.5 & .../L3.7 \intg & $15.46\pm0.02$ & $0.79\pm0.17$ & 0.071 & 1, 5 & 126 & 15 & 3.4 & 43 & 29 & 19 & $1.6\pm0.8$ & 49; 33 \\
PSO J001.1277$-$26.0675 & .../T4 & $16.18\pm0.02$ & $1.04\pm0.19$ & 0.019 & 4, 5 & 86 & 11 & 2.4 & 67 & 41 & 20 & $2.7\pm0.7$ & 53; 53 \\
WISEPC J000849.76$-$173922.6 & .../T6 & $16.85\pm0.02$ & $0.97\pm0.17$ & 0.017 & 10 & 92 & 11 & 2.4 & 80 & 62 & 25 & $3.2\pm0.5$ & 47; 47 \\
PSO J003.3437$-$11.7168 & .../T4 & $16.05\pm0.02$ & $0.86\pm0.15$ & 0.014 & 4, 5 & 108 & 12 & 2.9 & 71 & 50 & 22 & $1.7\pm0.4$ & 53; 53 \\
2MASS J00150206+2959323 & .../L7.5pec & $16.15\pm0.02$ & $0.92\pm0.35$ & 0.041 & 4, 5 & 120 & 14 & 3.4 & 158 & 126 & 67 & $1.2\pm0.2$ & 46; 46 \\
2MASSW J0015447+351603 & L2/L1.0 & $13.74\pm0.02$ & $0.78\pm0.14$ & 0.058 & 1, 5 & 105 & 14 & 3.1 & 97 & 76 & 61 & $0.9\pm0.3$ & 44; 44, 4 \\
PSO J004.1834+23.0741 & .../T0 & $16.50\pm0.02$ & $0.95\pm0.33$ & 0.043 & 4 & 195 & 11 & 3.1 & 118 & 87 & 40 & $1.2\pm0.1$ & 6; 6 \\
PSO J004.6359+56.8370 & .../T4.5 & $16.22\pm0.02$ & $0.90\pm0.15$ & 0.156 & 4, 5 & 97 & 11 & 2.5 & 1214 & 1165 & 528 & $0.1\pm0.1$ & 53; 53 \\
\enddata
\end{deluxetable*}
\end{longrotatetable}

\subsection{Observing Strategy}
\label{obs.strategy}
Parallaxes require multiple epochs of observations, ideally spread over many
months to maximally sample the parallax ellipse of an object. Our initial plan
was to observe each target 10 times over two years but not more than five times
within any UKIRT observing semester (February-July or August-January), to ensure
broad coverage of parallax ellipses while extending the time baselines long
enough to robustly distinguish parallactic motions from proper motions. The
actual completion rate per semester for our program was typically $\approx$70\%
because of queue competition, instrument changes, and stretches of poor
weather. We therefore needed to extend our observations beyond two years.

We conducted all observations in $J$-band where L and T~dwarfs are brightest
relative to the sky background, and copious (bluer) background stars are
available to establish the local astrometric reference frame. Stars and brown
dwarfs have different $J$-band spectra, so images of these objects in the same
target field taken at different airmasses will suffer from different amounts of
DCR. This systematic effect can be minimized by observing each target within a
narrow range of airmass around transit. DL12 estimate that keeping all
observations within an airmass range of 0.07 will limit the effect of DCR to
$\lesssim$1~mas for T~dwarfs, with an even smaller effect for L~dwarfs. A
0.07~airmass range is equivalent to $\pm1$~hour of the meridian for a target at
$\delta=-30\degr$ observed from Maunakea, marking the lowest transit elevation
(40\degr) for our target list. We therefore requested that all of our targets be
observed within $\pm1$~hour of transit. In practice, nearly all of our targets
were observed at higher elevations and over smaller ranges of airmasses, making
DCR insignificant for our parallaxes (Section~\ref{plx.dcr}).

For a typical observation, at each epoch, we obtained nine frames using UKIRT's
3$\times$3 large microstepping sequence. Microstepping uses dithers
corresponding to fractions of a pixel on the detector (in this case, $11+1/3$
pixels) along each axis, which can then be interleaved into single ``leavstack''
images with nominally higher angular resolution. (We ultimately chose to use the
individual microstepped frames rather than the leavstacks for our analysis;
Section~\ref{reduc.casu}.) For each observation of a target, we used the same
guide star and autoguider set-up in order to minimize systematic errors,
although for 10 objects, we were forced to switch to a brighter guide star after
the first epoch following difficulties observing with our initial choice of
guide star. Whenever possible, we placed the base position of the microstep
pattern at the center of Camera 3. For 38\% of our targets, we needed to use a
different base position in order to have a sufficiently bright guide star.

\begin{figure*}
  \includegraphics[width=1.08\columnwidth]{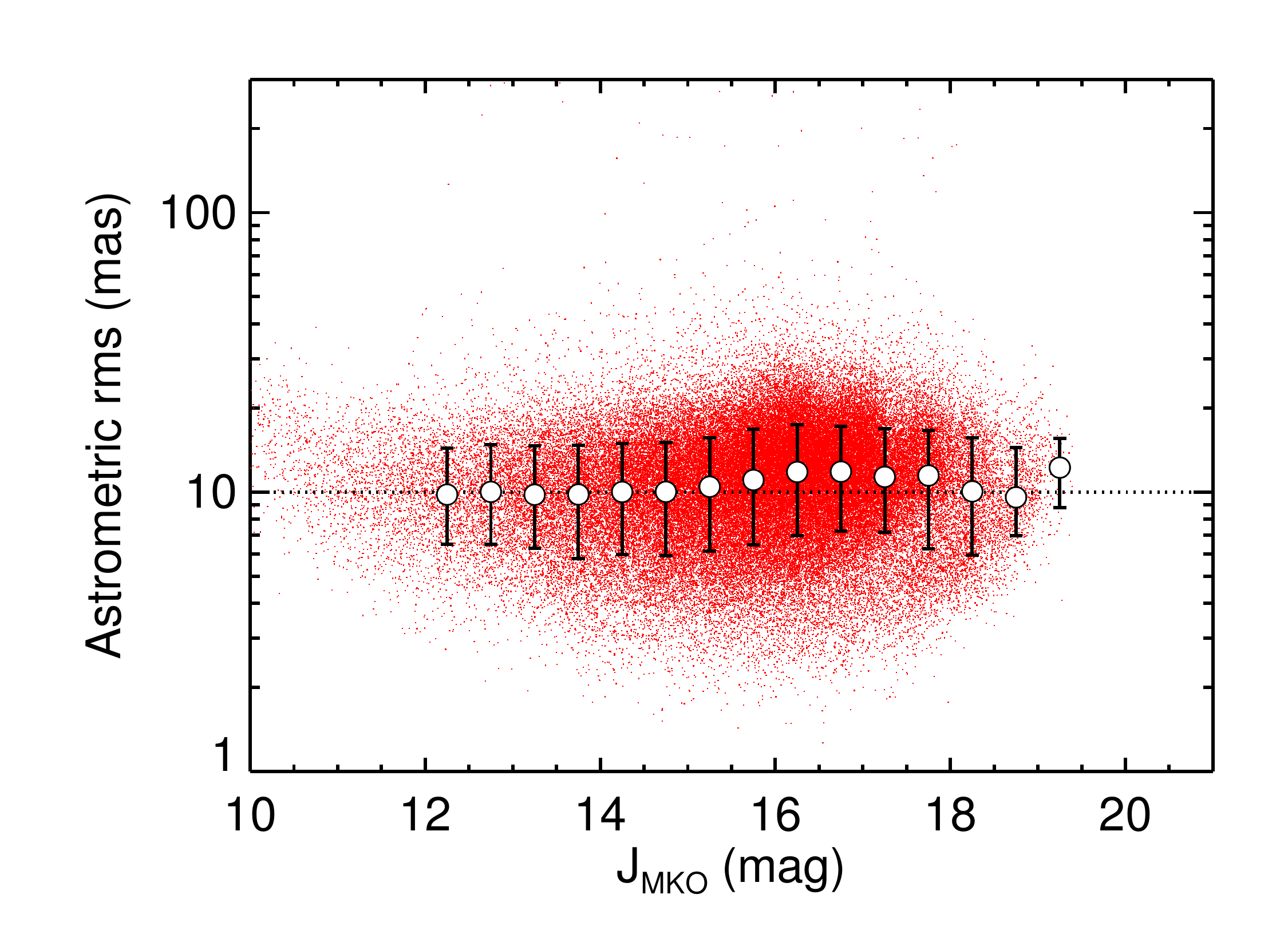}
  \hfill
  \includegraphics[width=1.08\columnwidth]{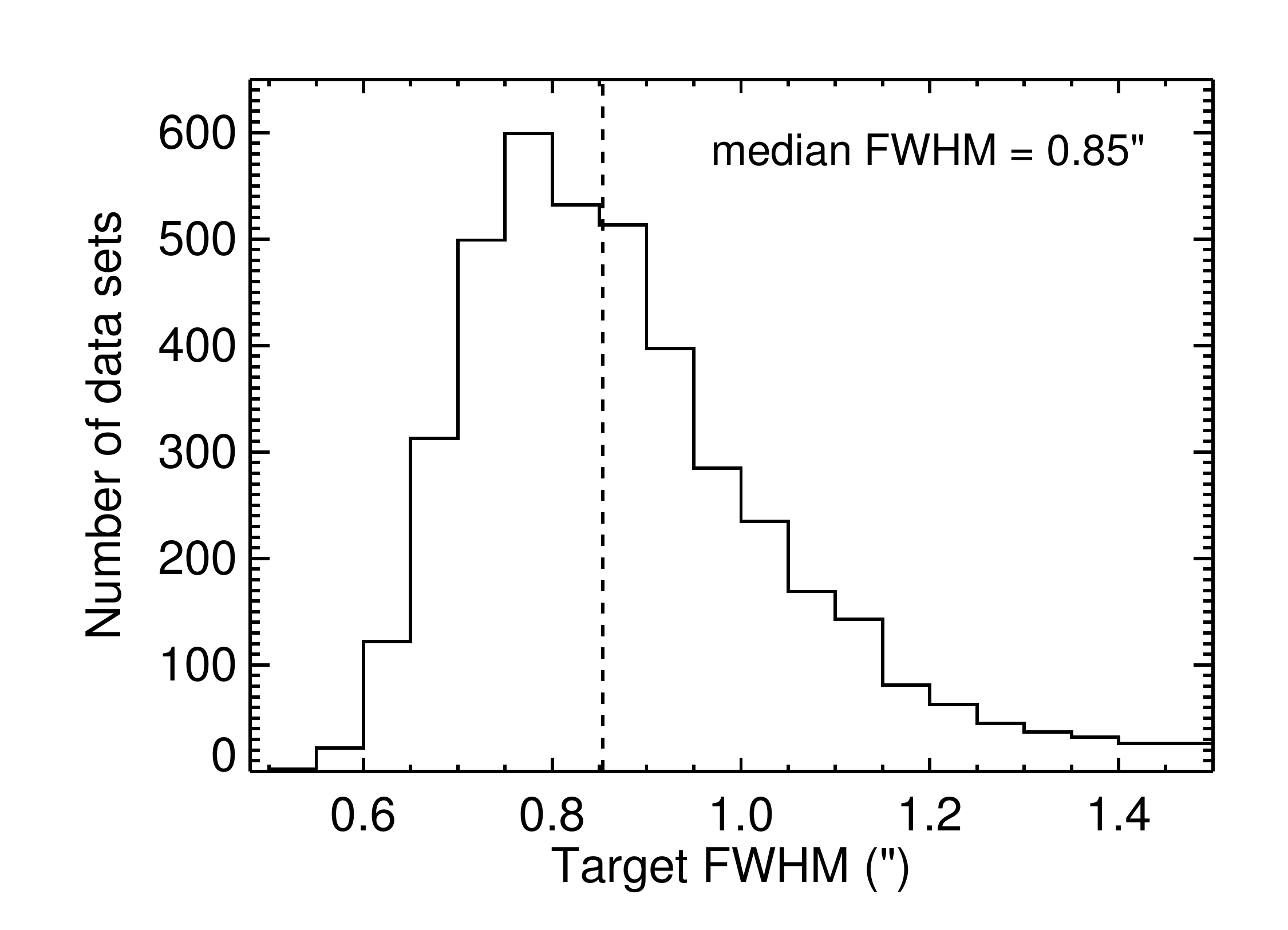}
  \includegraphics[width=1.08\columnwidth]{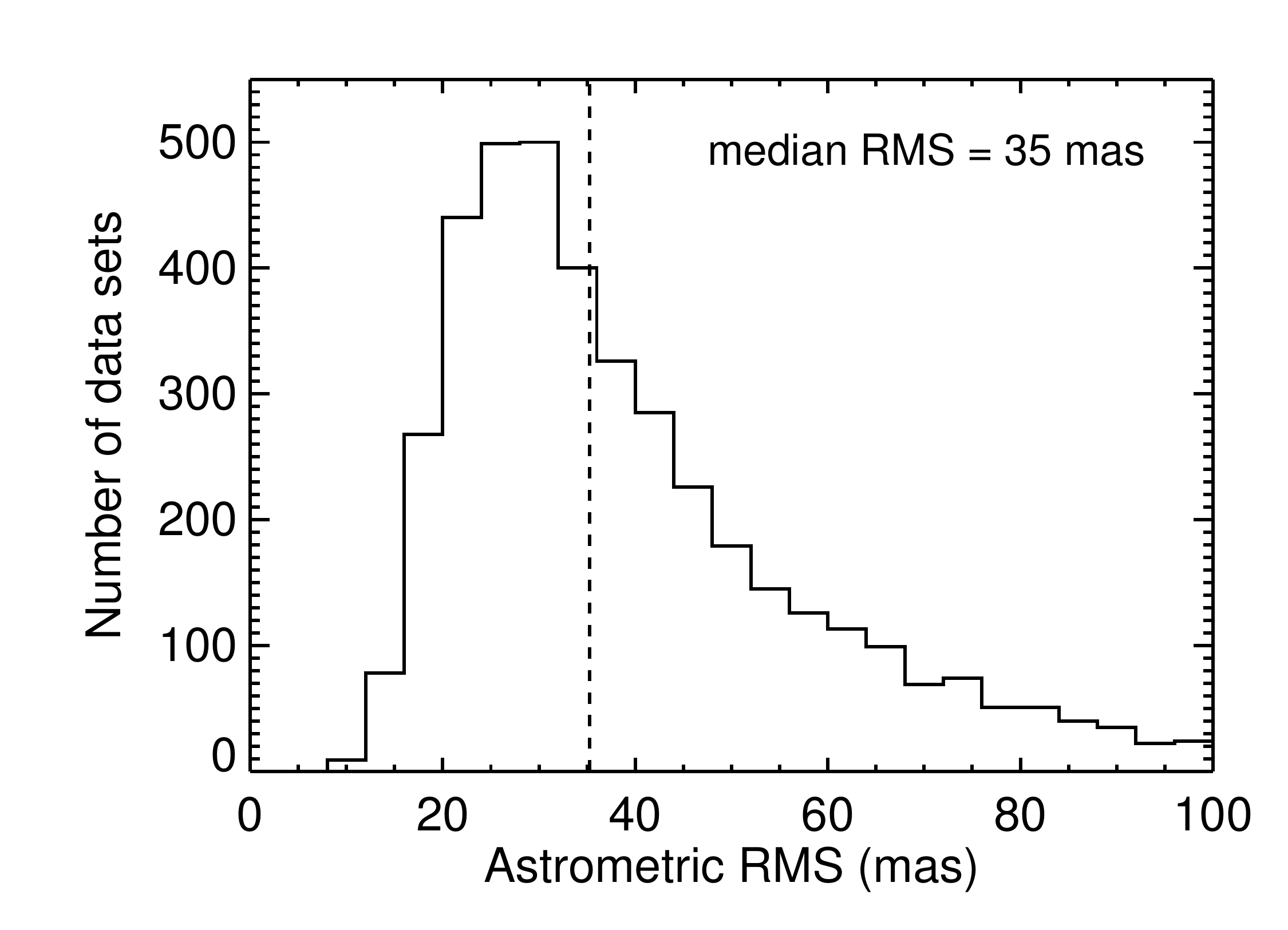}
  \hfill
  \includegraphics[width=1.08\columnwidth]{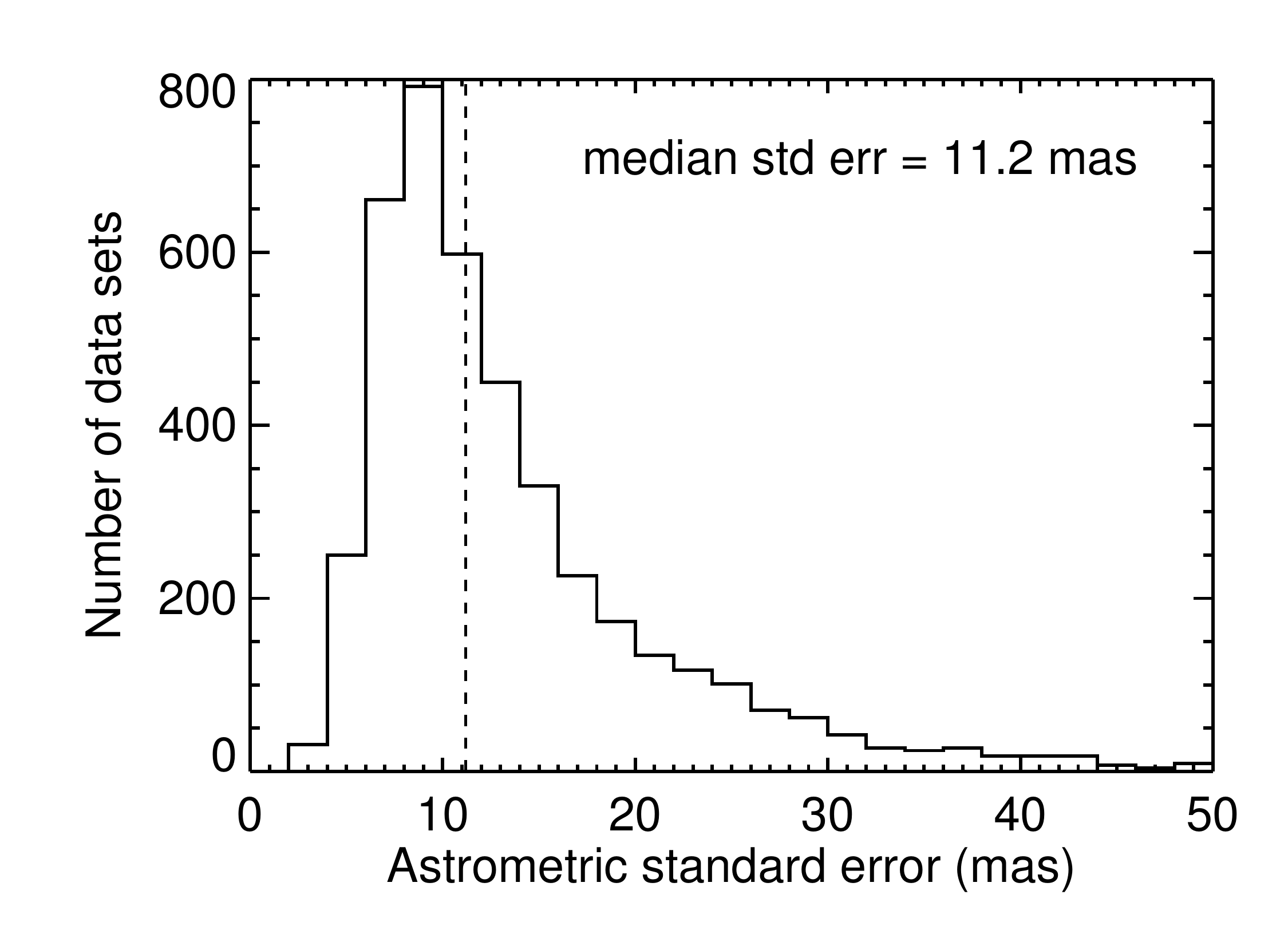}
  \caption{{\it Top left panel}: rms of the astrometric measurements of all
    detections matched across individual frames at each epoch, plotted as a
    function of \jmko. The white circles indicate the median and 68\% confidence
    limits for the rms in bins of 0.5~mag. The rms increases slightly with
    \jmko\ but overall is fairly flat as a function of magnitude, as expected,
    given that we increase the exposure times for fainter targets. For brighter
    detections ($<$15~mag), the rms is impacted by a floor of $\approx$15~mas in
    the short ($<$5~s) exposures taken before we established the minimum 5~s
    exposure time (Section~\ref{obs.strategy}). {\it Top right panel}:
    distribution of the FWHM of our parallax targets at each epoch of
    observation.  {\it Bottom panels}: distributions of the per-epoch
    astrometric rms ({\it left}) and standard error ({\it right}) for our
    parallax targets.  The standard errors, which we adopt as our final
    astrometric uncertainties, are consistent with our goal of $\lesssim$12~mas
    astrometric precision per epoch.}
  \label{fig.asterr}
\end{figure*}

We chose integration times long enough to achieve our goal of parallax
errors~$\le$10\% while sufficiently short to make observing our entire target
list feasible. An object at the 25~pc boundary of our sample has a parallax of
40~mas, requiring parallaxes with errors~$\le$4~mas to achieve 10\% or better
precision. With 10 planned observations per target, we required astrometric
uncertainties of~$\lesssim$12~mas per epoch. Approximating astrometric precision
as seeing divided by photometric signal-to-noise ratio (S/N), and assuming
$0.8"$~seeing (typical for UKIRT), we required a
\hbox{${\rm photometric\ S/N}\gtrsim\frac{0.8"}{12~{\rm mas}} = 67$} per epoch.
We used the WFCAM Integration Time Calculator and $J$-band photometry for our
targets from the literature ($J=12.0-19.2$~mag) to determine the individual
exposure times needed to obtain ${\rm S/N}=40$, expected to result in a total
${\rm S/N}=120$ for each epoch's nine-point microstepping sequence. The
individual exposure times ranged from 1 to 300~s, with the faint limit of our
target list ($J=19.2$~mag) determined by the ${\rm S/N}=40$ minimum for 300~s
exposures.  In addition, we required that all observations be taken with
${\rm seeing}\le1''$ and in clear or slightly cloudy skies ($\le$0.5~mag of
extinction) to ensure that we would detect our targets and ample reference stars
in all exposures. The median seeing for our observations was $0.85''$, as
measured by the FWHM of our parallax targets (Figure~\ref{fig.asterr}). Most of
our chosen integration times were standard for WFCAM, in the sense that we could
use regularly updated facility dark frames with those integration times for
calibration. For objects requiring nonstandard integration times, we repeated
the 3$\times$3 microstepping sequence with shorter standard integrations to
obtain the desired total integration times.

After the first year of observations, we assessed the astrometric precision of
our images and determined that we were reaching our target of $\lesssim$12~mas
per epoch only for exposure ${\rm times}\gtrsim$4~sec. Therefore, in 2016
February, we raised the minimum integration time for all targets to 5~s in order
to obtain more background stars to use as astrometric references and to better
average the astrometric jitter caused by Earth's turbulent atmosphere.

\section{Data Reduction}
\label{reduc}
We reduced our data using a modified version of our custom IDL astrometry
pipeline developed for WIRCam on CFHT, explained in detail in DL12 and
LDA16. Here, we summarize the components of the pipeline and describe our
modifications for UKIRT/WFCAM data.

\subsection{CASU pipeline and leavstacks}
\label{reduc.casu}
All data were initially reduced by the WFCAM Cambridge Astronomical Survey Unit
(CASU) pipeline \citep{Irwin:2004ej,Lawrence:2007hu}, which performs dark
subtraction, flat-fielding, and astrometric and photometric calibration. Our
nine-point microstepped frames were then interleaved at CASU to produce
leavstack images by reversing the dither offsets to align the frames and placing
the images on an interwoven grid such that each 0.4'' pixel was represented by a
0.133'' pixel centered at the same position. We retrieved the CASU-reduced
leavstack images and ``flat files'' (catalogs of detections) from the WFCAM
Science Archive \citep{Hambly:2008kr}. For objects with more than one nine-point
sequence of frames at each epoch, the resulting leavstacks were not themselves
stacked by the CASU pipeline, and in these cases, we received more than one
leavstack per epoch.

Visual inspection of the leavstack images revealed several problems. In many
leavstacks, the point sources were significantly oversampled with large
variations in pixel-to-pixel brightness (a known issue that can be addressed by
drizzling-type procedures). Many point sources contained brighter pixels in
every-third-pixel grids that were offset from the core of the point-spread
function (PSF). We also noticed trails of bright pixels extending from point
sources in some leavstacks, again in every-third-pixel grids. We learned that
the CASU interleaving process performs no alignment of individual objects or
removal of outlier pixels, so the success of the interleaving depends entirely
on the accuracy of the WFCAM guider and microstepper, which appeared to have
been inaccurate for some of our observations. (CASU performs outlier removal on
leavstacks taken at different points of larger dither patterns, but our
observations did not include such dither sequences.) In addition, the
astrometric measurement errors of sources reported in the flat files were
implausibly small given the size and appearance of the PSFs in the images. We
therefore decided to work directly with the original CASU-detrended images
rather than the leavstacks and flat files.

For each leavstack, we reconstructed the nine original microstepped frames by
extracting every-third-pixel grids from the leavstack. In other words, for
$n\!\in\!\{0,1,2,\ldots,2047\}$, we extracted pixels $(3n,3n)$ to form the first
image, pixels $(3n+1,3n)$ to form the second image, etc. For observations
producing $n>1$~leavstacks at each epoch, we reconstructed $n\times9$~individual
frames per epoch. For each image, we modified the World Coordinate System (WCS)
astrometry keywords from the leavstack and flat file FITS headers to compensate
for the change in pixel scale from the leavstacks to the individual frames and
for the microstepping offsets.

\subsection{Astrometric Measurements}
\label{reduc.measurement}
As described in DL12, we extracted $(x,y)$ position measurements from each
individual frame using \hbox{Source Extractor v2.19.5} \citep{Bertin:1996hf},
and began the process of associating detections across multiple frames at each
epoch by first cross-matching the detections with an external reference
catalog. We used PS1 Data Release 1 as our reference catalog because its
observations (2010 May -- 2014 March) preceded ours by only a few years, it is
deeper and/or more precise than other all-sky catalogs including 2MASS, AllWISE,
UCAC5 \citep{Zacharias:2017bk}, and \gaiat, and its astrometry is tied to the
\gaiao\ absolute reference frame
\citep{GaiaCollaboration:2016gd,Lindegren:2016gr}. We accessed PS1 via the {\it
  VizieR} catalog service using the IDL routine {\tt queryvizier.pro}.

\subsubsection{Astrometric Projection}
\label{reduc.projection}
In order to cross-match with PS1, we needed \radec\ coordinates for the
detections in each frame. The WFCAM FITS headers provide initial astrometric
solutions that convert the $(x,y)$ positions into \radec\ coordinates of each
detected object. The solutions includes standard WCS linear terms along with a
single cubic term for WFCAM's distortion, which is $\approx$10$''$ at the outer
edges of the cameras. As described in DL12, our pipeline uses the catalog
matching software \hbox{SCAMP v1.4.4} \citep{Bertin:2006vk} to improve the
astrometric solution for each image and thereby foster better matching of
detections across dithers (Section~\ref{reduc.registration}). However, the
astrometric projection used by the WFCAM image headers (zenithal polynomial
projection, or ZPN) is incompatible with SCAMP as well as other stages of our
pipeline. In addition, we found that the reference pixel in WFCAM header
astrometry changed by as much as 75~pixels ($30''$) between some epochs for most
targets, making it impossible to create a single astrometric solution uniting
multiple observations of a target without modifying the reference pixel and
other WCS keywords.

We resolved both of these problems by establishing a single reference pixel for
all frames at (1024, 1024), the center of WFCAM Camera 3, and calculating a new
tangential (TAN) projection for each frame as the starting point for our
astrometric solutions. We used the original ZPN astrometry to calculate the
\radec\ at (1024, 1024), discarded the ZPN astrometric solution, and used SCAMP
to calculate a TAN projection with additional higher-order terms by matching our
detections to the 2MASS catalog, because the PS1 catalog was not available in
SCAMP.

\subsubsection{Registration of Dithers}
\label{reduc.registration}
The SCAMP solutions gave us \radec\ coordinates that were sufficiently accurate
to cross-match $>$90\% of our detections in each image with PS1 using a $3''$
matching radius.  We then calculated more precise astrometric solutions for each
frame as described in DL12, associating detections lying within $1''$ of each
other across multiple frames at a single epoch. We discarded objects detected in
fewer than half of the frames.

We optimally registered the remaining cross-identified objects as in DL12.
After obtaining an initial registration solution, we used the IDL routine {\tt
  robust\_sigma.pro} to clip positional measurements more than 3$\sigma$
discrepant with the mean position in order to eliminate inaccurate measurements
or artifacts incorrectly associated with detected objects. After robust
clipping, we recalculated a final registration solution. We note that we did not
use the Source Extractor-reported positional uncertainties for individual
detections to weight the solution.

\subsubsection{Distortion Solution for WFCAM}
\label{reduc.distortion}
\begin{figure*}
    \includegraphics[width=2\columnwidth]{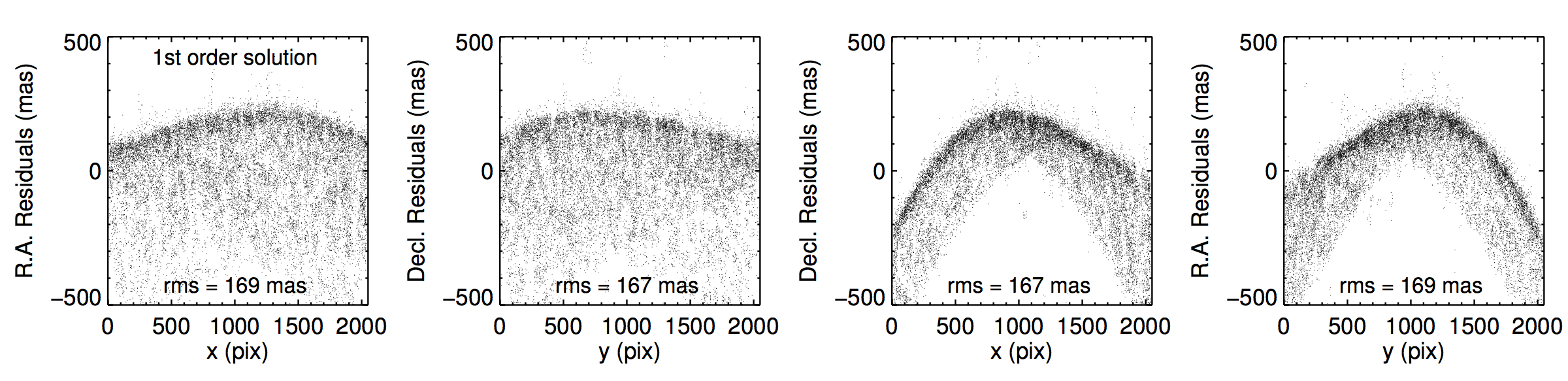}
    \includegraphics[width=2\columnwidth]{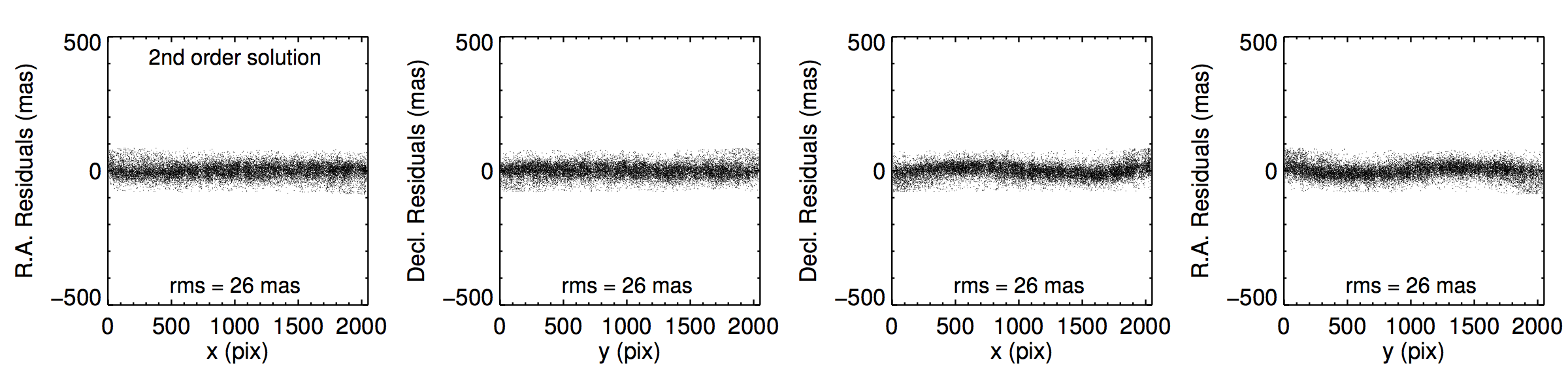}
    \includegraphics[width=2\columnwidth]{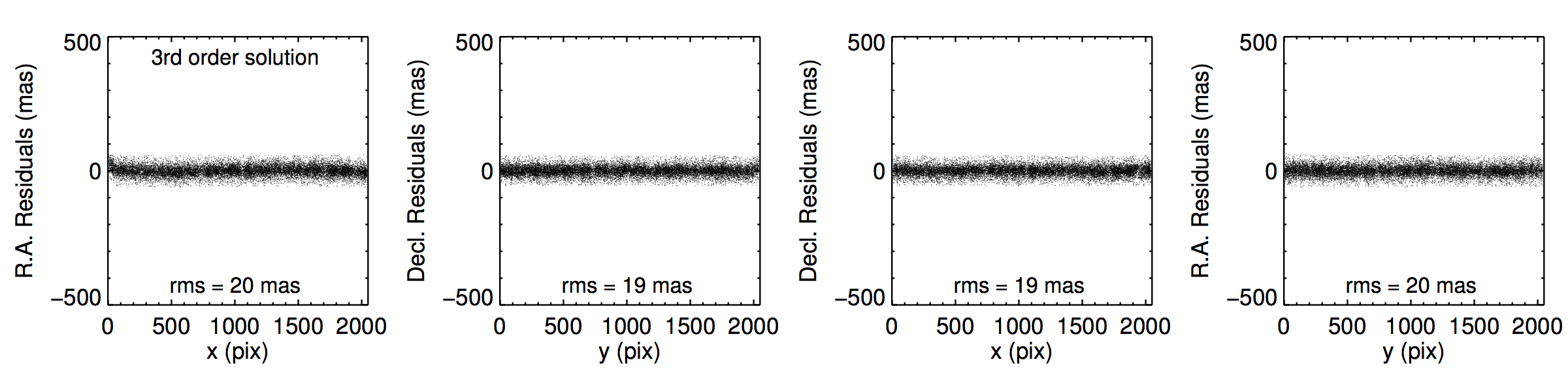}
    \caption{Residuals of measured star positions on WFCAM Camera 3 compared
      with the PS1 catalog after applying linear and higher-order distortion
      corrections for WFCAM, as a function of $x$ and $y$ positions.  These
      plots show residuals for $\approx$2100 stars in images targeting
      WISE~J174640.78$-$033818.0 observed overnine9 dithers with offsets spanning
      9.1'' along each axis.  Second- and third-order terms both reduce the rms
      of the residuals.  There is no obvious structure in the third-order
      residuals, and higher-order solutions do not lower the rms, so our
      third-order solution is sufficient for WFCAM.}
  \label{fig.distort}
\end{figure*}

DL12 corrected their astrometry for distortion in CFHT/WIRCam using a
third-order polynomial function in $x$ and $y$ applied to all frames. Their
polynomial was calculated by fitting multiple dithers of a dense star field to
the SDSS DR7 catalog. We used the same approach to correct for distortion in
WFCAM Camera 3. We calculated a polynomial solution by fitting one epoch of
observations (nine dithers, from 2015 July 25 UT) of WISE~J174640.78$-$033818.0
($\approx$2500~detections per frame) to the PS1
catalog. Figure~\ref{fig.distort} shows the residuals of our fits using first-,
second-, and third-order polynomials, clearly indicating that third-order terms
are needed to capture the distortion. We also fit fourth- and fifth-order
polynomials but found no improvement: these higher-order fits have rms residuals
of 19--21~mas, the same as the third-order fit.  We therefore adopted our
third-order polynomial (presented in Table~\ref{tbl.distort}) as the distortion
solution for WFCAM Camera 3, and we applied this correction to our
cross-identified detections at each epoch prior to optimal registration
(Section~\ref{reduc.registration})\footnote{We originally calculated this
  distortion solution before the more precise reference frame of \gaiat\ was
  available.  Following its release, we also calculated distortion solutions
  from the same images using \gaiat\ as a reference catalog, and we found that
  the {\gaia}-based solutions had rms $\approx$3~mas smaller than the PS1-based
  solutions for polynomial orders 2, 3, 4, and~5.  As with the PS1-based
  distortion solutions, the rms did not improve beyond the third order for the
  {\gaia}-based solutions.  We also re-ran 48 objects through our parallax
  pipeline using the third-order {\gaia}-based distortion solution, and we found
  no significant difference in the final parallaxes and their uncertainties
  compared with our results using the PS1-based solution: the mean difference in
  the parallaxes was $0.012\pm0.021$~mas (quoting the standard error on the
  mean), and the mean difference in the uncertainties was $0.007\pm0.010$~mas.
  We therefore opted to retain our original PS1-based distortion solution.}.

\begin{figure}
    \includegraphics[width=1\columnwidth]{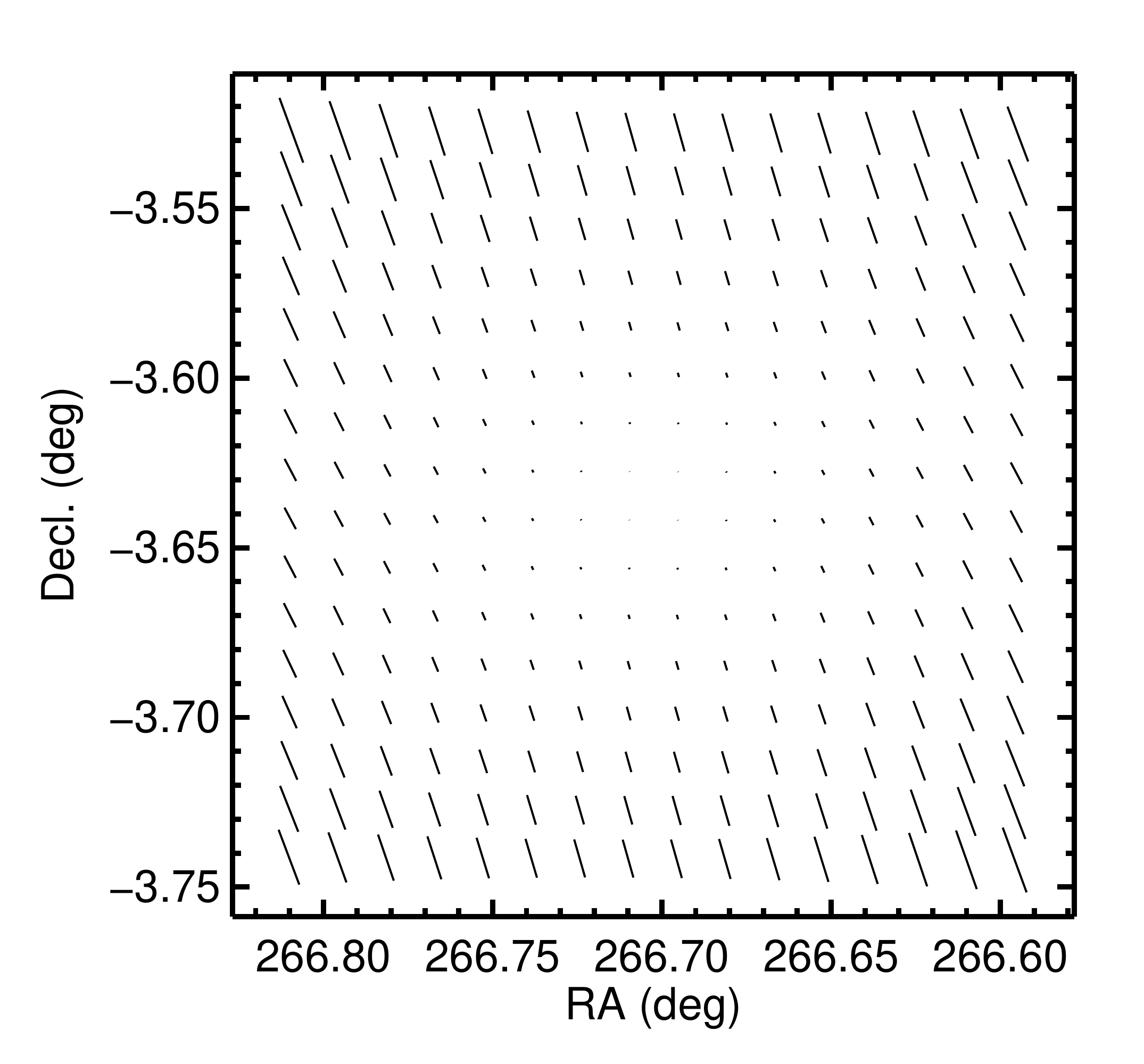}
    \caption{Map showing the difference in positions of a uniform grid of
      $(x,y)$ points after the application of two different third-order distortion
      solutions, which are based on WFCAM images of the target
      WISE~J174640.78$-$033818.0 taken at different epochs (2015 July 25 UT and
      2016 March 18 UT).  The differences are presented following conversion to
      \radec.  The positional differences have been multiplied by 3600 so that
      their sizes in arcsec may be read directly from the axes (which have
      degree scales).  The largest differences (at the corners) are
      $\approx$20~mas, i.e., the same size as the rms in the individual
      distortion solutions, so the solutions calculated from different epochs
      are statistically equivalent.}
  \label{fig.distort.epochs}
\end{figure}

We also investigated the temporal stability of our distortion solution by
calculating third-order solutions from observations of the same object
(WISE~J174640.78$-$033818) at five other epochs, spanning 26~months encompassing
the original 2015 July 25 UT epoch and including two instances when WFCAM was
removed from the telescope and then remounted.  The rms for these other
third-order distortion solutions was 20--26~mas in all but one case, similar to
the 19--20~mas for our adopted solution.  (The latest-date solution, from 2017
July 6 UT, had rms 34--35~mas.)  We tested the impact of using different
distortion solutions by applying each of them to a uniform grid of $(x,y)$
points and converting the corrected points to \radec\ coordinates centered on
WISE~J174640.78$-$033818.  Figure~\ref{fig.distort.epochs} shows a
representative example comparing the original 2015 July 25 UT epoch and 2016
March 18 UT, between which WFCAM was removed from UKIRT and remounted.  The
largest differences in \radec\ between different distortion solutions (at the
corners of the grid) were $\approx$20~mas in almost all cases. As the
$\lesssim$20~mas offsets are no larger than the rms of the individual distortion
solutions, we conclude that a single distortion solution is representative for
all epochs in our program.  We also recalculated parallaxes for 49 of our
targets using the distortion solutions from these five other epochs and, again,
found no significant variation: typically, parallaxes and uncertainties changed
by less than 0.05~mas (i.e., a few percent of the uncertainty).  We conclude
that the distortion solution is stable enough across epochs, including
instrument changes, for our pipeline to compensate for any minor variations.

We also corrected for differential aberration and refraction caused by the
Earth's atmosphere in the same manner as DL12 and DL17.

\begin{deluxetable}{lCcC}
\tablecaption{Distortion Coefficients for WFCAM Northeast Array (Camera 3) \label{tbl.distort}}
\tabletypesize{\small}
\tablewidth{0pt}
\tablehead{   
  \colhead{Term\phm{xxxx}} &
  \colhead{$a_{ij}$} &
  \colhead{\phm{xx}} &
  \colhead{$b_{ij}$}
}
\startdata
$x^2$ & \phm{$-$}1.486\times10^{-7\phn} &  & \phm{$-$}4.832\times10^{-7\phn} \\
$xy$ & -2.643\times10^{-7\phn} &  & -2.766\times10^{-7\phn} \\
$y^2$ & \phm{$-$}5.007\times10^{-7\phn} &  & \phm{$-$}1.321\times10^{-7\phn} \\
$x^3$ & -6.436\times10^{-14} &  & -6.496\times10^{-11} \\
$x^2y$ & \phm{$-$}5.640\times10^{-11} &  & -6.172\times10^{-11} \\
$xy^2$ & \phm{$-$}6.094\times10^{-11} &  & -4.511\times10^{-11} \\
$y^3$ & \phm{$-$}6.589\times10^{-11} &  & -2.088\times10^{-11} \\
\enddata
\tablecomments{
  To use this distortion correction, first define the origin as the central
  pixel (1024,1024):
  \begin{equation*}
    x'=x-1024 \hspace{20pt} y'=y-1024
  \end{equation*}
  where $x$ and $y$ are the pixel positions for a
  detection. Distortion-corrected positions may then be computed as
  \begin{align*}
    x'' &= x'+a_{20}x'^2 + a_{11}x'y' + a_{02}y'^2 + a_{30}x'^3 + ...\\
    y'' &= y'+b_{20}x'^2 + b_{11}x'y' + b_{02}y'^2 + b_{30}x'^3 + ...
  \end{align*}
}
\end{deluxetable}

\subsubsection{Astrometric Uncertainties}
\label{reduc.errors}
Cross-matching and registering the detections from each microstepped frame
produced mean and rms \radec\ measurements for each detection.
Figure~\ref{fig.asterr} shows the distribution of the combined rms (adding the
rms for $\alpha$ and $\delta$ in quadrature) for our parallax targets at each
epoch (median 35~mas), along with the rms positions of all detections as a
function of \jmko\ magnitude and the distribution of FWHM for our targets at
each epoch. The astrometric precision of each per-epoch measurement should scale
as 1/$\sqrt{N_{\rm frames}}$ (i.e., standard errors on the mean), and we adopt
these as our final uncertainties for each epoch. Figure~\ref{fig.asterr} also
shows the distribution of these standard errors, which are typically 7--12~mas
(median 11.2~mas). These uncertainties are somewhat larger than those from other
red-optical and infrared parallax programs discussed in DL12, but they are
consistent with our original goal of $\lesssim$12~mas astrometric precision per
epoch and sufficient to obtain $\approx$3--4~mas parallax precision with
$\approx$10~epochs of observation.

\subsubsection{Combining Epochs and Absolute Astrometric Calibration}
\label{reduc.combine}
For each target, we combined our final positions for all detections across all
epochs using the method described in DL12, clipping measurements with an R.A. or
decl. rms greater than 80~mas. We used a $1''$ matching radius to associate
objects across epochs, accounting for the proper motions of our targets using
measurements from PS1 \citep{Best:2018kw} or other literature sources. We
required objects to be detected in at least $N_{\rm ep}-2$ epochs (where
$N_{\rm ep}$ is the total number observed) to reject transient sources,
artifacts, and those whose detectability varied with observing conditions (i.e.,
very faint objects). We then calibrated the astrometry of the combined positions
to the absolute reference frame of PS1 as described in DL12.

\subsection{Photometry}
\label{reduc.phot}
In addition to astrometry, we calculated $J$-band photometry for sources in our
images.  For each epoch, we cross-matched the \radec\ positions of detections in
the flat files with the positions of detections from our pipeline using a $3''$
matching radius, subtracted the median offset in position from the flat file
positions, and repeated the cross-match.  For all of the matched objects (which
included the parallax targets and other objects in the fields), we computed the
magnitudes using the fluxes reported in the flat files and the zero-points,
aperture corrections, and extinction corrections from the flat file FITS
headers.  Our pipeline thus produced a \jmko\ magnitude for each detection at
each epoch.  For our parallax targets, we used the IDL routine {\tt
  resistant\_mean.pro} to calculate the outlier-resistant mean and standard
deviation of the photometry across epochs.  We added in quadrature a systematic
uncertainty of 0.015~mag \citep{Hodgkin:2009jr} to the standard deviations of
our measurements, which in the end was usually the dominant source of error.
Our photometry includes the first \jmko\ measurements for 193 of our targets and
improvements over previously published \jmko\ photometry for another 60~targets.

To assess the accuracy of our photometry, we identified the parallax targets
with previously published $J$-band photometry from WFCAM (e.g., UKIDSS) having
uncertainties less than 0.1~mag, and compared our new measurements to the
published values.  We found that our $J$ magnitudes were offset from the
literature WFCAM magnitudes by $-0.004\pm0.002$~mag (rms of 0.032~mag), so we
subtracted this value from our measurements (adding the uncertainties in
quadrature) to obtain our final $J$-band photometry.

\section{Calculation of Parallaxes}
\label{plx}

\subsection{Relative Parallax Solutions}
\label{plx.relative}

\subsubsection{Levenberg--Marquardt Least Squares}
\label{plx.relative.lmls}
Initially, we calculated parallaxes, proper motions, and positions for our
targets relative to our pipeline-generated astrometric reference frames using
the approach described in DL12. Briefly, DL12 used the Levenberg--Marquardt
least-squares (LMLS) method, as implemented in the IDL package {\tt mpfit}
\citep{Markwardt:2009wq}, to fit five parameters to the mean coordinates from
each epoch: parallax (\plx), proper motion (\mua, \mud), and position at the
first epoch of observation \radeco. With this solution as a starting guess, DL12
then used Markov Chain Monte Carlo (MCMC) to determine the posterior
distributions and calculate uncertainties for the five parameters (typically
1--4~mas for parallaxes). The MCMC trials produced Gaussian-like posteriors,
which could be robustly described by means and standard deviations. No automated
scheme was employed to mask outlier epochs that were biasing the fits. Rather,
each solution was inspected by eye, and in some cases, epochs with a low S/N or
significantly discrepant airmass were removed, with the goal of bringing the
reduced chi-squared (\rchi) of the solution closer to 1.

We found that the \rchi\ of our LMLS best-fit solutions tended to be larger
\hbox{than 1} (Figure~\ref{fig.comp.chisq}), with the astrometric errors of the
individual epochs often smaller than the scatter of the measurements about the
fit, suggesting the errors were somewhat underestimated. We therefore explored
additional methods for calculating astrometric solutions, in order to determine
the most accurate parallaxes and proper motions with uncertainties that fairly
represented the precision of our data.

\subsubsection{Bootstrap Resampling}
\label{plx.relative.boot}
First, we used bootstrap resampling to re-calculate parallaxes and proper
motions from the individual epochs, using LMLS fitting as before, and determined
median parameters and 68\% confidence limits from the bootstrap posteriors.
Bootstrapping typically yielded parameters consistent with those from the MCMC
while increasing both the range and average size of the uncertainties. A handful
of the bootstrap solutions yielded very large parallax errors ($\gtrsim10$~mas).

\subsubsection{Iteratively Re-weighted Least Squares}
\label{plx.relative.irls}
We then solved for parallax, proper motion, and position following an
iteratively re-weighted least-squares (IRLS) procedure. IRLS has been shown to
calculate astrometric solutions and uncertainties consistent with least-squares
solutions for well-detected objects in PS1 while minimizing the impact of
outlier measurements (E. Magnier et al., 2020, in preparation).  Briefly, using
an LMLS solution as a starting point, IRLS re-weighted the position measurements
at each epoch and repeated the LMLS fit in an iterative fashion.  We used Cauchy
weights of the form
\begin{equation}
W = \frac{1}{1 + (\frac{X}{2.385})^2}
\end{equation}
where $X$ is the deviation of an $\alpha$ or $\delta$ measurement from its
best-fit value, normalized by the measurement error, i.e., 
\begin{equation}
X = \frac{\alpha_{\rm meas}-\alpha_{fit}}{\sigma_\alpha}~~{\rm or}~~X = \frac{\delta_{\rm meas}-\delta_{fit}}{\sigma_\delta}.
\end{equation}
To re-weight, we multiplied our measurement errors by $\frac{1}{\sqrt{W}}$,
which effectively multiplied the inverse variance weights used in the LMLS
fitting by $W$. We iterated until the three parameters \plx, $\mu_\alpha$, and
\mud\ each varied by less than a factor of $10^{-3}$, capping the number of
iterations at 300. In this implementation, IRLS essentially increased the
measurement errors of more deviant epochs to bring those errors into concurrence
with the scatter of residuals around the best fit.

\begin{figure}
  \includegraphics[width=1\columnwidth]{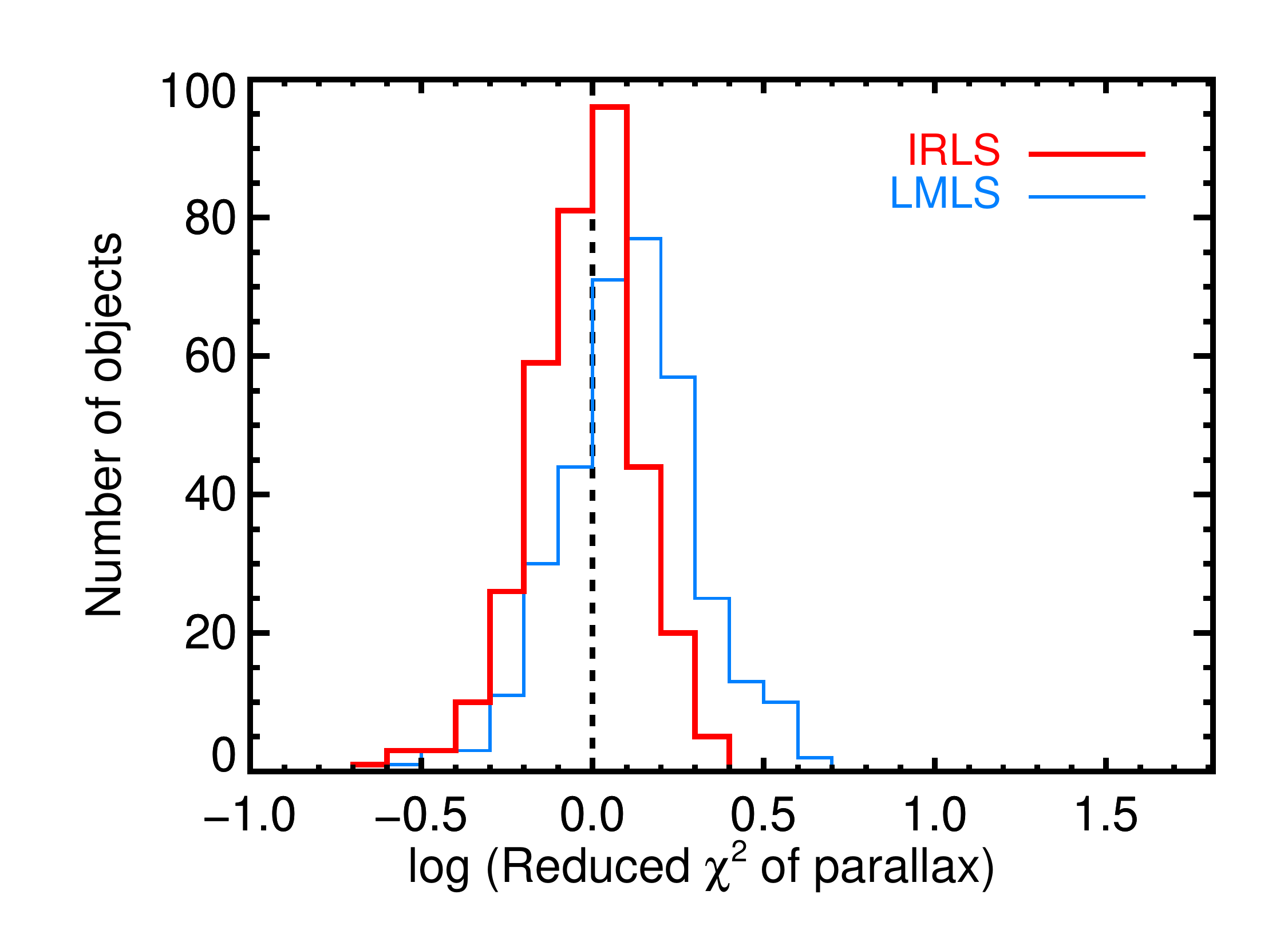}
  \caption{The distribution of \rchi\ for our IRLS solutions (thick red
    histogram) compared with the same distribution for our LMLS solutions (thin
    blue histogram). The IRLS \rchi\ values peak strongly near 1
    (${\rm median}=\varmedrchi$), indicating that the IRLS-re-weighted errors
    are commensurate with the scatter of measurements around the IRLS solutions.
    The \rchi\ values for the LMLS solutions peak higher than 1, implying that
    their uncertainties tend to be underestimated.}
  \label{fig.comp.chisq}
\end{figure}

Figure~\ref{fig.comp.chisq} compares the distribution of \rchi\ for our
astrometric solutions from IRLS to the original LMLS solutions. The \rchi\
distribution for our IRLS solutions peaks almost exactly at 1
(${\rm median}=\varmedrchi$), indicating that the scatter of data about our
parallax and proper motion fits is consistent with the re-weighted astrometric
errors at individual epochs for Gaussian-distributed data. The \rchi\
distribution for the original LMLS solutions peaks higher than 1, confirming
that their uncertainties tended to be somewhat underestimated.

After obtaining a final IRLS solution, we recalculated parallaxes, proper
motions, positions, and uncertainties from MCMC trials using the IRLS-determined
parameters and re-weighted astrometry as the starting point for the MCMC chains.
As an alternative method, we bootstrap-resampled the IRLS-re-weighted
measurements to obtain median parameters and confidence intervals.

In total, we obtained parallax and proper motions solutions for each target
using six distinct methods, which are listed in Table~\ref{tbl.method.gaia}.

\begin{deluxetable}{lccccc}
\tablecaption{Comparison of UKIRT Parallaxes to \gaiat\ Parallaxes \label{tbl.method.gaia}}
\tabletypesize{\small}
\tablewidth{0pt}
\tablehead{
  \colhead{} &
  \multicolumn{2}{c}{All \gaiat\tablenotemark{a}} &
  \colhead{} &
  \multicolumn{2}{c}{$\mathrm{RUWE}<1.4$\tablenotemark{b}} \\
  \cline{2-3}
  \cline{5-6}
  \colhead{Method} &
  \colhead{$\chi^2$} &
  \colhead{\rchi} &
  \colhead{} &
  \colhead{$\chi^2$} &
  \colhead{\rchi}
}
\startdata
LMLS              & 204.1 & 1.98 & & 199.4 & 2.08 \\
LMLS + MCMC       & 126.9 & 1.23 & & 125.1 & 1.30 \\
Bootstrap + LMLS  & 108.8 & 1.06 & & 107.4 & 1.12 \\
IRLS              & 161.0 & 1.56 & & 158.3 & 1.65 \\
IRLS + MCMC       & 109.8 & 1.07 & & 108.1 & 1.13 \\
Bootstrap + IRLS  & 129.4 & 1.26 & & 128.1 & 1.33 \\
\enddata
\tablecomments{For each of the six methods we used to calculate UKIRT
  parallaxes, this table shows the $\chi^2$ and reduced $\chi^2$ (\rchi) for a
  1:1 fit of our parallaxes to {\gaiat}'s, using two different samples of
  \gaiat\ parallaxes.} 
\tablenotetext{a}{There are \varngaia\ objects in our parallax sample that
  have \gaiat\ parallaxes, resulting in \varngaiadof\ degrees of freedom.}
\tablenotetext{b}{There are \varngaiagood\ objects in our parallax sample that
  have \gaiat\ parallaxes with $\mathrm{RUWE}<1.4$, resulting in
  \varngaiagooddof\ degrees of freedom.}
\end{deluxetable}

\subsubsection{Choosing the Final Solution}
\label{plx.relative.best}
We compared our UKIRT parallax solutions from our six methods to \gaiat\
parallaxes for the \varngaia~targets in common.  We tested the hypothesis that
our parallaxes are equal to {\gaiat}'s parallaxes by calculating \rchi\ for a
1:1 fit, i.e., using the \gaia\ parallaxes as a model for ours.  (Prior to this
comparison, we corrected our UKIRT parallaxes from relative to absolute
astrometry as described in Section~\ref{plx.absolute}).
Table~\ref{tbl.method.gaia} lists our methods and their $\chi^2$ and \rchi\
values for this fit.  In order to ensure that we were comparing to robust
\gaiat\ parallaxes, we also used the Renormalized Unit Weight Error (RUWE)
statistic \citep{Lindegren:2018vg}, selecting only the objects with \gaia\
parallaxes having $\mathrm{RUWE}<1.4$ \citep{Lindegren:2018vg}.  This turned out
to be \varngaiagood\ of the \varngaia\ targets in common.  We then repeated our
calculation of \rchi\ for a 1:1 fit of parallaxes from our six methods, this
time to the RUWE-selected subset of \gaiat\ parallaxes
(Table~\ref{tbl.method.gaia}).  The values of the \rchi\ for this comparison are
actually 0.06--0.10 higher than those of the \rchi\ for the full \gaiat\
comparison because the handful of RUWE-excluded objects have Gaia parallaxes
very similar to ours.

\begin{figure*}
  \centering
  \begin{minipage}[t]{0.54\textwidth}
    \includegraphics[width=1\columnwidth, trim = 10mm 0 0 0]{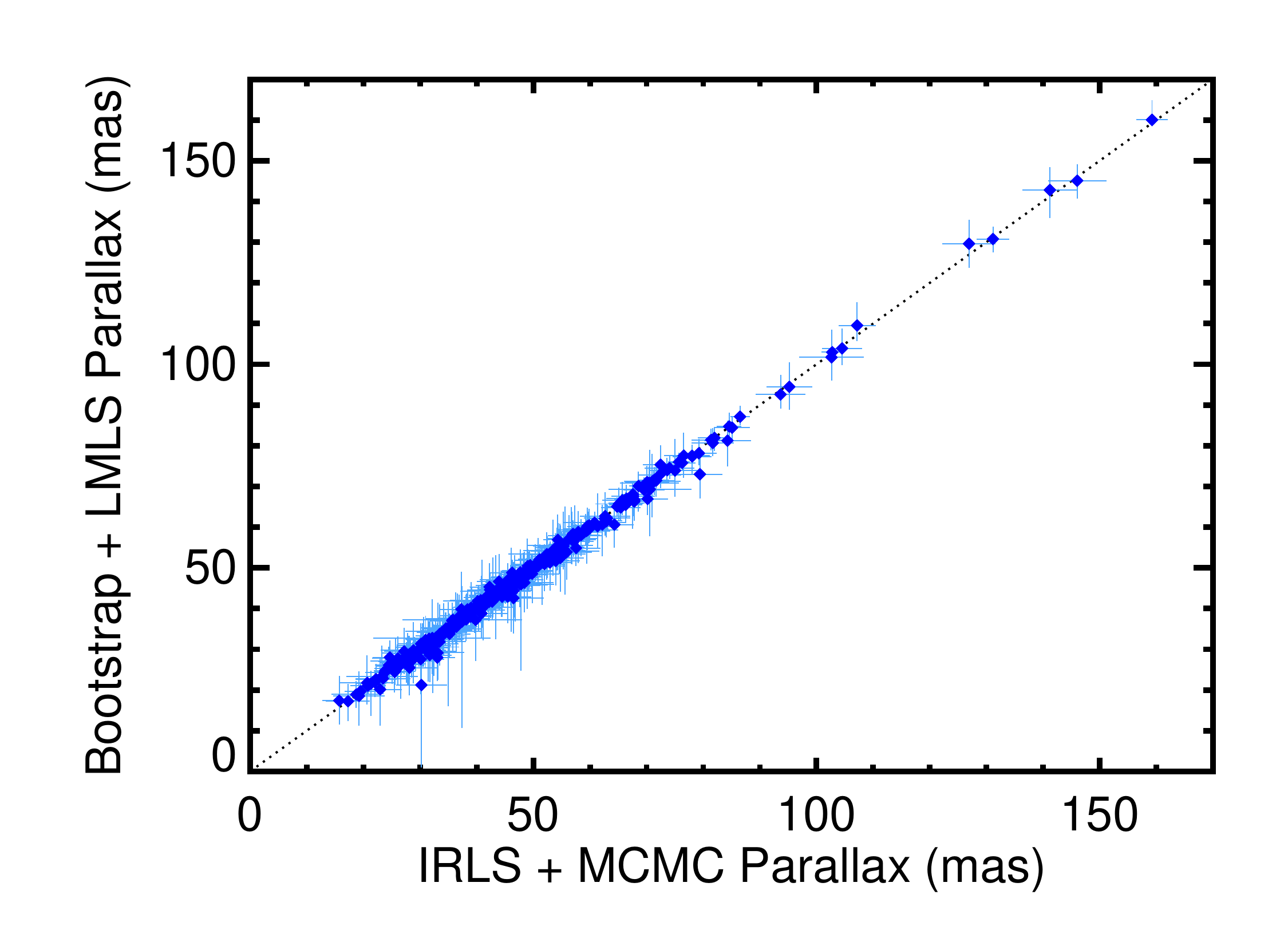}
  \end{minipage}
  \hfill
  \begin{minipage}[t]{0.44\textwidth}
    \includegraphics[width=1\columnwidth, trim = 10mm -30mm 0 0]{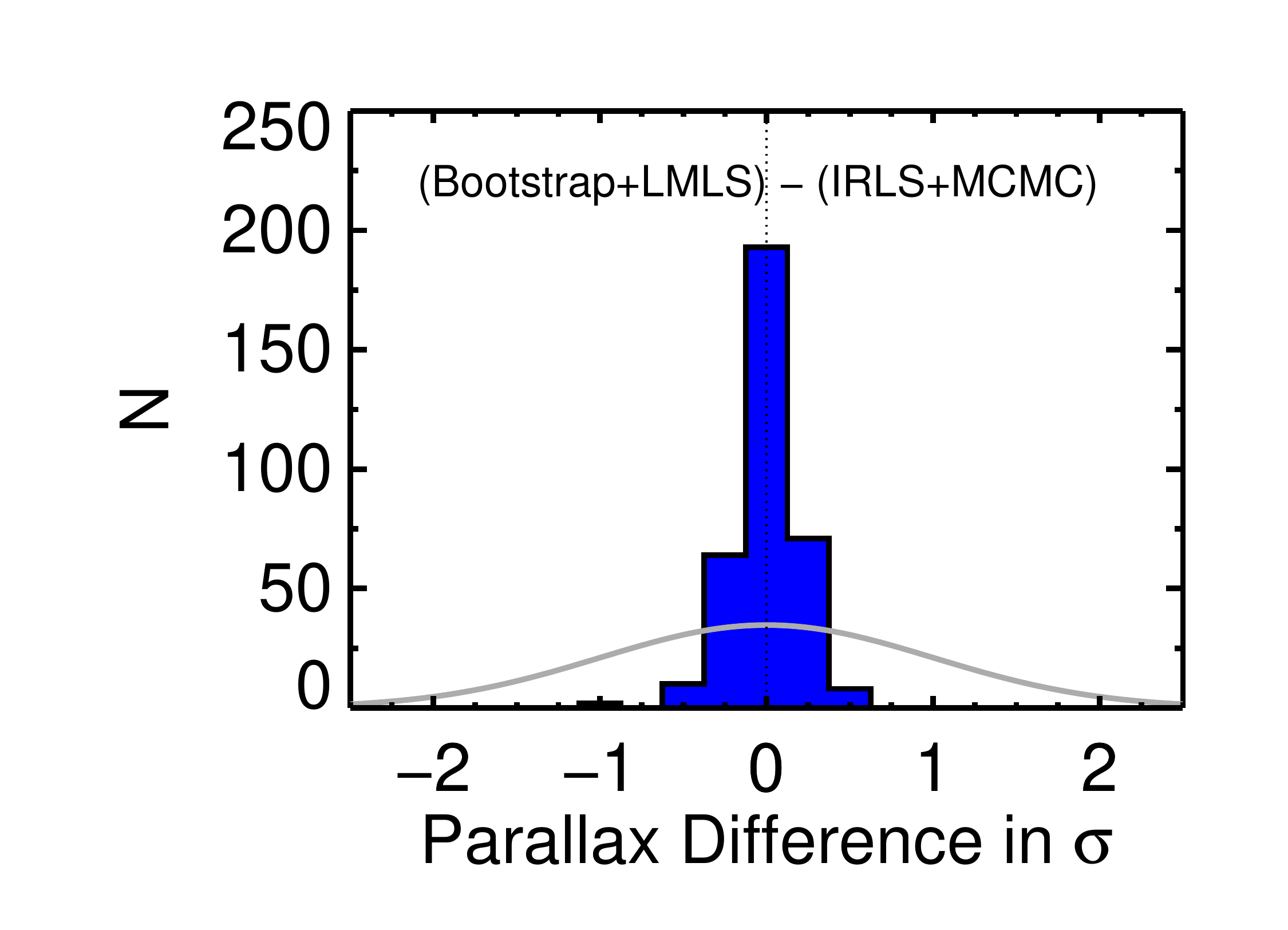}
  \end{minipage}
  \begin{minipage}[t]{0.54\textwidth}
    \includegraphics[width=1\columnwidth, trim = 10mm 0 0 0]{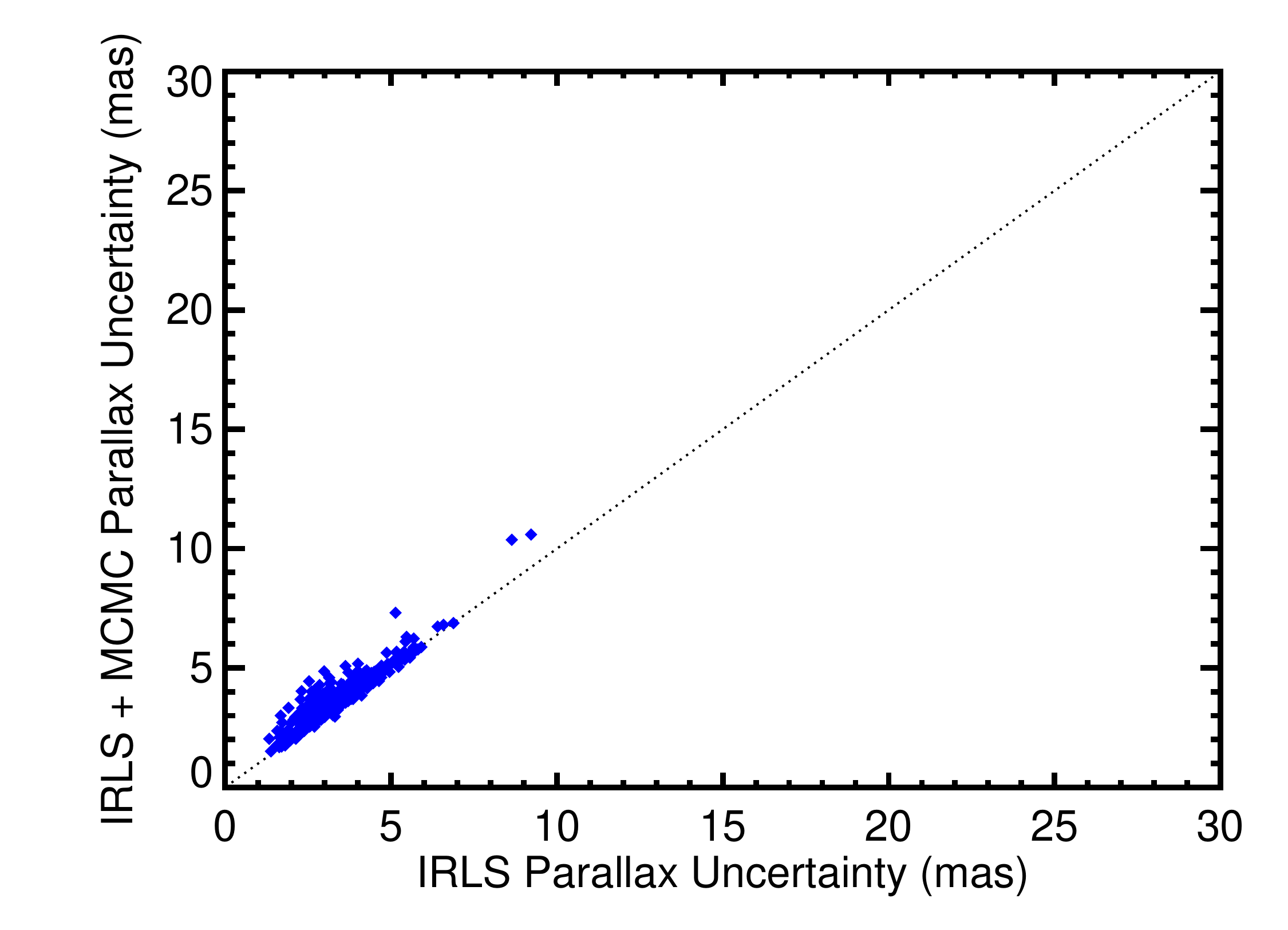}
  \end{minipage}
  \hfill
  \begin{minipage}[t]{0.44\textwidth}
    \includegraphics[width=1\columnwidth, trim = 10mm -30mm 0 0]{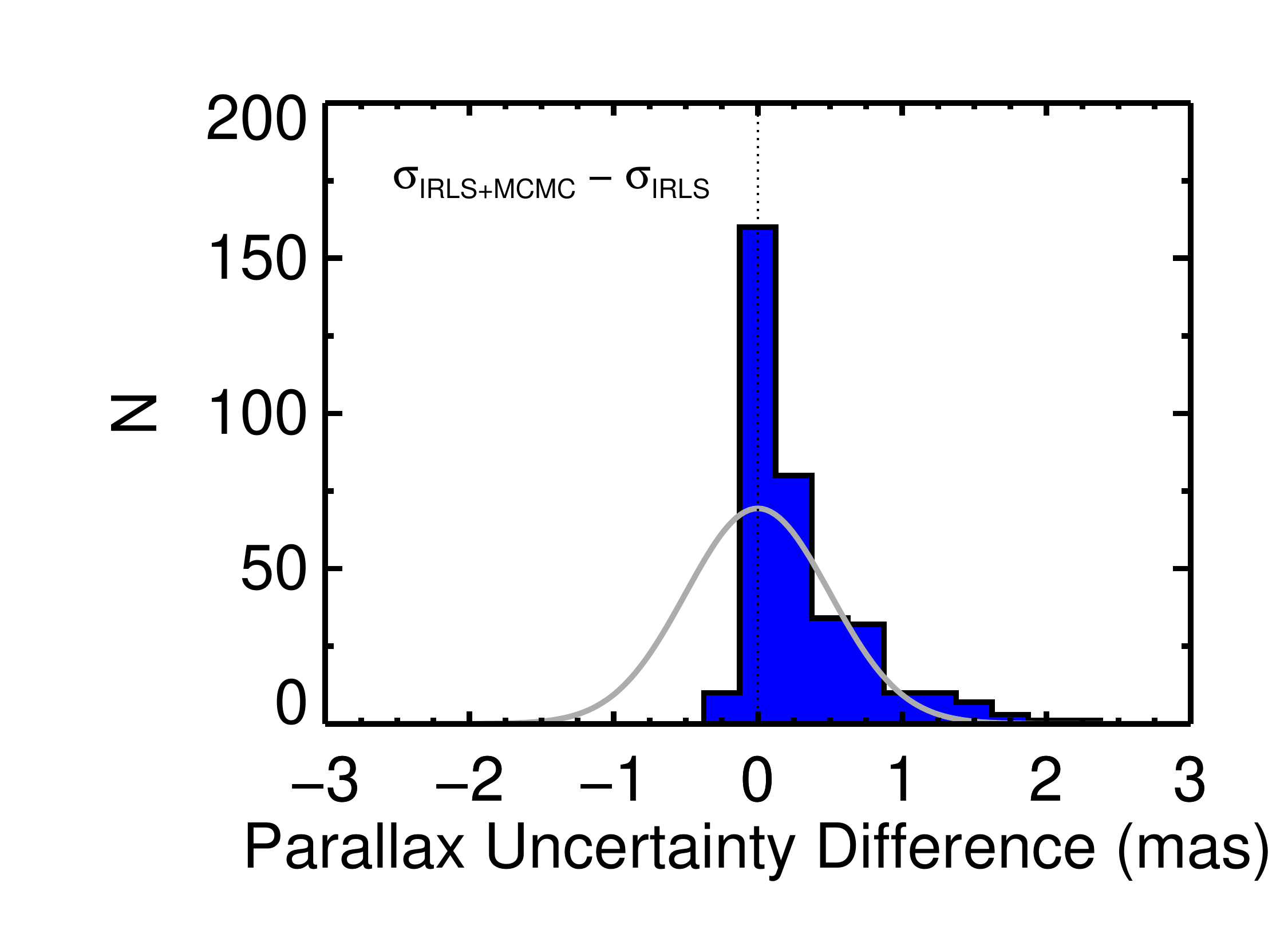}
  \end{minipage}
  \begin{minipage}[t]{0.54\textwidth}
    \includegraphics[width=1\columnwidth, trim = 10mm 0 0 0]{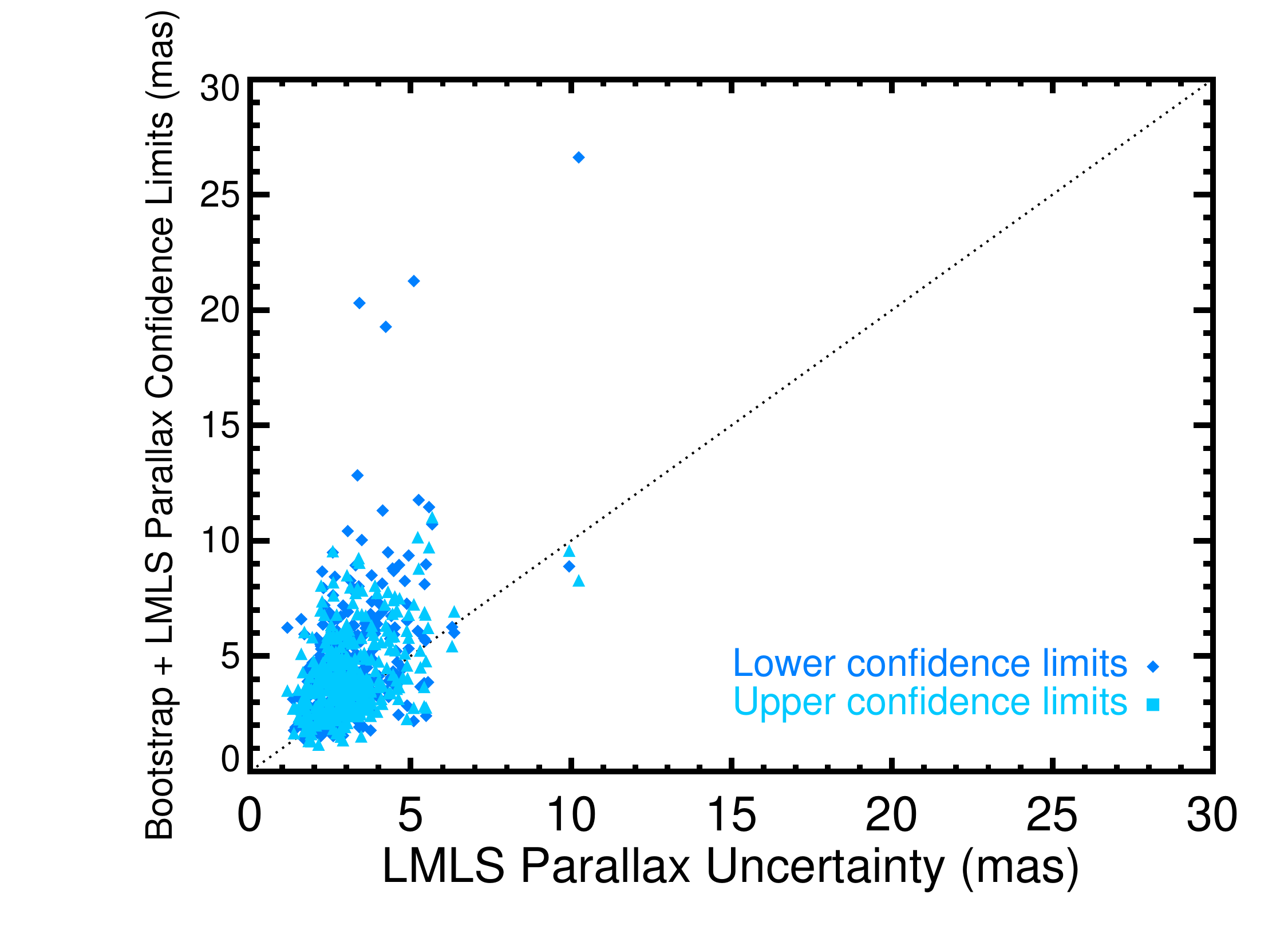}
  \end{minipage}
  \hfill
  \begin{minipage}[t]{0.44\textwidth}
    \includegraphics[width=1\columnwidth, trim = 10mm -30mm 0 0]{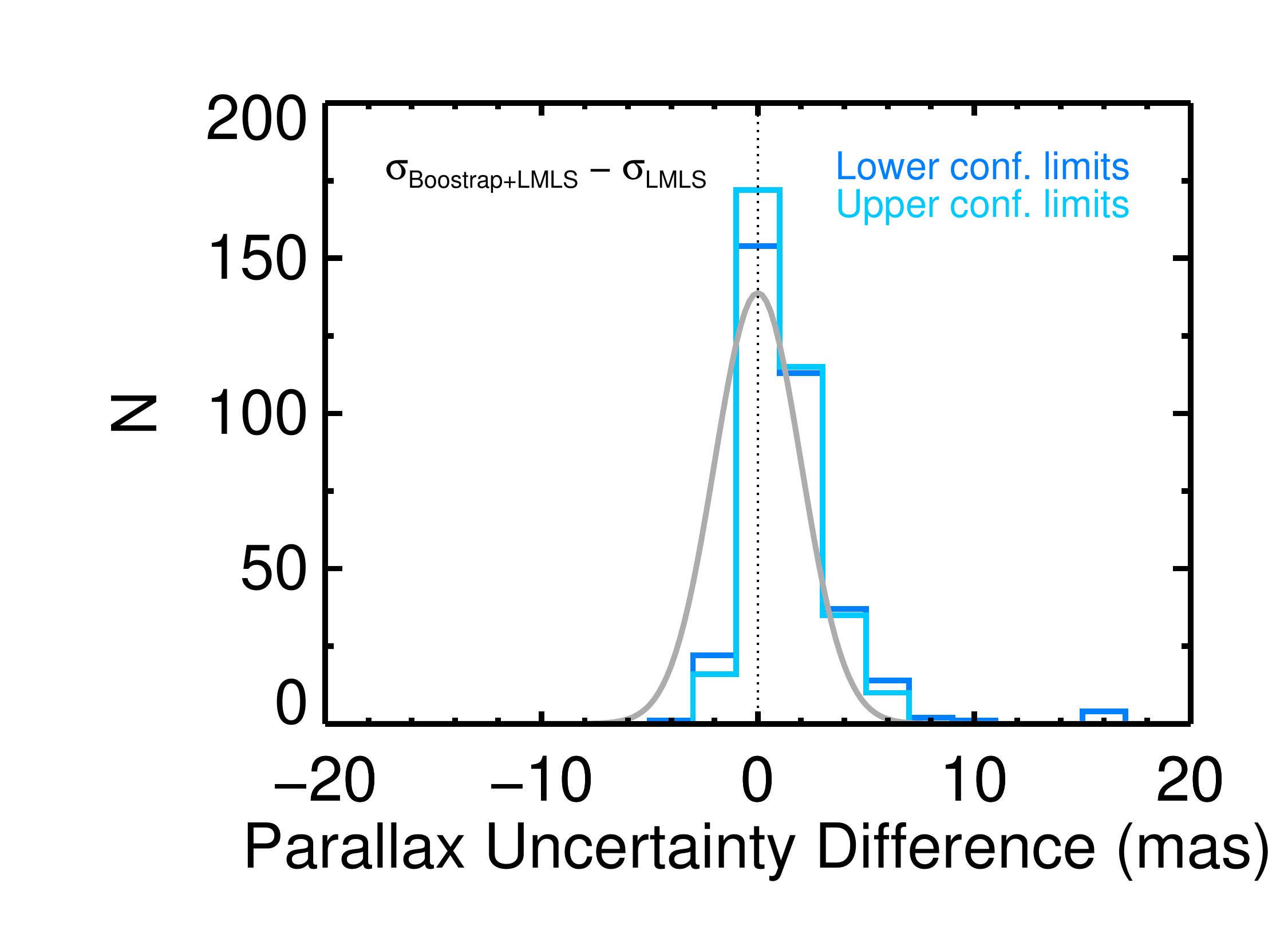}
  \end{minipage}
  \caption{{\it Top panels}: Comparison of parallaxes from our two methods with
    the lowest \rchi\ when compared to \gaiat\ parallaxes: bootstrapping+LMLS
    and IRLS+MCMC. The dotted black line marks equality, and the histogram at
    the right shows the differences in parallaxes, with the gray curve
    indicating a Gaussian distribution centered at 0 whose standard deviation is
    a $1\sigma$ difference in parallax, normalized to the number of objects in
    the histogram. The parallaxes from the two methods are all consistent within
    the uncertainties.  However, while IRLS+MCMC usually produces slightly
    larger uncertainties than IRLS alone ({\it middle panels}, with a Gaussian
    of standard deviation 0.5~mas), bootstrapping+LMLS generates a wide range of
    uncertainties that are often large increases over the LMLS uncertainties
    ({\it bottom panels}, with a Gaussian of standard deviation 2~mas).}
  \label{fig.boot.vs.irmc}
\end{figure*}

In the end, we obtained $\rchi>1$ for all six methods, implying that our
uncertainties or the \gaiat\ uncertainties are in general underestimated for
these ultracool dwarfs. Two methods clearly have the lowest \rchi\ in both
comparisons: bootstrapping of the original LMLS astrometry and IRLS+MCMC. We
examine the results from these methods more closely.  The parallaxes calculated
using the two methods are very consistent, within $0.5\sigma$ for 98\% of the
targets and within $1.0\sigma$ in all cases. However, the uncertainties produced
by the two methods differ significantly(Figure~\ref{fig.boot.vs.irmc}).
IRLS+MCMC typically obtains slightly larger parallax uncertainties than those
from the initial IRLS solution (0--2~mas increase for 86\% of our targets),
while bootstrapping produces errors with a broader range of differences compared
to the LMLS uncertainties ($-1$ to 4~mas increase for 85\% of targets).

\begin{figure*}
  \includegraphics[width=0.52\columnwidth]{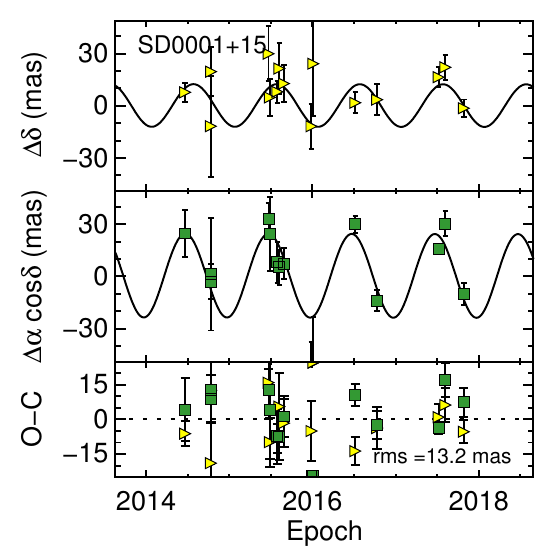}
  \includegraphics[width=0.52\columnwidth]{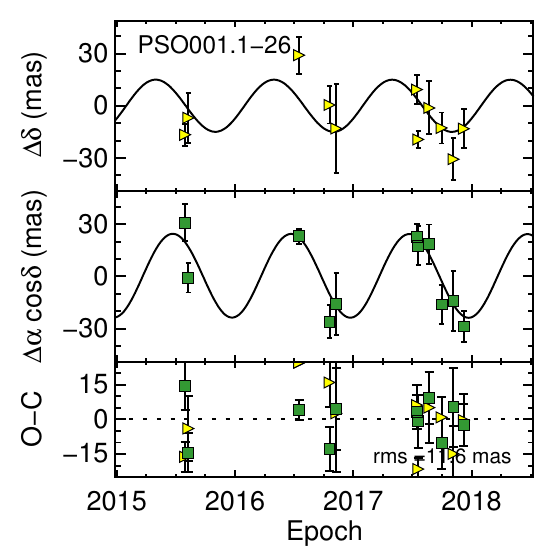}
  \includegraphics[width=0.52\columnwidth]{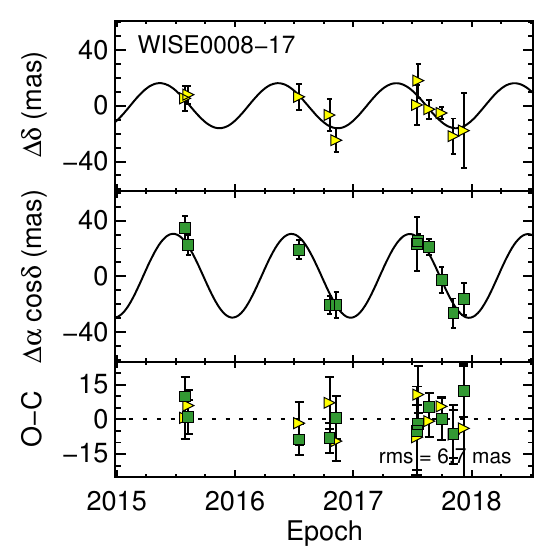}
  \includegraphics[width=0.52\columnwidth]{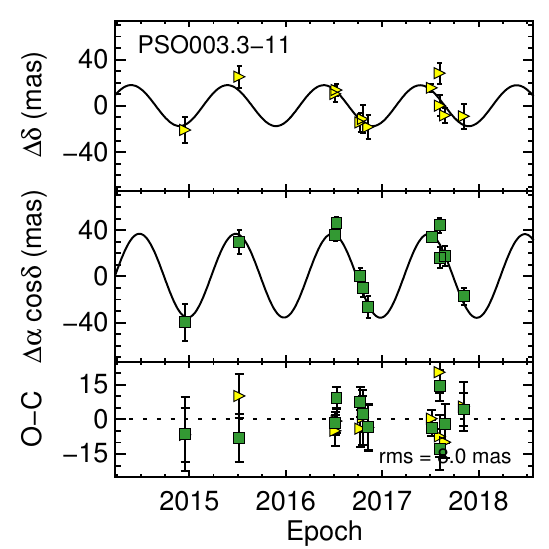}
  \includegraphics[width=0.52\columnwidth]{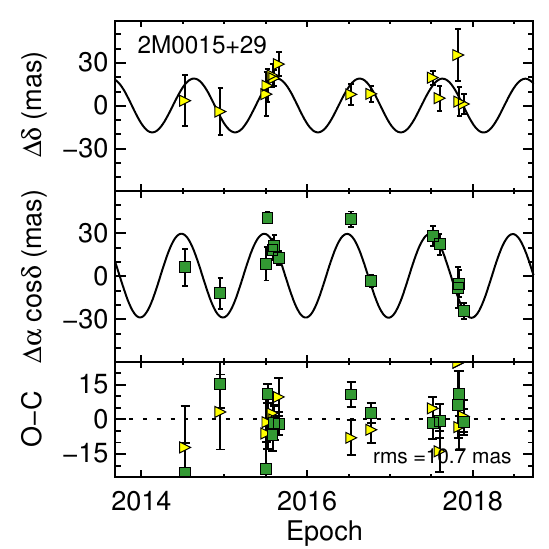}
  \includegraphics[width=0.52\columnwidth]{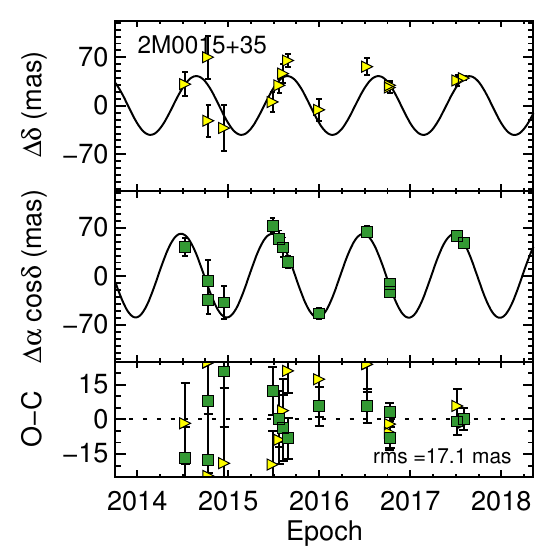}
  \includegraphics[width=0.52\columnwidth]{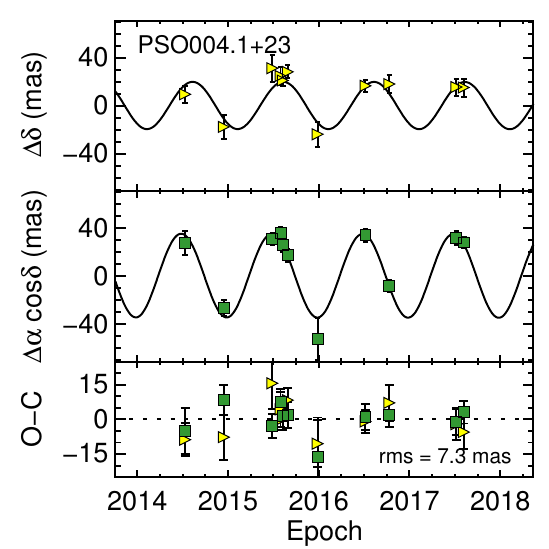}
  \includegraphics[width=0.52\columnwidth]{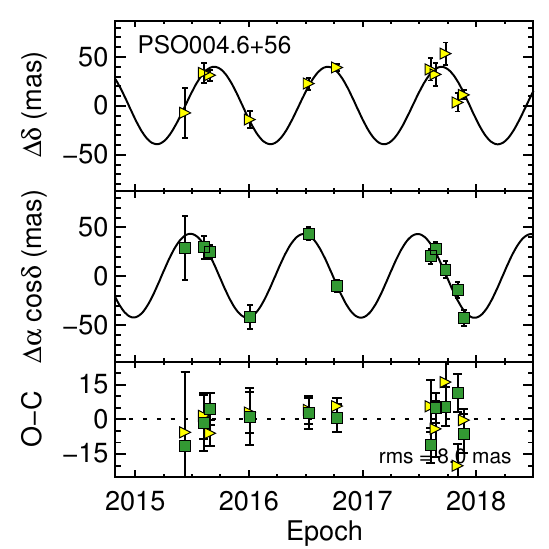}
  \caption{Parallax fits and residuals for our targets. For each object, the top
    and middle panels show relative parallax fits in decl. ($\delta$) and R.A.
    ($\alpha\ {\rm cos}\,\delta$), respectively, after subtracting the best-fit
    proper motion, as a function of Julian year. The bottom panels show the
    residuals after subtracting both the proper motion and the parallax.  (A
    portion of the plots are shown here as examples. The complete set of 348
    plots is available in the online journal.)}
  \label{fig.curves}
\end{figure*}

Considering also that the IRLS solutions have \rchi\ closer to 1
(Section~\ref{plx.relative.irls}), we find a preference for IRLS+MCMC over the
bootstrapping results. We therefore adopt the IRLS+MCMC parameters and
uncertainties as our final solutions. Figure~\ref{fig.curves} shows our data
along with the best-fit R.A. and decl. parallax curves and residuals for each of
our targets. We compare our solutions to \gaiat\ in more detail in
Section~\ref{results.compgaia}.

\subsection{Excluded Epochs}
\label{plx.exclude}
While analyzing the six methods for obtaining astrometric solutions described in
Section~\ref{plx.relative}, we noted a handful of targets whose parallax
confidence limits from bootstrapping were significantly larger than the
uncertainties from other methods, an indication that one or two data points may
be significantly biasing those parallax fits. Many of these targets also had an
LMLS parallax that differed significantly from a \gaiat\ or literature
measurement, and/or parallaxes that varied significantly across the six methods
we tested.

We therefore inspected the per-epoch astrometry and parallax solutions for all
targets having (1) $\rchi>2$ from the IRLS fit or (2) one or both confidence
limits from bootstrapping+IRLS more than 4~mas greater than the parallax
uncertainty from IRLS+MCMC. Five of these targets had eight or fewer epochs of
observations spanning $\le$2~yr, suggesting that more observations are needed to
identify possible outliers and securely measure their parallaxes.

For most of the rest of these targets, we found one of two types of outliers
among the epochs. The first type was an $\alpha$ or $\delta$ measurement at one
epoch with an implausibly small uncertainty ($<$2~mas, compared to a median of
$\approx$8~mas), likely a statistical consequence of using a small number of
dithers ($\le$9) to compute the rms. The second type was an $\alpha$ or $\delta$
measurement deviating by $\gtrsim$50~mas from the IRLS solution, often carrying
a similarly large measurement error. We found that excluding these epochs
resolved the problems with the parallax fits, i.e., reduced the \rchi\ to
$\approx$1, lowered the bootstrapping uncertainties into consistency with other
methods, and/or brought the parallax into agreement with the \gaiat\ or
literature value.

We also attempted to correct the fits in a more automated fashion by adding a
noise floor (up to 4~mas) to the astrometry at each epoch and by clipping epochs
with low weights from the IRLS fitting, but neither of these was successful in
systematically improving the parallax fits. We therefore adopted the parallaxes
with the epochs excluded by hand for 16 of our targets.

\subsection{Differential Chromatic Refraction}
\label{plx.dcr}
Our observation strategy strove to minimize any impact from DCR on our
astrometry by observing each target over a narrow airmass range around transit
(i.e., minimum airmass; Section~\ref{obs.strategy}). A handful of observations
were nevertheless taken at higher airmasses. To assess the impact of these
observations, we recalculated parallaxes after removing observations taken at
airmasses more than 0.1 greater than the minimum airmass at which each target
was observed. In total, this meant we removed one epoch from 62 targets (18\% of
all targets) and two or more epochs from 24 targets (7\%). The expected impact
of DCR on the measurements of these epochs was $\approx$2~mas for later-T~dwarfs
at elevation 40\degr\ (the lowest elevation for our observations; DL12), and
less for earlier-type objects and observations at higher elevations. We found
that the parallaxes and proper motions for the objects with removed epochs
changed by less than the uncertainties for the final parallaxes
($\approx$2--5~mas; Section~\ref{results}), and in general, the uncertainties
increased slightly as expected when removing one or two out of $\sim$10
measurements. The epochs removed due to discrepant airmass also did not
correspond systematically to outlier measurements in our data. We conclude that
DCR contributes insignificantly to our parallaxes and proper motions and their
uncertainties, which are dominated by the variation in position measurements
across individual frames. We therefore did not remove the higher-airmass epochs
from our final astrometric solutions.

\subsection{Orbital Motion}
\label{plx.orbit}
Stars and brown dwarfs that are binaries will exhibit orbital motion in addition
to parallax and proper motions. Sufficiently precise and frequent astrometric
measurements can detect such orbital motion even when a companion is unresolved,
and they can provide useful constraints on the masses of the components
\citep[e.g.,][]{Sahlmann:2013ky,Dupuy:2015gl}. The $\approx$11~mas precision of
our per-epoch astrometry and $\approx$2--5~mas precision of our parallaxes are
insufficient to detect orbital motions that are typically $\approx$5~mas in
amplitude, manifest over multiple-year periods, and alter parallaxes by less
than 1~mas (e.g., DL12, DL17). The residuals of our parallax fits shown in
Figure~\ref{fig.curves} also revealed no obvious trends suggesting orbital
motion. We therefore did not attempt to correct for orbital motion of any of our
targets, including known binaries.

\subsection{Correction to Absolute Parallax and Proper Motion}
\label{plx.absolute}
Up to this point, we have calculated parallaxes and proper motions relative to
field objects in our images, most of which are background stars. These reference
stars, while more distant, nevertheless have nonzero parallaxes and proper
motions. The mean positions of these stars were calibrated to the PS1 absolute
reference frame (Section~\ref{reduc.measurement}), but individual epochs are
still impacted by the parallactic motions of the background stars, which
systematically reduce the parallaxes measured for our targets. To convert these
relative parallaxes to absolute astrometry, DL12 used Besan\c{c}on galaxy models
\citep{Robin:2003jk} to estimate the mean parallaxes of field objects in their
images and apply this as a correction. LDA16 included proper motions in their
conversions from relative to absolute astrometry. Others have used photometry to
estimate distances to the reference stars
\citep[e.g.,][]{Dahn:2002fu,Faherty:2012cy}. With the advent of \gaiat, we can
use actual parallax measurements for reference stars in a target's field to
determine the correction to absolute astrometry for our UKIRT parallaxes.

\begin{figure}
  \includegraphics[width=1\columnwidth]{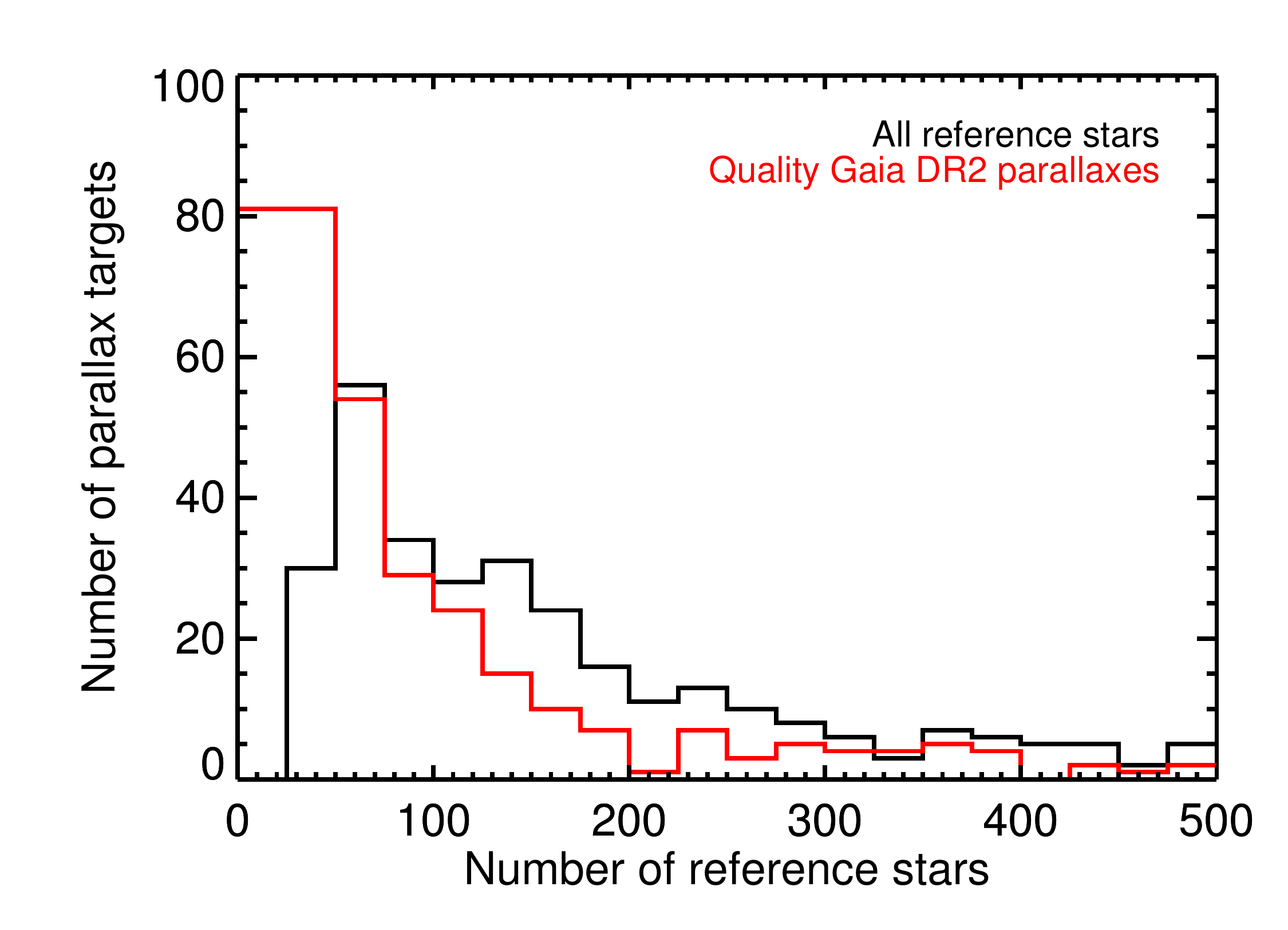}
  \includegraphics[width=1\columnwidth]{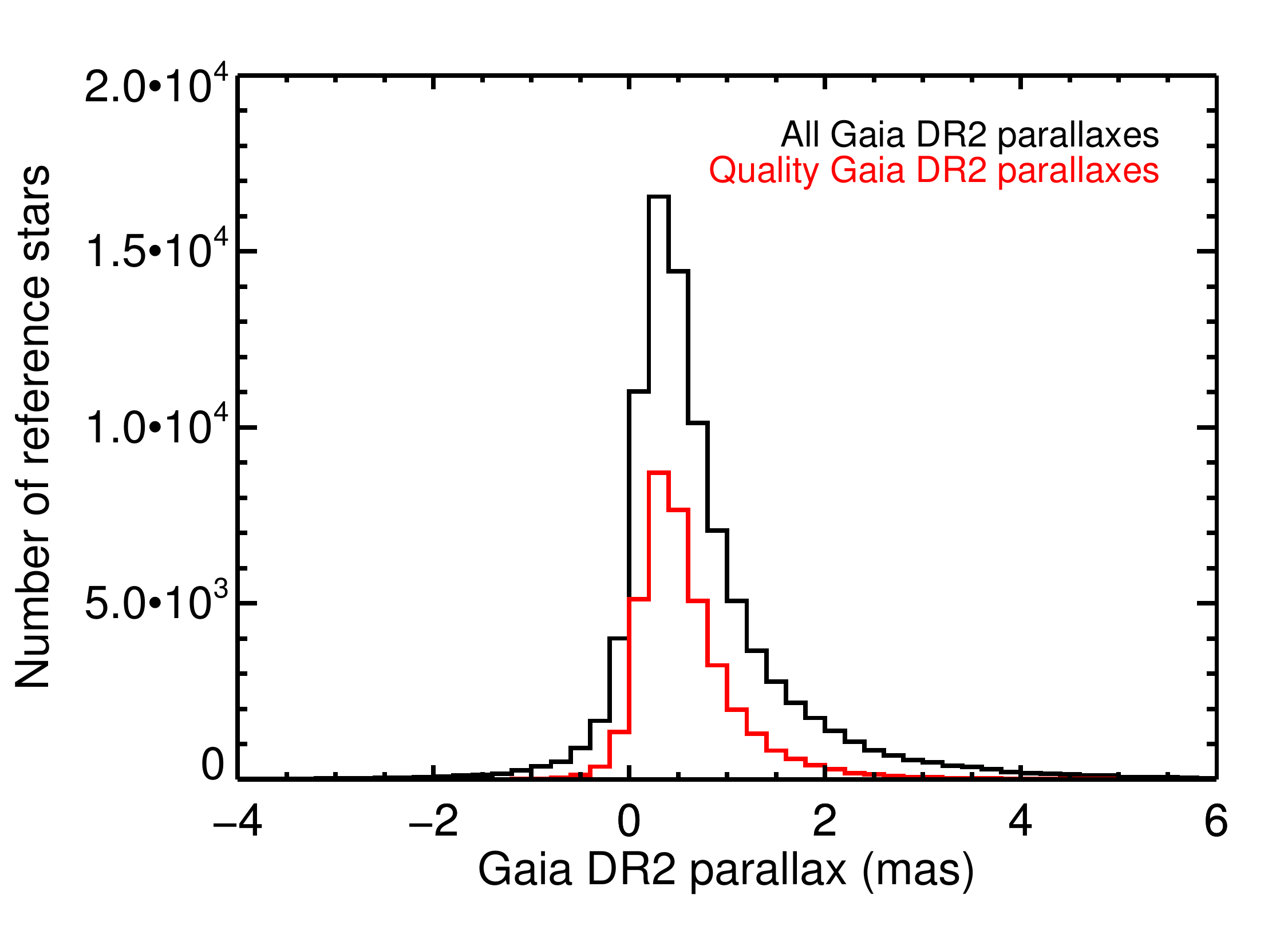}
  \includegraphics[width=1\columnwidth]{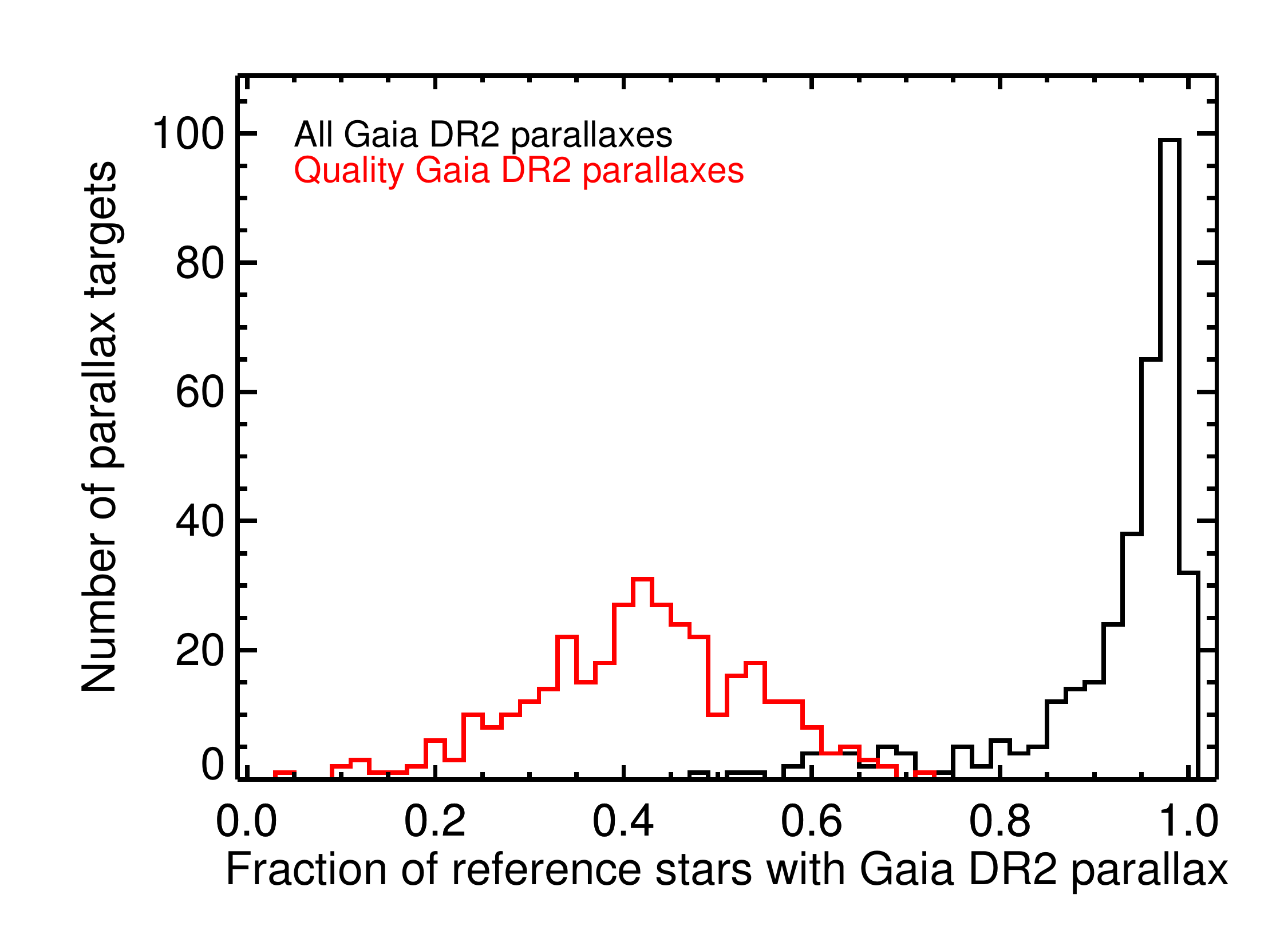}
  \caption{{\it Top panel}: distributions of the total number of reference frame
    stars for each parallax target (black) and the number of reference stars
    having ``quality'' \gaiat\ parallaxes (38\% of all reference stars; see the
    text for definition).  {\it Middle panel}: distributions of all \gaiat\
    parallaxes (black) and quality \gaiat\ parallaxes (red; 42\% of all \gaia\
    parallaxes) for the reference stars. Nearly all quality \gaia\ parallaxes
    are between $-0.5$ and 2~mas.  {\it Bottom panel}: distribution of the
    fraction of stars used to build the astrometric references that have a
    \gaiat\ parallax (black) and a quality \gaiat\ parallax (red).  We used only
    the quality \gaia\ parallaxes to correct our ultracool parallaxes from
    relative to absolute reference frames.}
  \label{fig.ref.gaiaplx}
\end{figure}

For each target in our sample, we cross-matched the set of field stars used for
the astrometric reference frame with \gaiat\ using a $1''$ matching radius. We
retained only the ``quality'' matches, namely those having \gaia\ parallax
errors less than 0.5~mas, DR2 parameter ${\tt astrometric\_excess\_noise}=0$,
and proper motions less than 20~\my\ (faster-moving objects were also excluded
from our UKIRT reference frames)\footnote{Following the release of the RUWE
  statistic for \gaiat\ \citep{Lindegren:2018vg}, we recalculated the
  corrections to absolute parallaxes and proper motions for 50 of our targets,
  this time using $\mathrm{RUWE}<1.4$ instead of the
  ${\tt astrometric\_excess\_noise}=0$ requirement.  This increased the number
  of Gaia reference stars in every case, typically by 40--100\%, and decreased
  the uncertainties in the corrections by $\approx$30--40\%.  However, (perhaps
  surprisingly) the corrections themselves changed by less than the
  uncertainties on the relative parallaxes and proper motions, and with no
  overall increase or decrease, so we concluded that changing to RUWE-selected
  Gaia DR2 parallaxes for the correction to absolute astrometry did not
  significantly impact our measurements. We have therefore retained our original
  values.}. These ``quality'' \gaia\ matches comprised on average 38\% of all
reference stars and 42\% of those with \gaia\ matches
(Figure~\ref{fig.ref.gaiaplx}). We added the 29~\uas\ zero-point offset from
\citet{Lindegren:2018gy} to the \gaiat\ parallaxes and the 43~\uas\ systematic
noise floor from \citet{Lindegren:2018gy} in quadrature to the \gaiat\ parallax
uncertainties (although the impact of these on our final parallaxes is
negligible).

Our pipeline calculates LMLS parallaxes and proper motions for all objects
(science targets {\it and} reference stars) matched across at least
$N_{\rm ep}-2$ epochs, so we also have UKIRT relative parallaxes and proper
motions for the \gaia-matched reference stars. The difference between the \gaia\
and UKIRT parallax for each object, $\plx_{Gaia}-\plx_{\rm UKIRT}$, is the
amount of parallax motion we ignore when defining our reference frames. For each
target, we computed the weighted mean and weighted standard error of
($\plx_{Gaia}-\plx_{\rm UKIRT}$) for all quality reference stars as the
correction for turning the relative parallaxes into absolute parallaxes
(Table~\ref{tbl.obs}). We also calculated and applied analogous corrections to
our UKIRT relative proper motion measurements, incorporating the 66~\uy\ noise
floor from \citet{Lindegren:2018gy}.

We investigated whether we could obtain more accurate astrometric reference
frames and parallaxes by using \gaiat\ instead of PS1 as our external reference
catalog throughout the reduction process. In theory, this approach might provide
more accurate corrections to absolute astrometry, since we would be using the
same stars for the reference frames and the correction to absolute astrometry.
But we might also lose astrometric precision since PS1 is a deeper catalog with
more objects than \gaiat. We selected 50 targets for which the fractions of
reference stars with quality \gaia\ parallaxes spanned the full range for our
target list, and we ran these 50 objects through our pipeline using \gaiat\ as
the reference catalog instead of PS1. The final parallaxes and uncertainties and
the corrections to absolute parallaxes were nearly identical whether PS1 or
\gaiat\ was used as the reference catalog. We opted to use the deeper PS1 as the
external catalog.

\vspace{-10pt}

\subsection{Targets with No Astrometric Solution}
\label{plx.nosolution}
We were unable to calculate a parallax for \varnnoplx\ of our \varntargets\
targets. Six of these were observed at fewer than 6 epochs, making a robust
astrometric solution impossible. One target, WISE~J033605.05$-$014350.4, proved
to be too faint to be detected in our longest 300~s frames. For the final
target, 2MASS~J21371044+1450475, we calculated a parallax of $19.8\pm19.7$~mas
using IRLS+MCMC and a wide range of other solutions (including 0 and a negative
parallax) using the other methods described in Section~\ref{plx.relative}. This
object is relatively bright at $J_{\rm 2MASS}=14.13$~mag and has a \gaiat\
parallax of $42.39\pm0.32$~mas, so it should have been a straightforward target
for our program. The scatter in our data suggest that several of the epochs were
corrupted due to poor observing conditions or guider failure, but it is unclear
from the UKIRT data alone which epochs are affected. We therefore do not include
2MASS~J21371044+1450475 in our final results.

\begin{longrotatetable}
\begin{deluxetable*}{lcCcCCCcCCCr}
\centering
\tablecaption{UKIRT/WFCAM Parallaxes and Proper Motions \label{tbl.results}}
\tabletypesize{\scriptsize}
\setlength{\tabcolsep}{0.065in}
\tablewidth{0pt}
\rotate
\tablehead{   
  \multicolumn{4}{c}{} &
  \multicolumn{3}{c}{Relative} &
  \colhead{} &
  \multicolumn{3}{c}{Absolute} &
  \colhead{} \\
  \cline{5-7}
  \cline{9-11}
  \colhead{Object} &
  \colhead{$\alpha_{\rm J2000}$} &
  \colhead{$\delta_{\rm J2000}$} &
  \colhead{Epoch} &
  \colhead{$\plx_{\rm rel}$} &
  \colhead{$\mu_{\alpha,{\rm rel}}{\rm cos}\,\delta$} &
  \colhead{$\mu_{\delta,{\rm rel}}$} &
  \colhead{} &
  \colhead{$\plx_{\rm abs}$} &
  \colhead{$\mu_{\alpha,{\rm abs}}{\rm cos}\,\delta$} &
  \colhead{$\mu_{\delta,{\rm abs}}$} &
  \colhead{$\chi^2$/dof} \\
  \colhead{} &
  \colhead{(deg)} &
  \colhead{(deg)} &
  \colhead{(MJD)} &
  \colhead{(mas)} &
  \colhead{(\my)} &
  \colhead{(\my)} &
  \colhead{} &
  \colhead{(mas)} &
  \colhead{(\my)} &
  \colhead{(\my)} &
  \colhead{}
}
\startdata
SDSS J000112.18+153535.5 & 000.3013707 & $+15.5925657$ & 56829.63 & $25.7\pm3.4$ & $123.2\pm2.1$ & $-175.3\pm1.9$ &  & $27.2\pm3.5$ & $132.5\pm2.2$ & $-180.3\pm1.9$ & 26.5/25 \\
PSO J001.1277$-$26.0675 & 001.1278586 & $-26.0676516$ & 57234.57 & $30.4\pm4.6$ & $-7.2\pm4.2$ & $-237.8\pm3.8$ &  & $33.1\pm4.7$ & $-0.9\pm4.3$ & $-243.7\pm3.8$ & 26.2/17 \\
WISEPC J000849.76$-$173922.6 & 002.2072022 & $-17.6569863$ & 57234.57 & $32.8\pm3.9$ & $-178.0\pm3.1$ & $-372.5\pm3.1$ &  & $36.1\pm3.9$ & $-171.1\pm3.1$ & $-373.1\pm3.1$ & 11.3/17 \\
PSO J003.3437$-$11.7168 & 003.3439192 & $-11.7168434$ & 57007.20 & $38.7\pm3.1$ & $209.0\pm3.2$ & $-17.7\pm3.0$ &  & $40.3\pm3.1$ & $213.7\pm3.3$ & $-26.7\pm3.1$ & 20.0/19 \\
2MASS J00150206+2959323 & 003.7604536 & $+29.9914289$ & 56850.60 & $30.4\pm3.3$ & $372.3\pm2.2$ & $-225.9\pm2.3$ &  & $31.6\pm3.3$ & $372.5\pm2.2$ & $-231.8\pm2.3$ & 23.8/23 \\
2MASSW J0015447+351603 & 003.9369026 & $+35.2664116$ & 56850.61 & $64.8\pm3.5$ & $47.2\pm2.8$ & $-255.0\pm2.7$ &  & $65.7\pm3.5$ & $47.2\pm2.8$ & $-257.8\pm2.7$ & 23.5/23 \\
PSO J004.1834+23.0741 & 004.1837991 & $+23.0741642$ & 56850.60 & $37.2\pm3.3$ & $402.5\pm2.0$ & $66.6\pm2.4$ &  & $38.4\pm3.3$ & $405.4\pm2.1$ & $61.8\pm2.5$ & 12.9/17 \\
PSO J004.6359+56.8370 & 004.6369722 & $+56.8371228$ & 57183.63 & $46.4\pm3.9$ & $376.5\pm3.2$ & $12.4\pm2.7$ &  & $46.5\pm3.9$ & $375.9\pm3.2$ & $10.4\pm2.7$ & 14.0/17 \\
\enddata
\tablecomments{
  This table gives all of the astrometric parameters derived from our MCMC analysis
  of the IRLS solution for each target.
  ($\alpha$, $\delta$, MJD): coordinates and epoch for our first observation of
  that target.
  ($\plx$, \mud, \mud): We list both relative and absolute parallaxes and proper
  motions.
  Absolute values incorporate a correction for the mean parallax and proper motion of
  our reference stars (Table~\ref{tbl.obs}), which includes an additional uncertainty
  added in quadrature to the relative astrometry's uncertainties
  (Section~\ref{plx.absolute}).
  $\chi^2$/dof: the smallest $\chi^2$ among the IRLS + MCMC chains along with the
  degrees of freedom (dof) for the data set.
}
\tablenotetext{a}{Observations for this object span only one year.  While our
  astrometric solution appears to be robust, the parallax and proper motions
  should be regarded as preliminary.}
\tablenotetext{b}{This known $0.9''$ binary \citep{Burgasser:2009gs} is resolved
  in our UKIRT images. Our astrometric solution appears reasonable
  (Section~\ref{results.binaries}) but should be treated with caution.}
\end{deluxetable*}
\end{longrotatetable}

\section{Results}
\label{results}

\begin{figure}
\begin{center}
    \includegraphics[width=1\columnwidth]{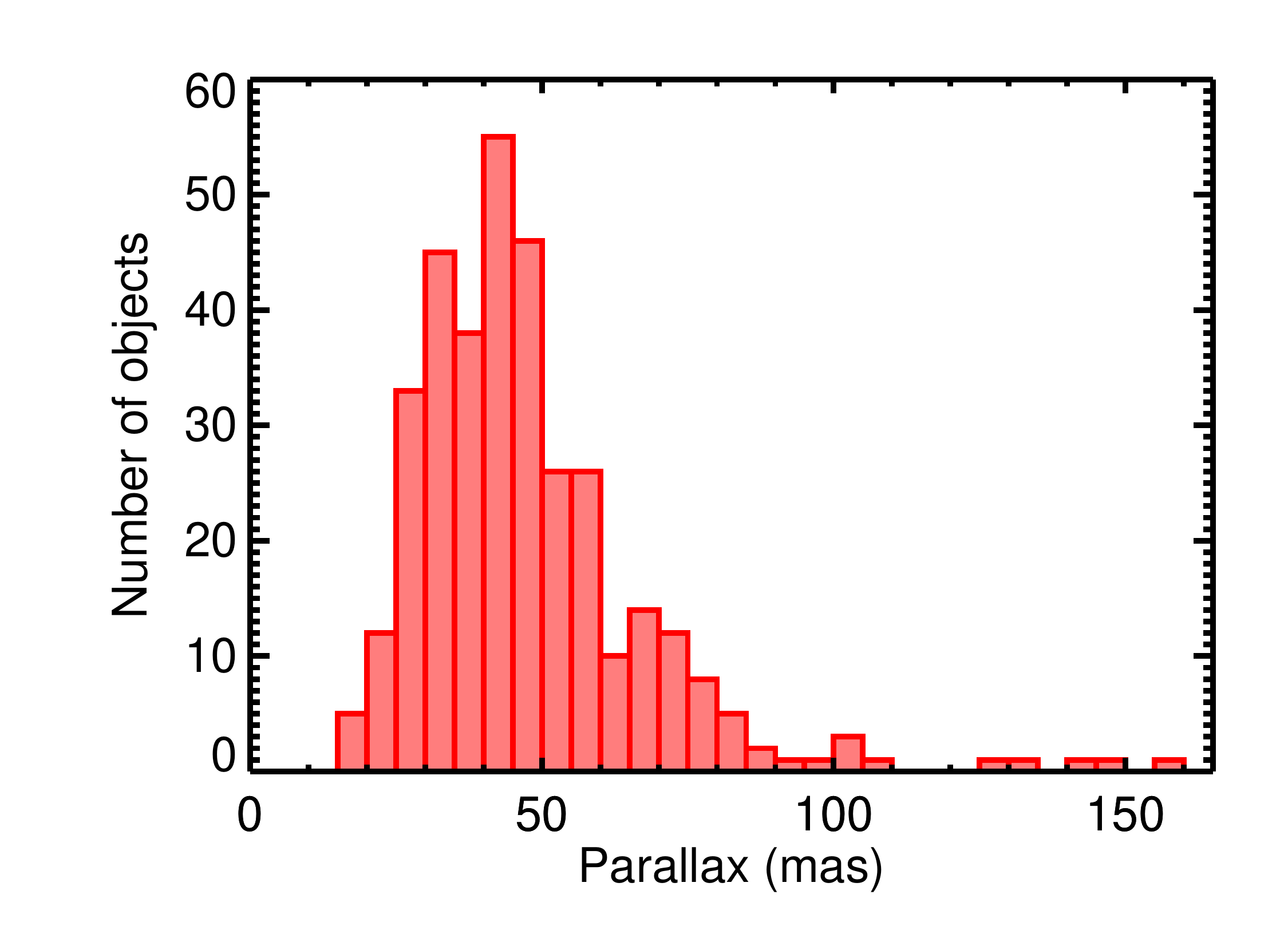}
    \includegraphics[width=1\columnwidth]{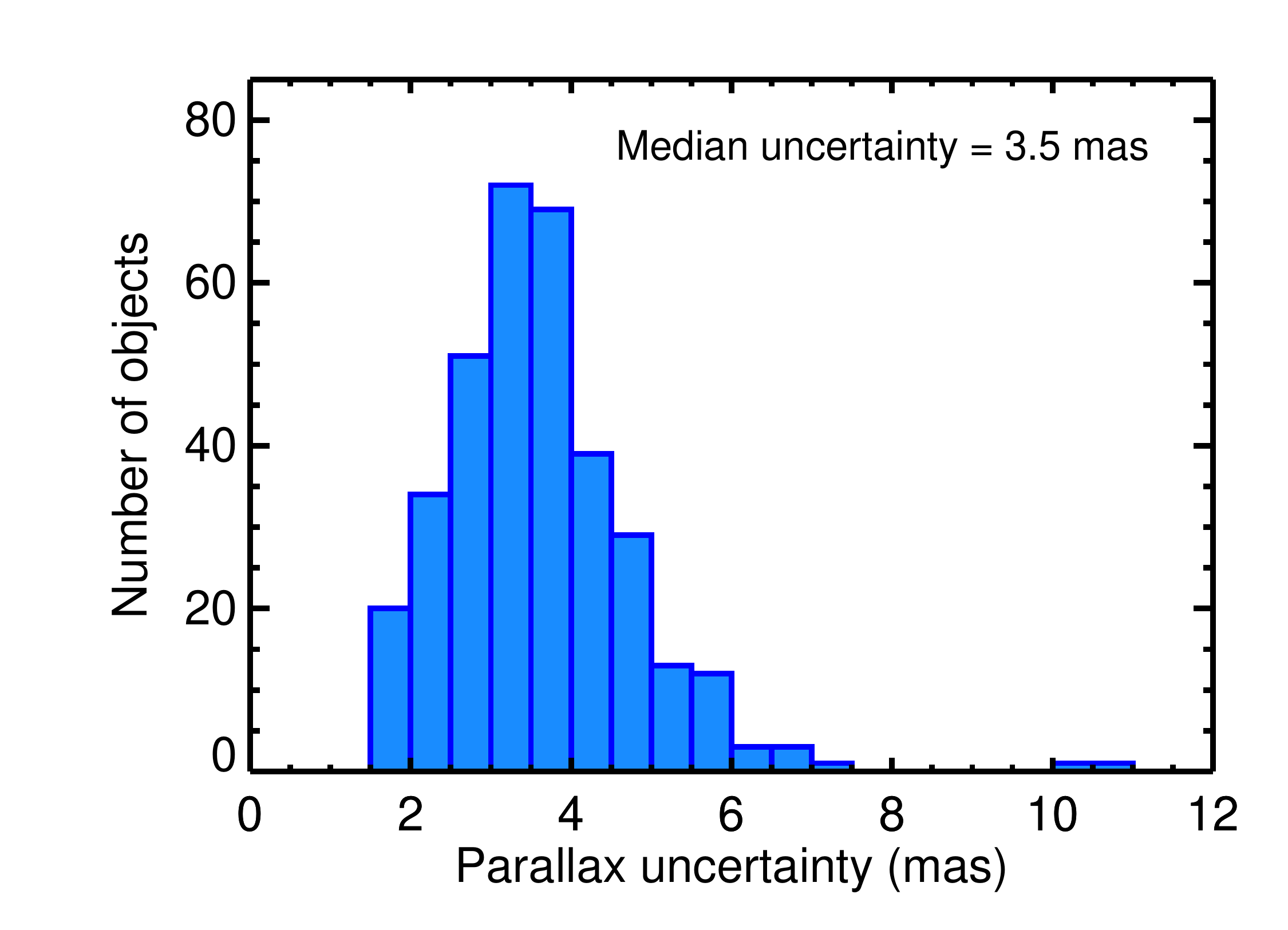}
    \includegraphics[width=1\columnwidth]{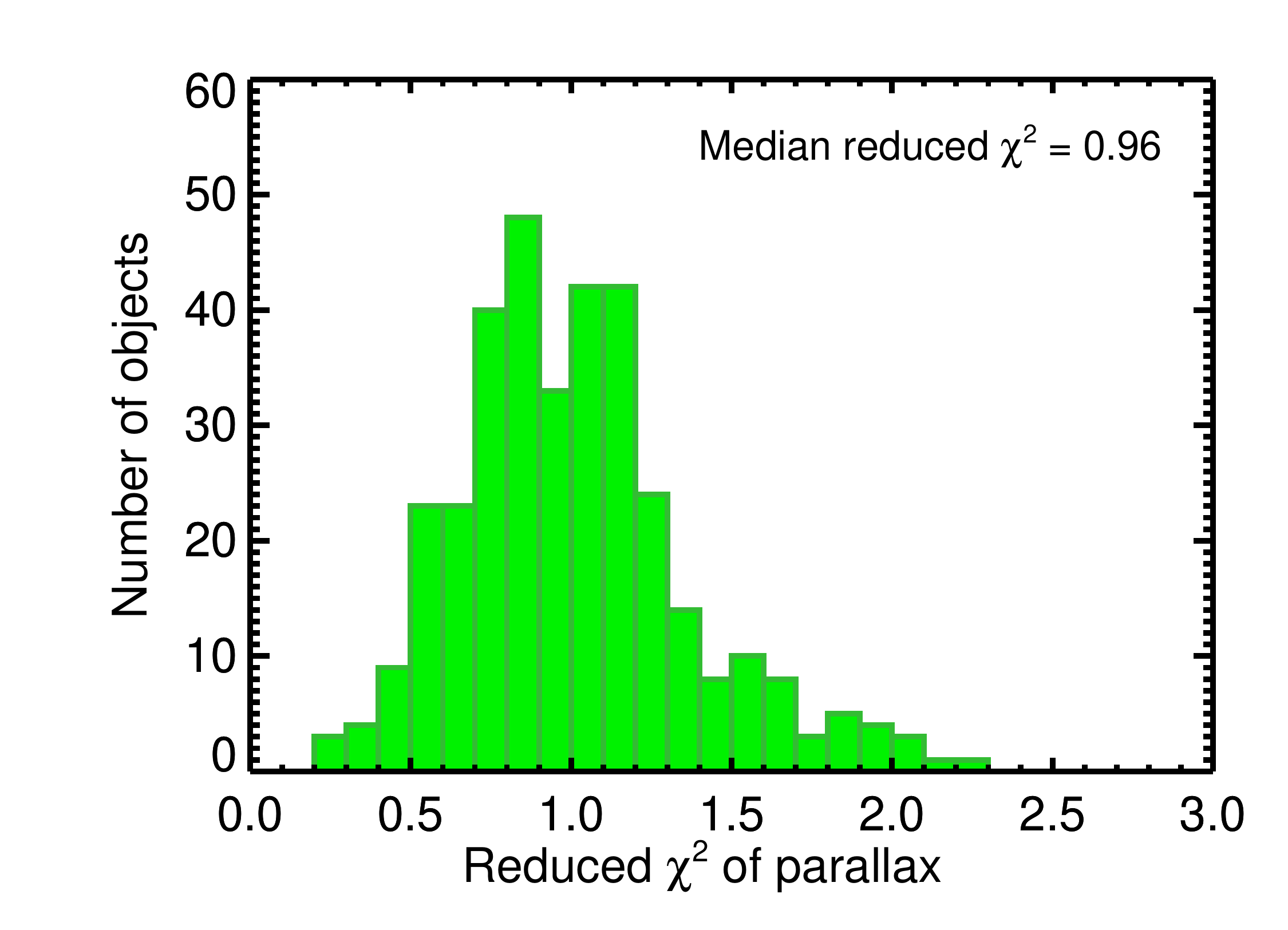}
    \caption{Distributions of our final parallaxes ({\it top panel}) and
      uncertainties ({\it middle panel}) including the corrections to absolute
      astrometry, and of the minimum \rchi\ for our IRLS+MCMC chains ({\it
        bottom panel}).  Objects with ${\rm parallaxes}\ge40$~mas are within our
      target 25~pc volume.  \varplxerrltsix\% of our uncertainties are less than
      6~mas, equivalent to a 15\% error at 25~pc.  Our \rchi\ distribution peaks
      near 1 (${\rm median}=\varmedrchi$), indicating that the scatter of data
      about our best-fit parallaxes is commensurate with the astrometric
      uncertainties.}
  \label{fig.ukirt.plx}
\end{center}
\end{figure}

 We present parallaxes, proper motions, and positions for \varngoodplx\ L and
T~dwarfs in Table~\ref{tbl.results}, representing the largest single sample of
infrared parallaxes published to date. Figure~\ref{fig.ukirt.plx} shows the
distributions for our absolute parallaxes, parallax uncertainties, and the
reduced $\chi^2$ of our final IRLS+MCMC solutions. We achieve a best parallax
precision of \varminplxerr~mas, and \varplxerrltsix\% of our uncertainties are
less than 6~mas. Our median precision of \varmedplxerr~mas is comparable to the
$\approx$1--4~mas precision of other ground-based infrared parallax observations
(LDA16). Our results also improve the parallaxes for \varnimprove\ objects with
previously published values.

Three targets in our sample --- ULAS~J085910.69+101017.1, PSO~J135.0395+32.0845,
and UGPS~J20480024+503821.9 --- have observations spanning only one year. While
our solutions for these objects appear to be robust, their parallaxes and proper
motions should be regarded as preliminary.

\subsection{Spectral Type Distribution and \gaiat\ Overlap}
\label{results.spt.gaia}
\begin{figure}
\begin{center}
    \includegraphics[width=1.04\columnwidth]{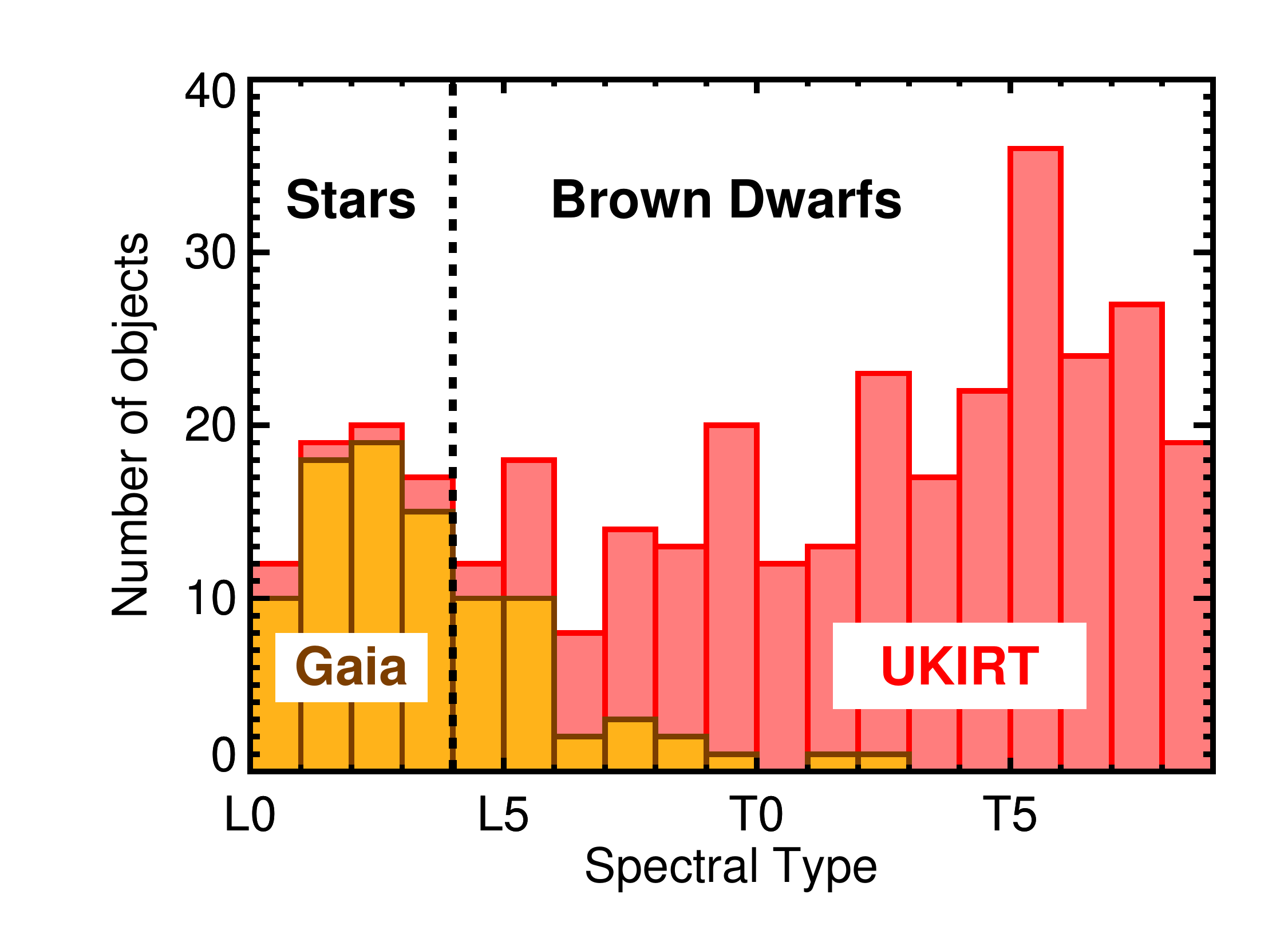}
    \caption{Spectral types for all of our {\varngoodplx}~UKIRT targets with
      parallax measurements (red).  We highlight the 94 targets that also have
      parallaxes from \gaiat\ (orange), which mostly have spectral types L0--L5.
      (\gaiat\ has also measured parallaxes for 10 primary stars of UKIRT
      wide-companion targets.)  UKIRT has obtained the first parallax
      measurements for \varnfirst\ targets, which are chiefly L6 and later
      types.  The black dashed line at L4 marks the approximate substellar
      boundary for field-age objects (early-L dwarfs can be either stars or
      younger brown dwarfs, but later types are exclusively brown dwarfs; DL17)
      indicating that \gaia\ is primarily a survey of stars, while most brown
      dwarfs require infrared observations for astrometry.}
  \label{fig.targspt}
\end{center}
\end{figure}

Figure~\ref{fig.targspt} shows the distribution of spectral types among our
targets with parallax measurements, highlighting the {\varngaiadirect} targets
(\varpgaia\%~of our sample) that also have a parallax from \gaiat. When both
optical and NIR spectral types are available for an object, we use the optical
types for L~dwarfs and the NIR types for T~dwarfs. While our UKIRT targets are
distributed across L0--T8 types, \gaiat\ is largely limited to earlier spectral
types; \varpgaiaearly\% of the \gaiat\ parallaxes are for spectral types earlier
than L7.

Including the \varngaiacomp\ UKIRT wide-companion targets for which \gaia\ has
observed the primary star, \gaiat\ provides more precise parallaxes than our
measurements for all \varngaia\ of our overlapping targets, along with
parallaxes for two more of our targets for which we did not measure a parallax
(2MASS~J09352803$-$2934596 and 2MASS~J21371044+1450475;
Section~\ref{plx.nosolution}).

Our UKIRT measurements include the first and only parallaxes for \varnfirst\
objects, most of which have spectral types L6 and later. Among these are
\hbox{\varntfirst\ T dwarfs}, comprising the largest sample of T dwarf
parallaxes published to date.

\subsection{Current Parallax Census}
\label{results.census}
\begin{figure}
\begin{center}
    \includegraphics[width=1.04\columnwidth]{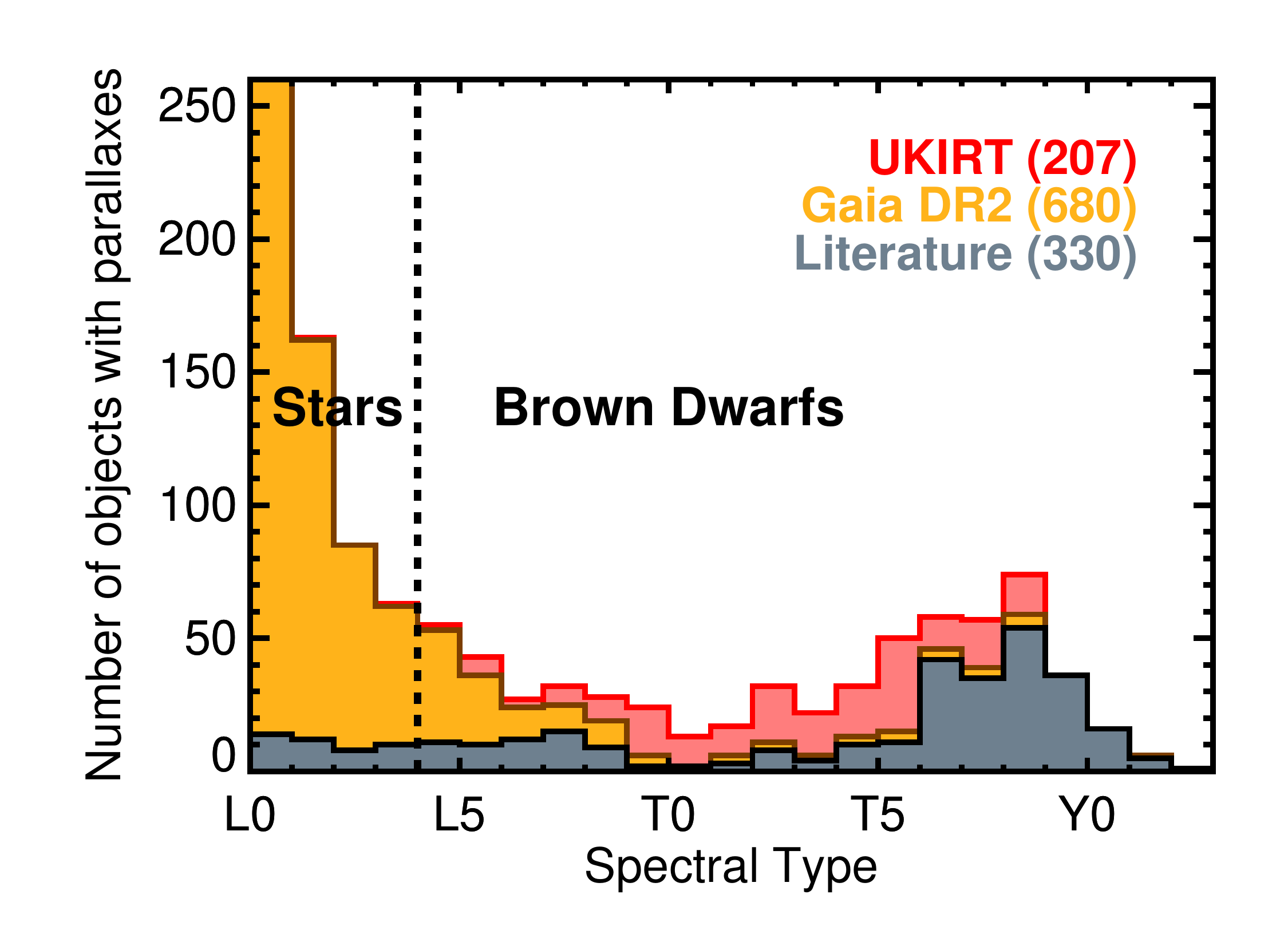}
    \includegraphics[width=1.04\columnwidth]{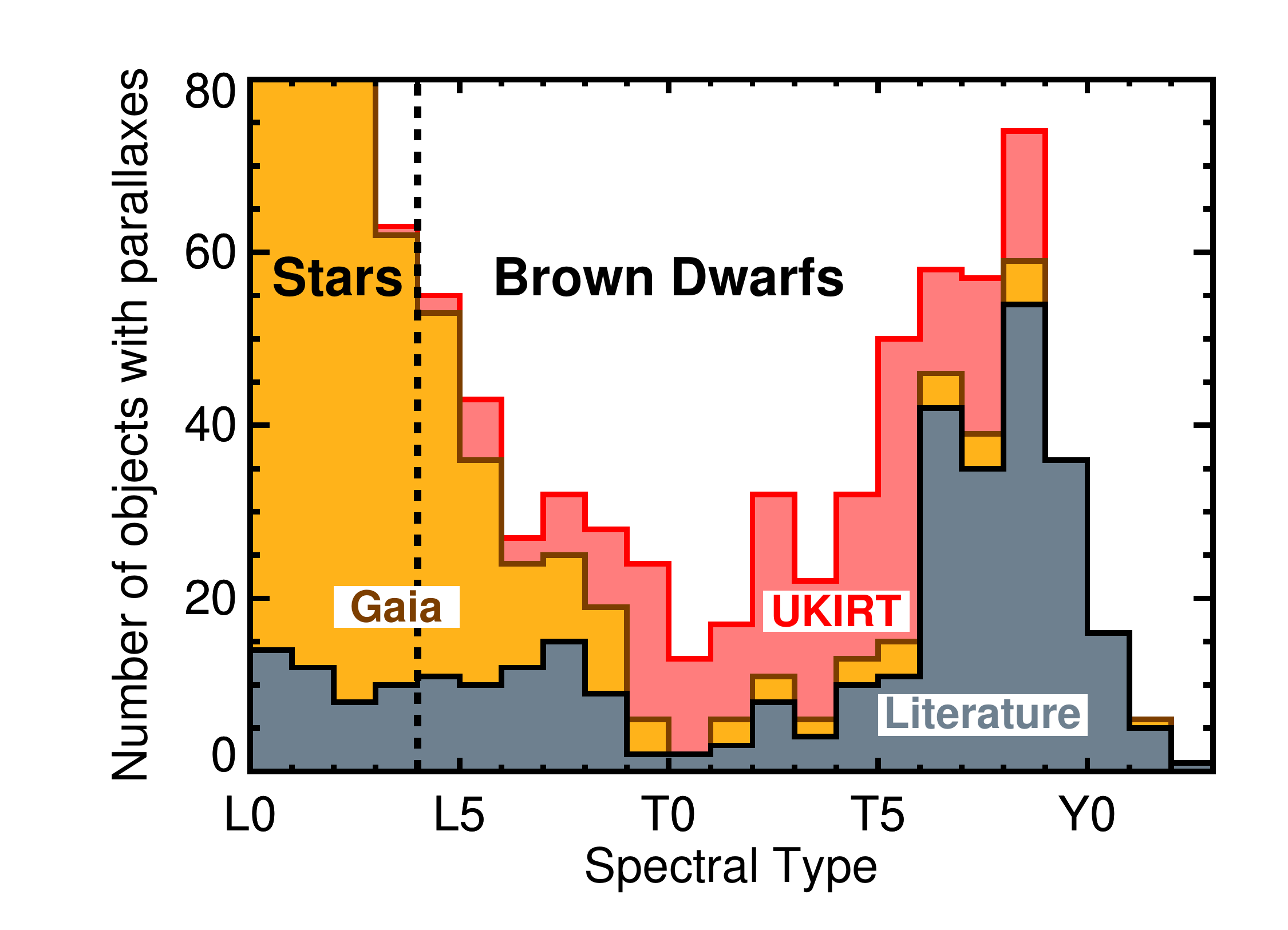}
    \caption{{\it Top panel}: distribution of spectral types for all L0 and
      later dwarfs with parallax measurements, stacking the contributions of
      parallaxes from UKIRT (red), \gaiat\ (orange), and other literature
      sources (gray).  For objects with more than one parallax, we show only the
      source of the most precise measurement.  The black dashed line at L4 marks
      the approximate substellar boundary for field-age objects as in
      Figure~\ref{fig.targspt}.  {\it Bottom panel}: same plot as top panel but
      with a smaller vertical axis range to highlight spectral types L5 and
      later.  \gaiat\ dominates spectral types L0--L5 but is mostly limited to
      observing stars.  Our UKIRT observations have filled in the L/T
      transition, tripling the number of L9--T5.5 parallaxes. The literature
      measurements for T6 and later types are chiefly new \spitzer\ parallaxes
      from K19.}
  \label{fig.plx.sources}
\end{center}
\end{figure}

\begin{figure}
\begin{center}
    \includegraphics[width=1.04\columnwidth]{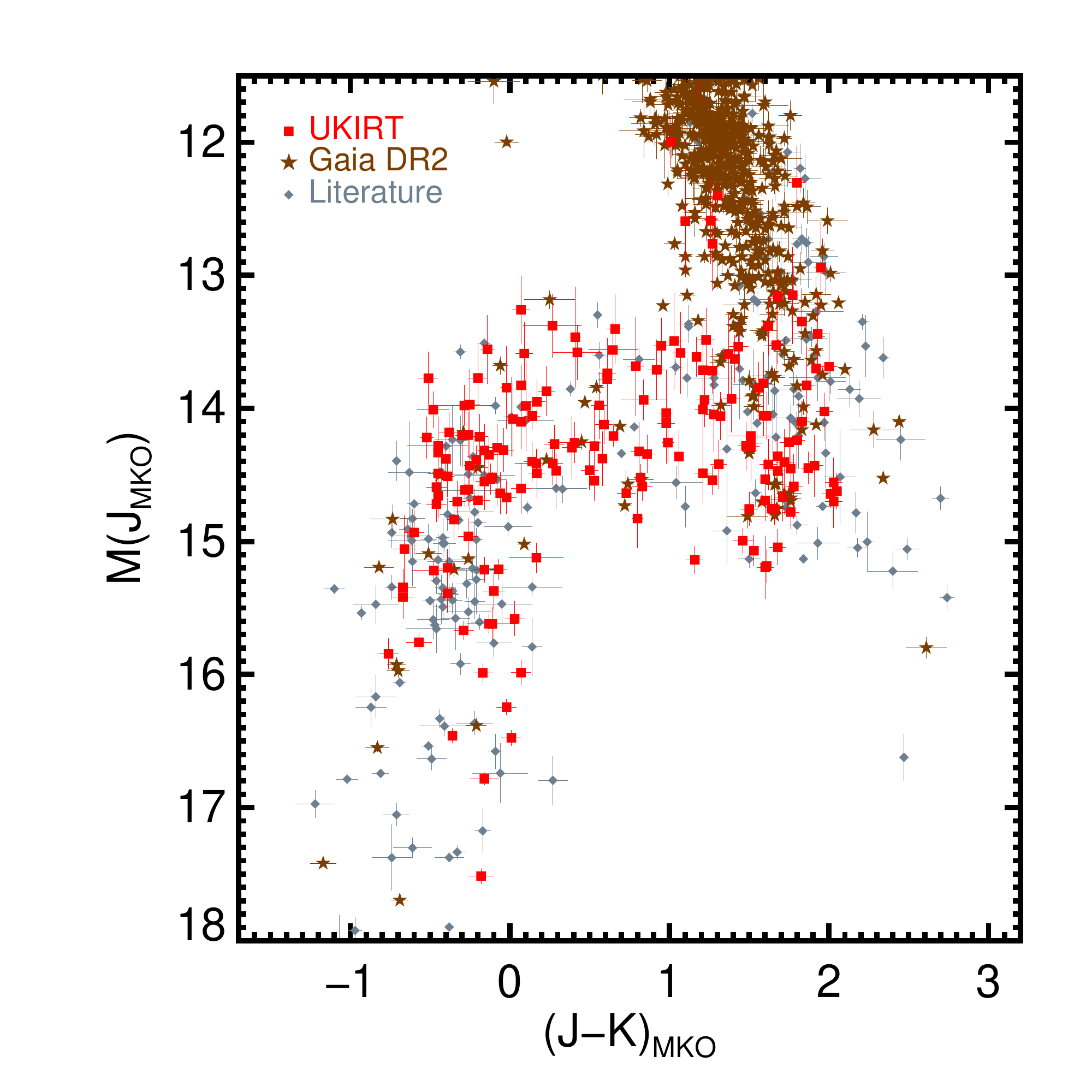}
    \caption{M$_J$ vs. $J-K$ (MKO) color-magnitude diagram for ultracool dwarfs
      having a parallax measurement with error~$\le20\%$.  We highlight the
      parallax contributions of UKIRT (red), \gaiat\ (brown), and other
      literature sources (gray), using only the most precise parallax available
      for an object.  Cooling brown dwarfs move down the L~dwarf sequence on the
      right, from right to left across the L/T transition, and down the
      later-T~dwarf sequence on the left-hand side.  Our UKIRT parallaxes have
      contributed \varntfirst\ new T~dwarf parallaxes.}
  \label{fig.cmd.all}
\end{center}
\end{figure}

We combined the Database of Ultracool Parallaxes (DL12; LDA16) and recent
parallaxes from the literature with our new UKIRT results and \gaiat\ parallaxes
for known L and T~dwarfs to create an updated census of ultracool parallaxes.
There are now published parallaxes for over 1100 L, T, and Y~dwarfs.
Figure~\ref{fig.plx.sources} shows the contributions from our UKIRT program,
\gaiat, and other sources to this parallax census as a function of spectral
type, choosing only the most precise measurement when more than one is available
for an object and excluding objects for which even the most precise parallax has
error~$>20\%$. Figure~\ref{fig.cmd.all} incorporates the parallaxes into an NIR
color-magnitude diagram showing the L and T dwarf population, using \jmko\ and
\kmko\ photometry from the literature along with \jmko\ photometry from our
UKIRT observations (Section~\ref{reduc.phot}).  For the 33 objects having \ktwo\
but not \kmko\ photometry, we used the M$_{K_S}$-based polynomials of DL17 to
convert \ktwo\ photometry into the MKO system.

With hundreds of measurements, \gaiat\ now provides the most precise parallaxes
for 90\% of L0--L5 dwarfs. Beyond L5, ground-based infrared measurements
dominate the L/T transition, with UKIRT now contributing the majority of L6--T6
parallaxes. In particular, we have tripled the number of parallaxes for spectral
types L8--T5.5. At these L/T transition spectral types, evolutionary and
atmospheric models have struggled to reproduce observed luminosities and
effective temperatures for objects with known masses and ages
\citep[e.g.,][]{Dupuy:2009ga,Dupuy:2015gl}. Our parallaxes will provide a wealth
of new empirical constraints for L/T transition modeling. For the coolest late-T
and Y dwarfs, CFHT (NIR; DL12, DL17), Magellan \citep[NIR;][]{Tinney:2014bl},
and \spitzer\ \citep[mid-IR;][K19]{Dupuy:2013ks,Beichman:2014jr,Martin:2018hc}
have produced most of the parallax measurements.

The mid-L spectral types at which UKIRT/WFCAM and other infrared instruments
begin to contribute a large fraction of the parallaxes is remarkably close to
the field substellar boundary at $\approx$L4 (DL17). Earlier spectral types will
contain a mixture of very low-mass stars and young brown dwarfs, while later
spectral types will be all brown dwarfs. Figure~\ref{fig.plx.sources} makes
clear that \gaia\ is largely a survey of the field population of stars,
measuring parallaxes for only a handful of brown dwarfs, while infrared
parallaxes will continue to be essential for studies of brown dwarfs in the era
of \gaia.

\subsection{Comparison of UKIRT and Literature Parallaxes}
\label{results.complit}

\begin{figure*}
  \centering
  \includegraphics[width=1.46\columnwidth, trim = 10mm 4mm 0 0]{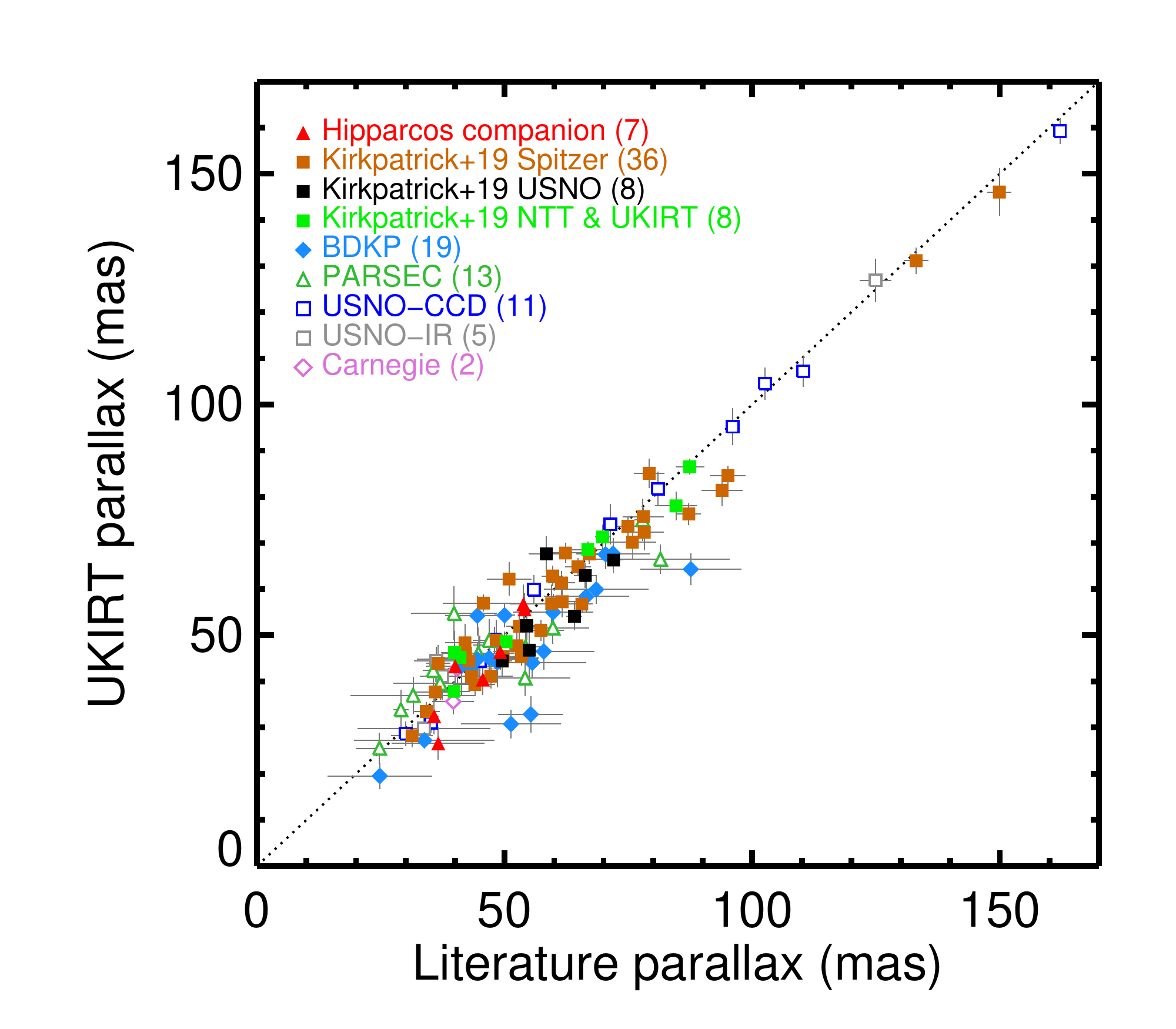}
  \begin{minipage}[t]{0.4\textwidth}
    \centering
    \includegraphics[width=1.00\columnwidth, trim = 12mm 18mm 0 0]{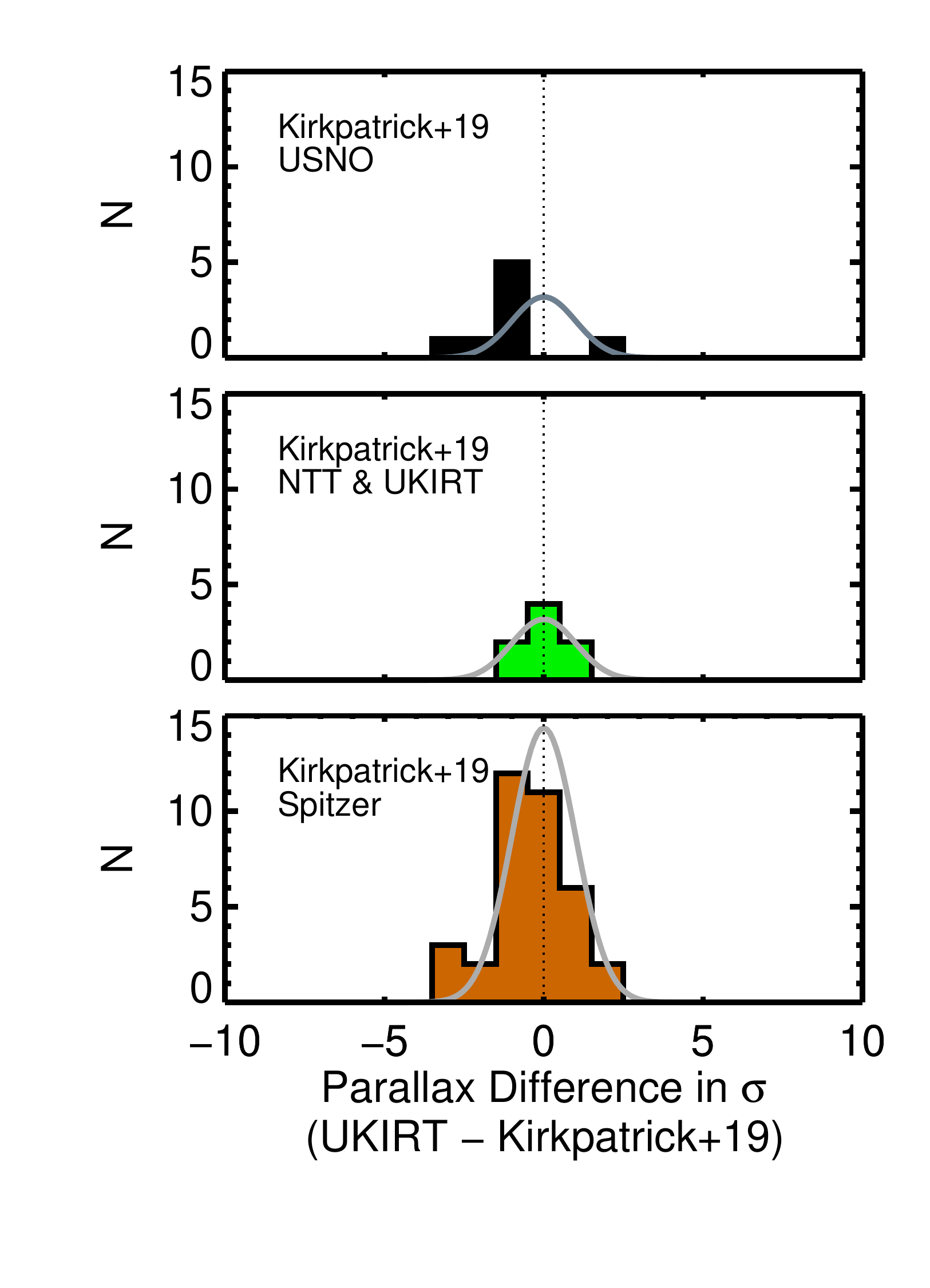}
  \end{minipage}
  \begin{minipage}[t]{0.4\textwidth}
    \centering
    \includegraphics[width=1.00\columnwidth, trim = 12mm 18mm 0 0]{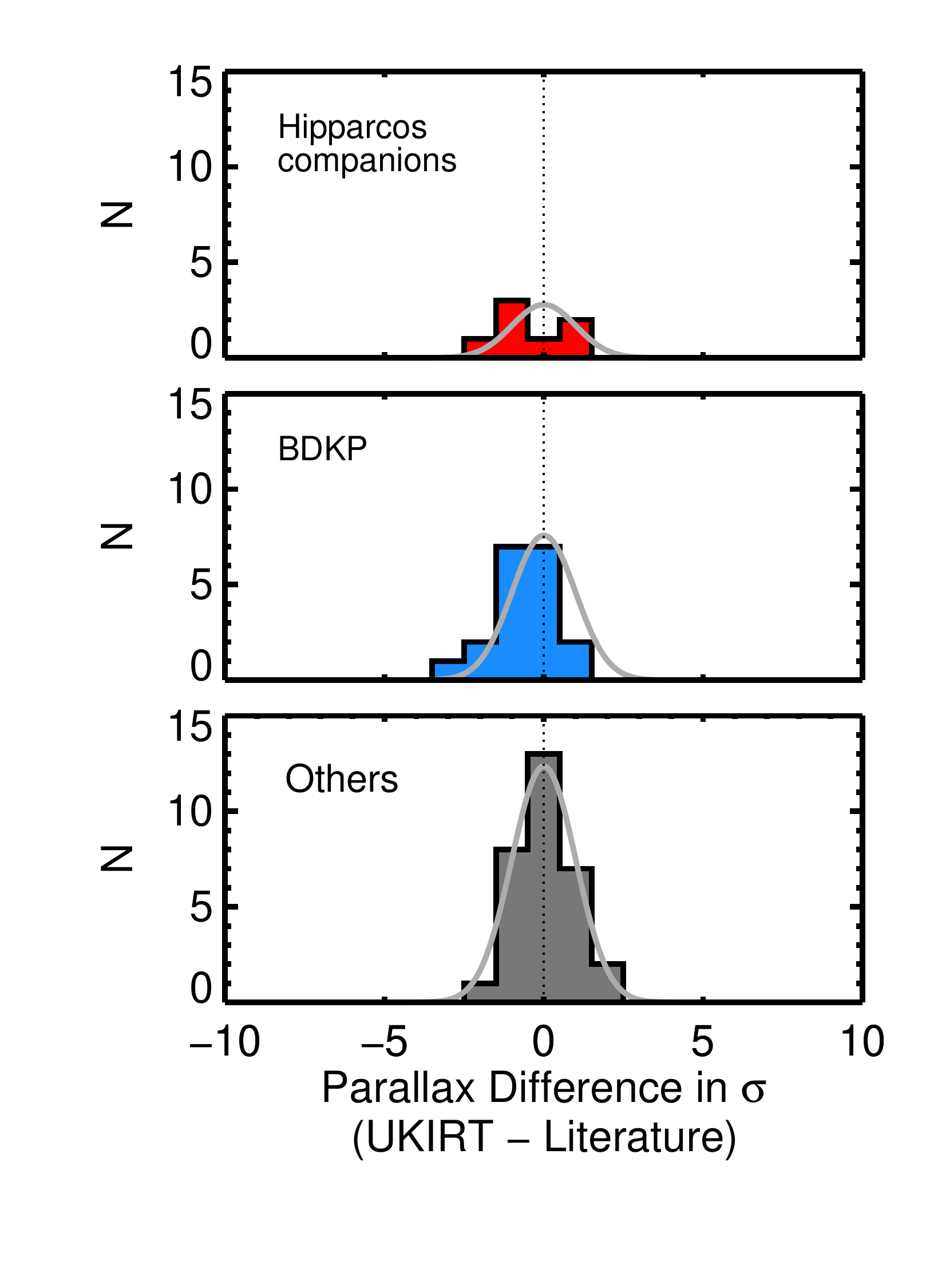}
  \end{minipage}
  \caption{{\it Top panel}: our UKIRT parallaxes compared with literature
    parallaxes for objects having both.  Parallaxes from specific programs are
    identified by color and symbol, with the number of objects shown in
    parentheses in the legend.  {\it Bottom panels}: histograms showing
    differences in parallaxes (UKIRT value -- literature value).  The dotted
    lines mark equality, and the gray curves indicate Gaussian distributions
    centered at 0 whose standard deviation is a $1\sigma$ difference in
    parallax, normalized to the number of objects in each histogram. Our
    parallaxes are consistent with literature measurements, in particular with
    those having ${\rm uncertainties}<10\%$.  The offset of the BDKP parallaxes
    relative to ours (BDKP parallaxes are systematically larger) is consistent
    with the offset relative to CFHT parallaxes seen by LDA16.  We note a
    similar offset of the K19 USNO and \spitzer\ parallaxes, although not the
    K19 NTT and UKIRT parallaxes.
    {\bf References}: 
    Kirkpatrick: K19;
    Hipparcos: \citet{vanLeeuwen:2007du};
    BDKP: \citet{Faherty:2012cy,Faherty:2016fx}; 
    PARSEC: \citet{Smart:2018en};
    USNO-CCD: \citet{Dahn:2017gu};
    USNO-IR: \citet{Vrba:2004ee,Gizis:2015fa};
    Carnegie: \citet{Weinberger:2016gy}.}
  \label{fig.complit}
\end{figure*}

For UKIRT targets having previous parallaxes in the literature, we compare our
new measurements in Figure~\ref{fig.complit}. We identify objects from several
major ultracool parallax programs, including
K19,
BDKP \citep{Faherty:2012cy,Faherty:2016fx},
PARSEC and NPARSEC \citep{Smart:2013km,Smart:2018en},
USNO-CCD \citep{Dahn:2002fu,Dahn:2017gu},
USNO-IR \citep{Vrba:2004ee,Gizis:2015fa},
and Carnegie \citep{Weinberger:2013cc,Weinberger:2016gy},
as well as companions to stars with {\it Hipparcos} parallaxes \citep{vanLeeuwen:2007du}.
We find good agreement for our parallaxes with most values from the
literature. Exceptions are listed in Table~\ref{tbl.results.different},
including parallaxes that differ by at least 2.0$\sigma$ or 10~mas.  The
exceptions include a Hipparcos parallax for the primary of HIP~73169B (discussed
in Section~\ref{results.interest.cpm}), three parallaxes from
\citet{Smart:2018en}, five parallaxes from BDKP, and eight parallaxes from K19.
In most of these cases with significant discrepancies, the UKIRT parallax is
smaller, a trend we examine below.

Most of the BDKP parallaxes that are statistically inconsistent with ours have
large uncertainties ($\gtrsim$10~mas). Overall, the BDKP parallaxes tend to be
larger, with a mean offset of $+0.6\pm0.2$~standard deviations (quoting the
standard error on the mean) from ours. A similar trend was also noted by LDA16
when comparing BDKP to CFHT parallaxes. K19 presented preliminary parallaxes
from three different programs, all of which have targets in common with our
UKIRT observations: \spitzer\ (36 targets in common), USNO's infrared camera
(eight targets in common), and NTT and UKIRT (eight targets in common). The
parallax uncertainties in K19 are very similar to ours, with $\approx$40\% of
the K19 measurements quoting smaller uncertainties. The astrometric solutions
from the three programs in K19 are produced by different reduction pipelines,
with the \spitzer\ pipeline described in K19, the USNO-IR pipeline in
\citet{Vrba:2004ee} and K19, and the pipeline used for NTT and UKIRT targets
described in \citet{Smart:2010gd,Smart:2013km}. The K19 parallaxes from
\spitzer\ and USNO are systematically larger than ours by $+0.5\pm0.2$~standard
deviations, similar to the offset seen with BDKP parallaxes. We note that the
K19 parallaxes from NTT and UKIRT show no significant offset relative to ours
($-0.09\pm0.32$~standard deviations). Compared to all literature parallaxes
other than those from BDKP and the K19 \spitzer\ and USNO programs, we calculate
essentially no offset ($+0.03\pm0.13$~standard deviations, with the literature
values being slightly larger).

\citet{Kirkpatrick:2011ey} discovered the unusually red L/T transition dwarf
WISEPA~J164715.59+563208.2, assigning it a spectral type of L9~pec~(red). They
also reported a $116\pm29$~mas parallax based on only five epochs of astrometry
from 2MASS, WISE, and \spitzer, all of which have astrometric uncertainties
$\ge$ 45~mas. Our $42.9\pm2.1 $~mas parallax is based on 11~epochs all measured
with UKIRT/WFCAM for which our median astrometric uncertainty is 11.2~mas, so we
prefer our parallax measurement.

\setcitestyle{citesep={,}}
\setcitestyle{citesep={;}}

\floattable
\begin{deluxetable}{lCCcCC}
\tablecaption{Objects with Discrepant Parallaxes \label{tbl.results.different}}
\tablecolumns{6}
\tabletypesize{\small}
\tablewidth{0pt}
\tablehead{
\colhead{Object} &
\colhead{Our \plx} &
\colhead{Published \plx} &
\colhead{References} &
\colhead{$\Delta\plx$} &
\colhead{$\Delta\plx / \sigma_{\Delta\plx}$} \\
\colhead{} &
\colhead{(mas)} &
\colhead{(mas)} &
\colhead{} &
\colhead{(mas)} &
\colhead{}
}
\startdata
2MASS~J02132062+3648506C\tablenotemark{a} & 70.1\pm4.0 &  45.0\pm10.0 & 1 &  25.1\pm10.8 & +2.3\sigma \\
SDSS~J032553.17+042540.1                  & 44.0\pm3.2 &  55.6\pm10.9 & 2 & -11.6\pm11.4 & -1.0\sigma \\
WISE~J043052.92+463331.6                  & 81.4\pm3.5 &  93.9\pm4.1  & 4 & -12.5\pm5.4  & -2.3\sigma \\
WISEPA~J050003.05$-$122343.2              & 84.6\pm2.2 &  95.1\pm3.6  & 4 & -10.5\pm4.2  & -2.5\sigma \\
WISE~J061437.73+095135.0                  & 56.7\pm2.0 &  65.6\pm2.2  & 4 &  -8.9\pm3.0  & -3.0\sigma \\
WISEPA~J085716.25+560407.6                & 76.3\pm2.4 &  87.2\pm2.4  & 4 & -10.9\pm3.4  & -3.2\sigma \\
2MASS~J09490860$-$1545485                 & 32.9\pm4.0 &  55.3\pm6.6  & 2 & -22.4\pm7.7  & -2.9\sigma \\
WISE~J103907.73$-$160002.9                & 45.3\pm2.0 &  53.4\pm2.6  & 4 &  -8.1\pm3.3  & -2.5\sigma \\
SDSS~J104409.43+042937.6                  & 30.7\pm3.2 &  51.3\pm10.1 & 2 & -20.6\pm10.6 & -1.9\sigma \\
SDSS~J115553.86+055957.5                  & 46.5\pm3.9 &  57.9\pm10.2 & 2 & -11.4\pm10.9 & -1.0\sigma \\
WISE~J125715.90+400854.2                  & 57.1\pm1.8 &  45.7\pm8.2  & 4 &  11.4\pm8.4  & +1.4\sigma \\
2MASSW J1326201$-$272937                  & 54.6\pm5.9 &  39.8\pm8.7  & 5 &  14.8\pm10.5 & +1.4\sigma \\
VHS~J143311.46$-$083736.3                 & 62.2\pm3.6 &  50.9\pm4.5  & 4 &  11.3\pm5.8  & +2.0\sigma \\
HIP~73169B\tablenotemark{a}               & 26.6\pm3.5 & 36.61\pm9.38 & 6 & -10.0\pm10.0 & -1.0\sigma \\
WISEPA~J164715.59+563208.2                & 42.9\pm2.1 &   116\pm29   & 3 & -73.1\pm29.1 & -2.5\sigma \\
2MASS~J17545447+1649196                   & 64.3\pm3.4 &  87.6\pm10.2 & 2 & -23.3\pm10.8 & -2.2\sigma \\
WISEPA~J190624.75+450808.2                & 54.1\pm3.0 &  64.1\pm1.6  & 4 & -10.0\pm3.4  & -2.9\sigma \\
2MASS J22092183$-$2711329                 & 40.6\pm3.9 &  54.1\pm9.2  & 5 & -13.5\pm10.0 & -1.4\sigma \\
2MASS J23185497$-$1301106                 & 66.4\pm3.2 &  81.5\pm14.0 & 5 & -15.1\pm14.3 & -1.0\sigma \\
\enddata
\tablecomments{This table includes all objects with literature and UKIRT
  parallax measurements that differ by at least 2$\sigma$ or 10~mas.}
\tablenotetext{a}{Discussed in Section~\ref{results.interest.cpm}.}
\tablerefs{(1) \citet{Deacon:2017kd}, (2) \citet{Faherty:2012cy}, (3)
  \citet{Kirkpatrick:2011ey}, (4) K19, (5) \citet{Smart:2018en}, (6)
  \citet{vanLeeuwen:2007du}}
\end{deluxetable}

\subsection{Comparison of UKIRT and \gaiat\ Parallaxes}
\label{results.compgaia}

As described in Section~\ref{results.spt.gaia}, \gaia\ has measured parallaxes
mostly for ultracool dwarfs with spectral types~$\le$L5. All but three of our
\varngaia\ UKIRT targets with \gaiat\ parallaxes are L~dwarfs. Two of the
remaining T~dwarfs are very nearby and bright: SIMP~J013656.57+093347.3
\citep[T2.5;][]{Artigau:2006bh} at 6~pc and WISE~J203042.79+074934.7
\citep[T1.5;][]{Mace:2013jh} at 10~pc. In the third case, \gaiat\ measured a
parallax for the M~dwarf binary 2MASS~J02132062+3648506AB, which has a wide
T~dwarf companion 2MASS~J02132062+3648506C \citep{Deacon:2017kd} in our
UKIRT sample.

\begin{figure*}
  \centering
  \begin{minipage}[t]{0.62\textwidth}
    \includegraphics[width=1.00\columnwidth, trim = 10mm 0 0 0]{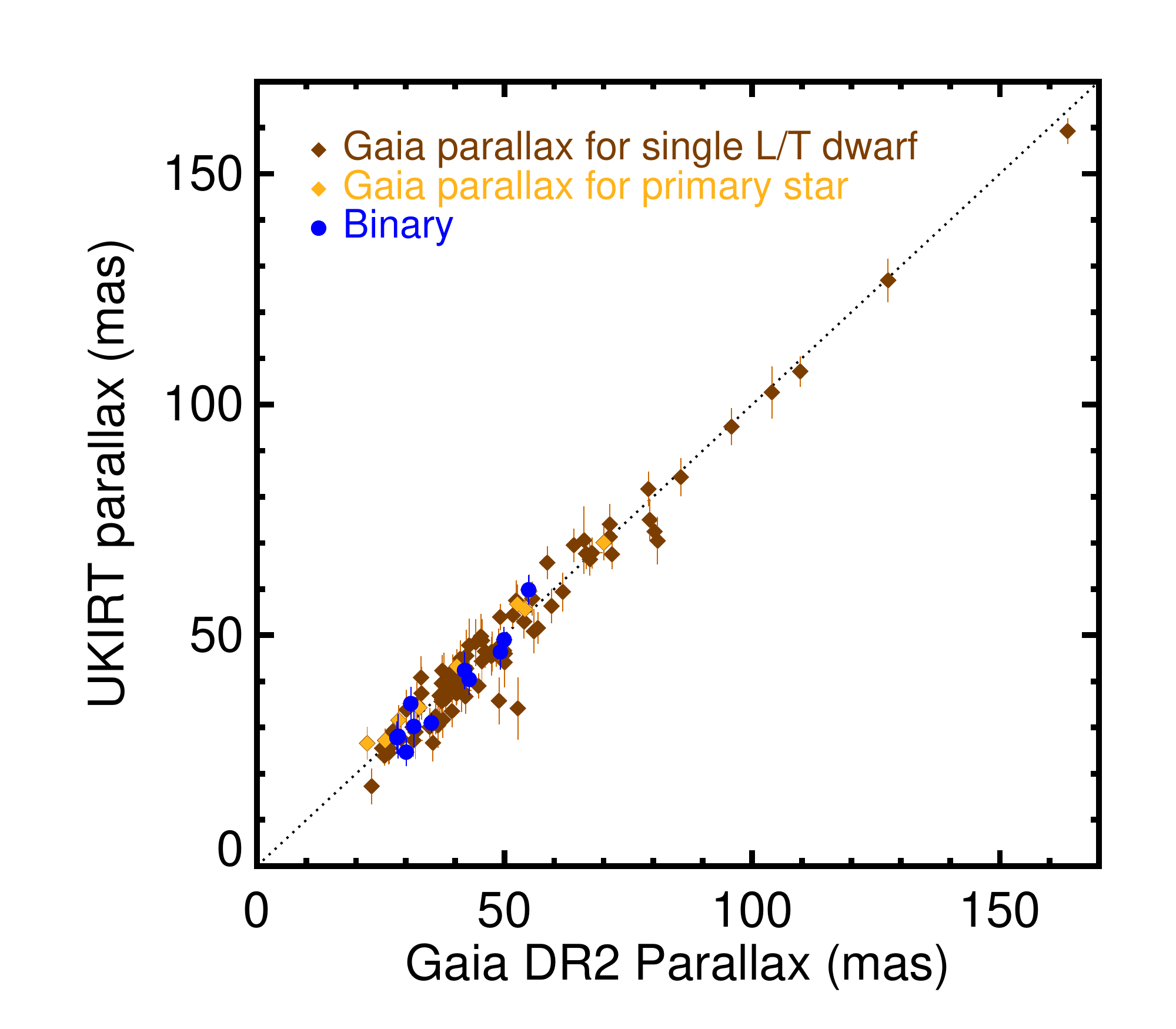}
  \end{minipage}
  \hfill
  \begin{minipage}[t]{0.36\textwidth}
    \includegraphics[width=1.00\columnwidth, trim = 16mm 0 0 0]{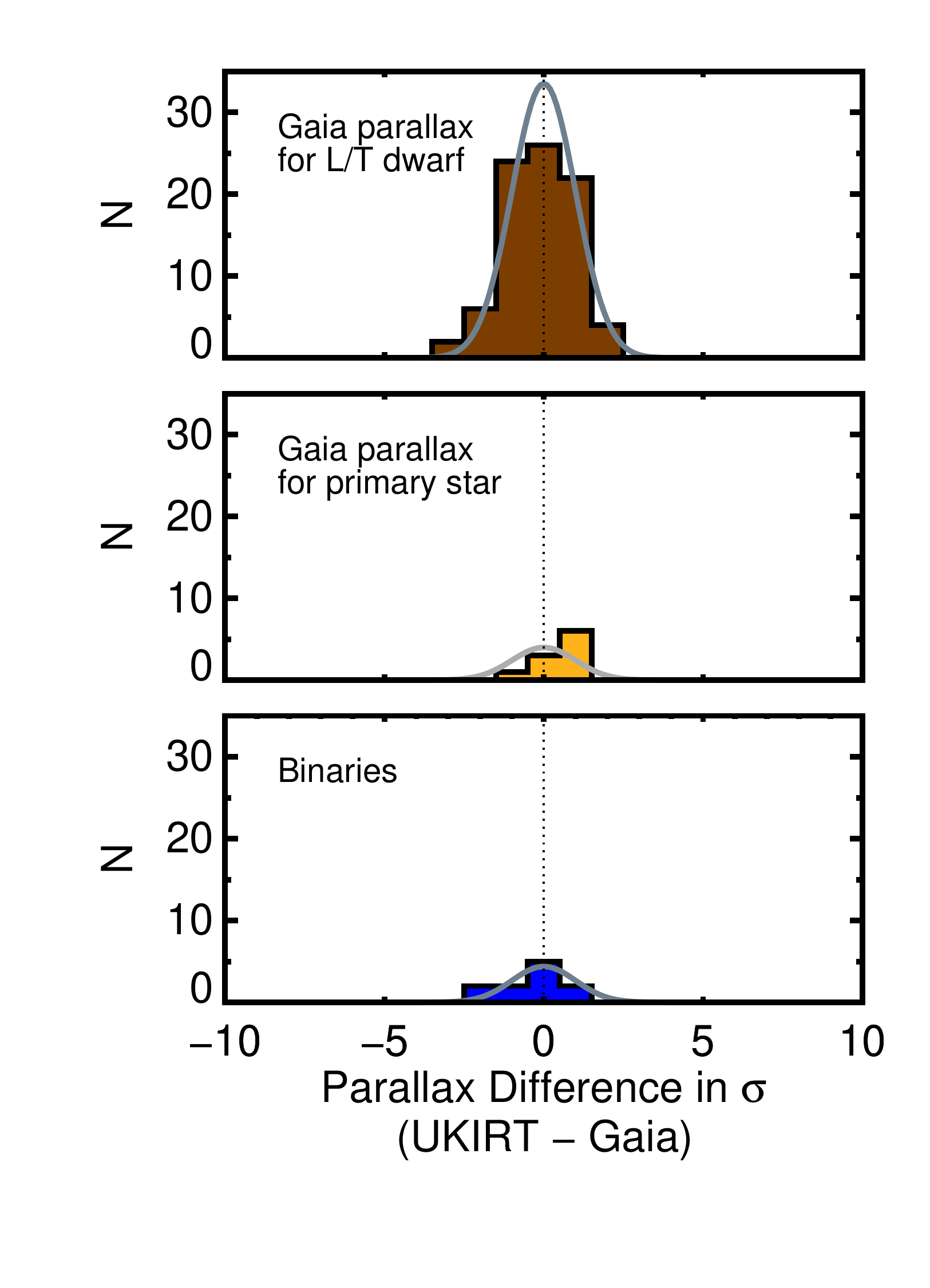}
  \end{minipage}
  \caption{Our UKIRT parallaxes compared with \gaiat\ parallaxes for the
    \varngaia~objects common to both data sets, presented in the same format as
    Figure~\ref{fig.complit}.  The brown diamonds mark single objects observed
    directly by \gaia, the orange diamonds indicate wide companions with \gaia\
    parallaxes from their host stars, and the blue circles mark known binaries
    observed directly by \gaia.  The UKIRT parallaxes agree very well with
    \gaiat, with $\rchi=1.09$ for the 1:1 model.  The correspondence between
    UKIRT and \gaia\ parallaxes for binaries is consistent with the
    correspondence for all objects, implying that our parallax measurements are
    not biased by partially resolved targets in our images.}
  \label{fig.comp.gaia}
\end{figure*}

We compare our parallaxes with those of \gaiat\ in Figure~\ref{fig.comp.gaia},
which show good overall agreement. We calculate a mean offset of
$-0.16\pm0.10$~standard deviations (quoting the standard error on the mean),
with our parallaxes being smaller than {\gaiat}'s, similar to the offset of
UKIRT parallaxes relative to some literature values although only marginally
statistically significant ($1.6\sigma$). The more discrepant parallaxes show no
overall trend with parallax. The offset suggests that our UKIRT measurements may
be systematically slightly smaller than those from \gaiat. However, since there
is evidence that the \gaiat\ parallax uncertainties or our own uncertainties are
slightly underestimated for the objects overlapping our UKIRT sample
(Section~\ref{plx.relative.best}), the offset with our UKIRT parallaxes is
likely to be even less significant.

\subsection{Comparison of Literature and \gaiat\ Parallaxes}
\label{results.compgaialit}
\begin{figure*}
  \centering
  \includegraphics[width=1.5\columnwidth, trim = 10mm 4mm 0 0]{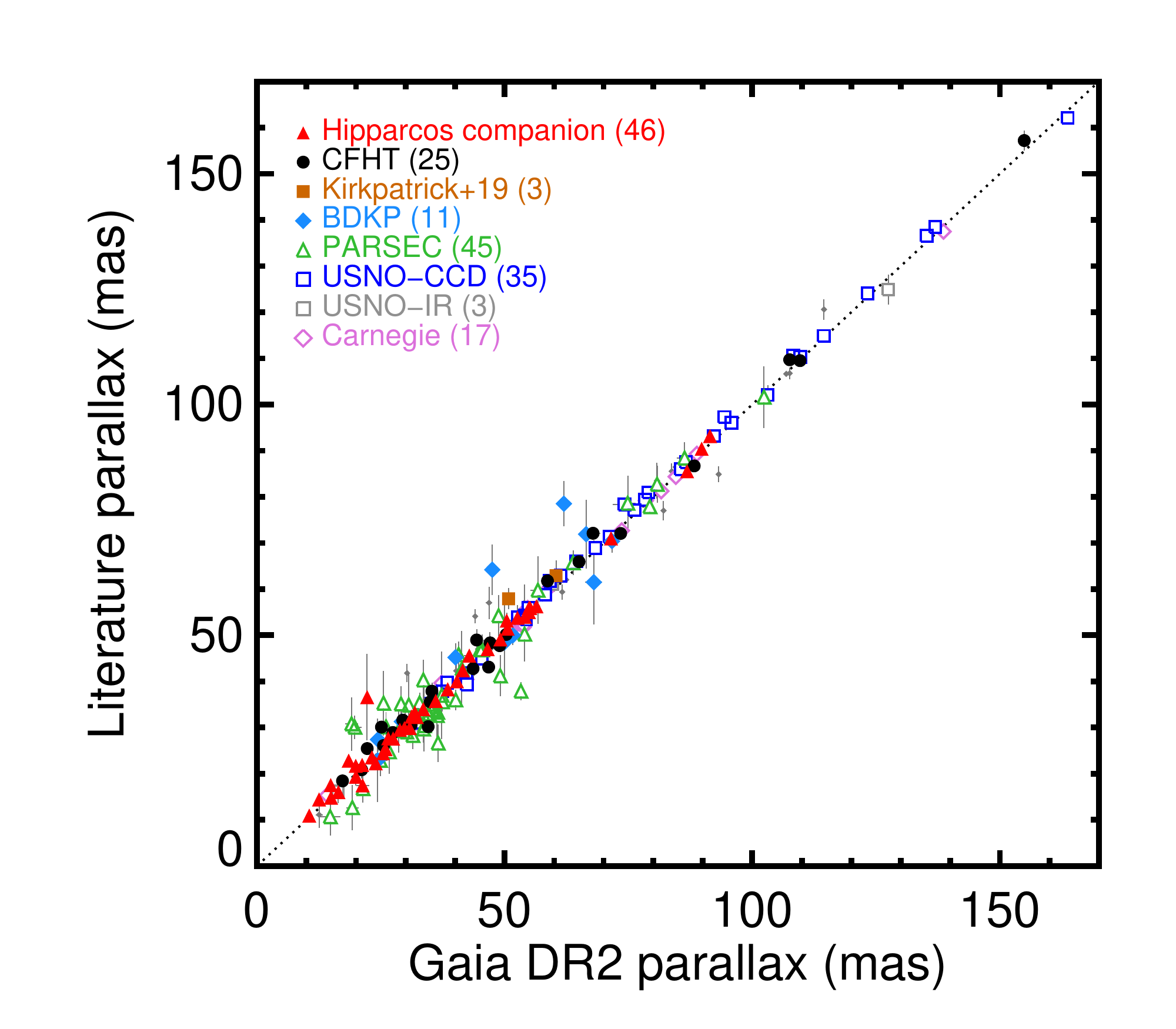}
  \begin{minipage}[t]{0.4\textwidth}
    \centering
    \includegraphics[width=1.00\columnwidth, trim = 12mm 18mm 0 0]{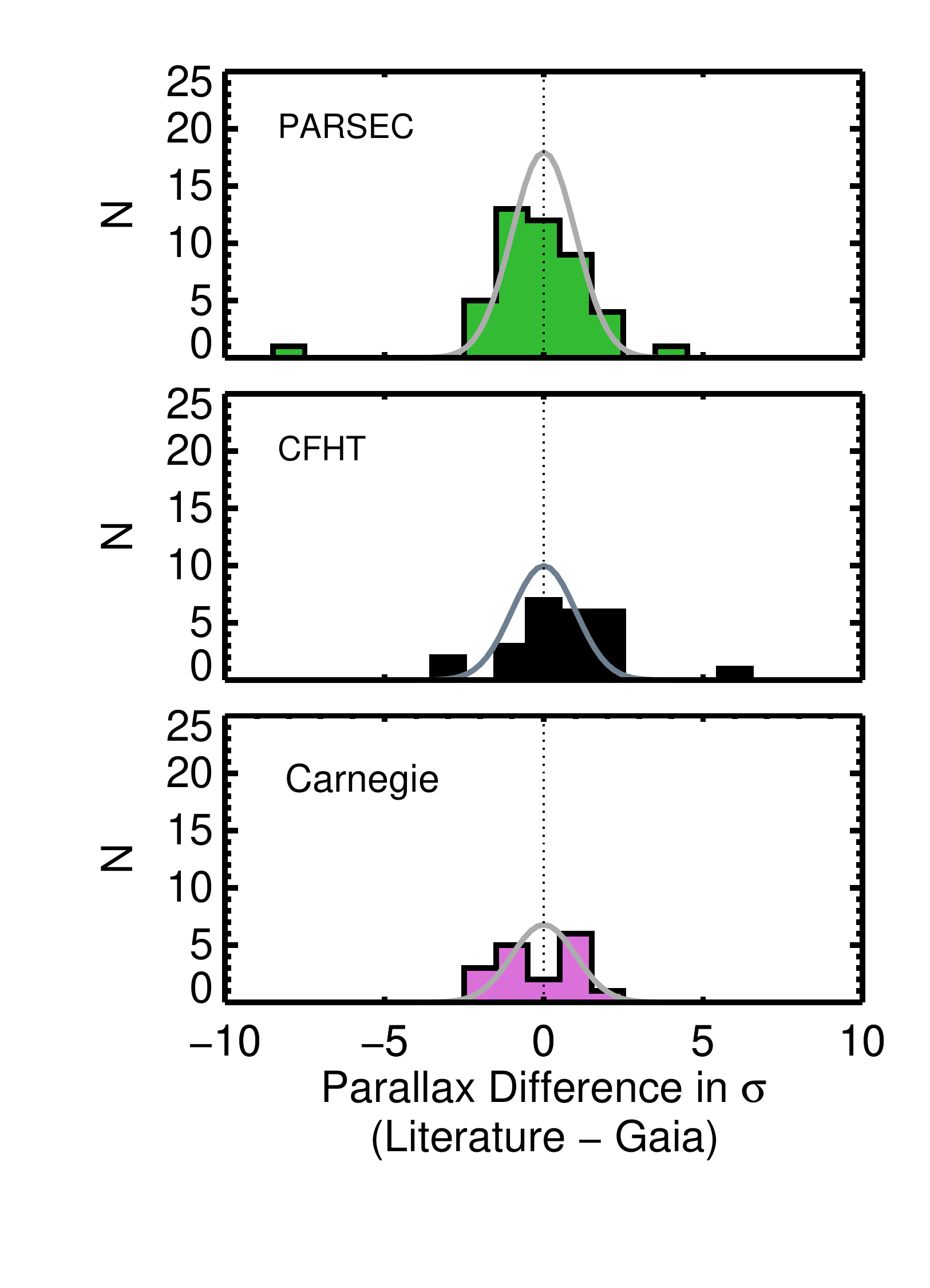}
  \end{minipage}
  \begin{minipage}[t]{0.4\textwidth}
    \centering
    \includegraphics[width=1.00\columnwidth, trim = 12mm 18mm 0 0]{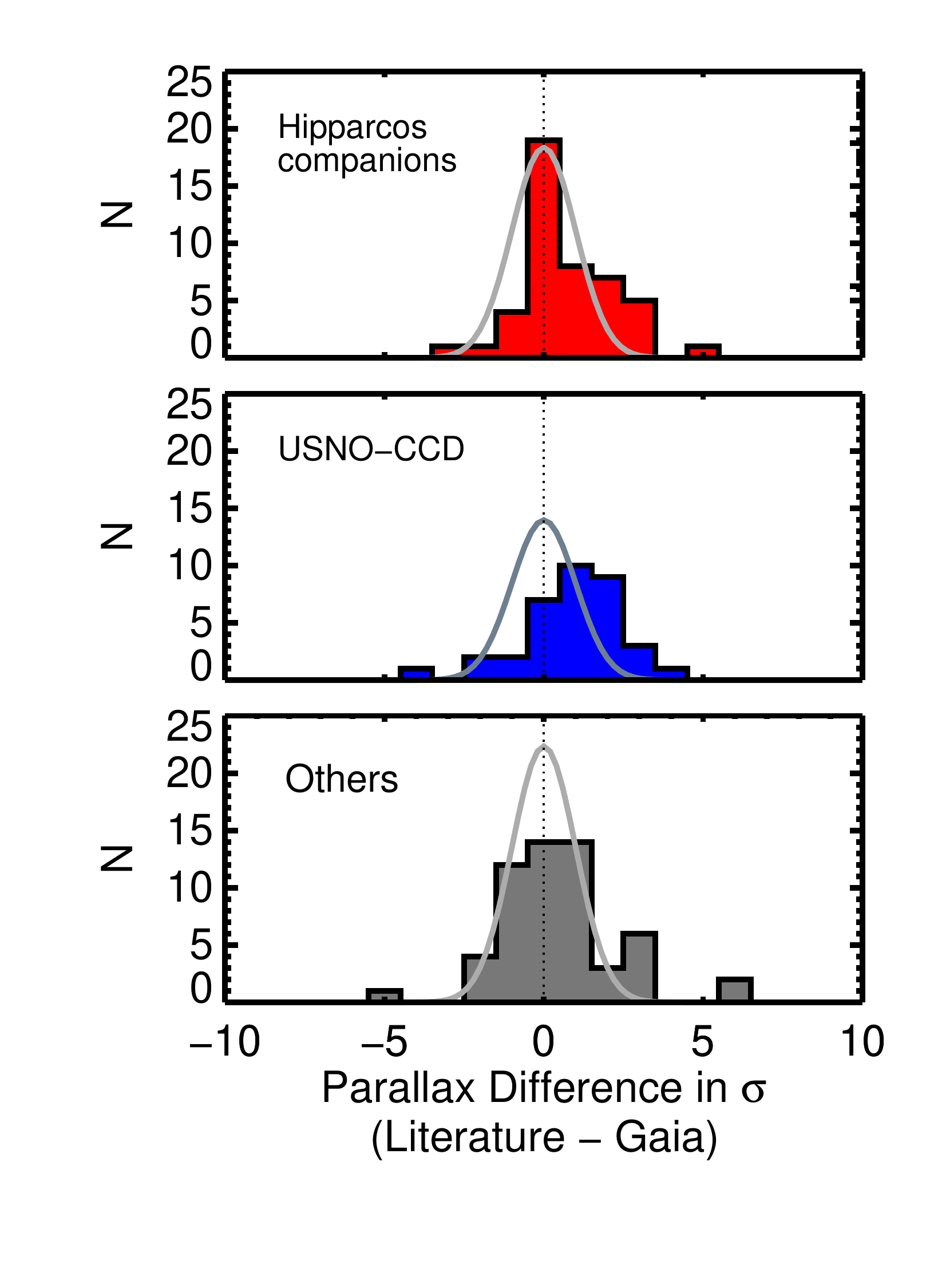}
  \end{minipage}
  \caption{Parallaxes measured by \gaiat\ for L and T~dwarfs also having
    parallax measurements in the literature, presented in the same format as
    Figure~\ref{fig.complit}.  Parallaxes from the programs overlapping with our
    UKIRT observations (Figure~\ref{fig.complit}) and from CFHT are identified
    by color and symbol, with the number of objects shown in parentheses in the
    legend.  Most literature measurements are consistent with \gaiat.  We note
    that literature parallaxes are systematically slightly larger than those
    from \gaiat, similar to the systematic offset we found with some literature
    parallaxes relative to our UKIRT measurements
    (Section~\ref{results.complit}; Figure~\ref{fig.complit}).
      {\bf References}: 
      Hipparcos: \citet{vanLeeuwen:2007du}; 
      CFHT: DL12, LDA16; 
      Kirkpatrick: K19;
      BDKP: \citet{Faherty:2012cy,Faherty:2016fx}; 
      PARSEC: \citet{Smart:2018en};
      USNO-CCD: \citet{Gizis:2015ey,Dahn:2017gu};
      USNO-IR: \citet{Vrba:2004ee,Gizis:2013ik,Gizis:2015fa};
      Carnegie: \citet{Weinberger:2016gy}.}
  \label{fig.compgaialit}
\end{figure*}

In addition, we cross-matched the updated census of most precise L and T~dwarf
parallaxes from Section~\ref{results.census} with \gaiat. We compare parallaxes
for matched objects in these catalogs in Figure~\ref{fig.compgaialit},
highlighting the programs that we compared to our UKIRT measurements in
Figure~\ref{fig.complit} as well as CFHT (DL12, LDA16). \gaiat\ confirms and
refines most previous parallax measurements. Exceptions tend to have large
literature uncertainties ($\gtrsim$10~mas), similar to what we saw in our
comparison with UKIRT values (Section~\ref{results.complit}). The largest
literature sample overlapping with our UKIRT program, K19, has only three
objects in common with \gaiat\ because K19 targeted late-T and Y dwarfs that are
mostly too red and faint for \gaia\ (and two of those three objects are in fact
companions to main-sequence stars that have \gaiat\ parallaxes).

We calculate an overall offset of $+0.40\pm0.11$~standard deviations (quoting
the standard error on the mean) for literature parallaxes with respect to
\gaiat, with the literature values tending to be larger. This offset is similar
to the offsets of some literature parallaxes relative to our UKIRT measurements
(Section~\ref{results.complit}), but is in the opposite direction of (and more
significant than) the $-0.16\pm0.10$ standard deviation offset of our UKIRT
measurements relative to \gaiat\ (Section~\ref{results.compgaia}). The offset
compared with \gaiat\ is the largest for measurements from
the BDKP program ($+0.8\pm0.4$ standard deviations),
CFHT ($+0.6\pm0.4$ standard deviations),
USNO-CCD ($+0.9\pm0.3$ standard deviations),
and Hipparcos ($+0.7\pm0.2$ standard deviations);
although, the small uncertainties quoted by the latter three programs enhance
the differences found by this analysis. Excluding the measurements from BDKP,
CFHT, USNO-CCD, and Hipparcos, the literature parallax offset relative to
\gaiat\ falls to $+0.02\pm0.17$ standard deviations, suggesting that the primary
drivers of the overall offset are a couple of systematically $\approx$10--20~mas
larger parallaxes from BDKP (visible in Figure~\ref{fig.compgaialit}),
underestimated uncertainties from CFHT, Hipparcos, USNO-CCD, and/or \gaiat, and
$\approx$1~mas larger parallaxes from CFHT, USNO-CCD, and Hipparcos. We note
also that the \gaiat\ {\tt astrometric\_excess\_noise} for most of the objects
with CFHT parallaxes is $\approx$1--4 mas, comparable to or larger than the
formal CFHT uncertainties. Future data releases from \gaia\ that include
parallax measurements from longer time baselines as well as refined astrometric
fitting may shed further light on these offsets that we have identified.

\setcitestyle{citesep={,}}
\setcitestyle{citesep={;}}

\subsection{Parallaxes for Binaries}
\label{results.binaries}
Our targets included 36 objects known or strongly suspected to be binaries, with
separations mostly requiring high-angular resolution imaging or high-resolution
spectroscopy to resolve. Our reduction pipeline assumed that each target was a point
source and did not attempt to fit for binary sources in the images. However, it
is possible for binaries with separations similar to the seeing during
observations ($\approx$0.5$''$ on our best nights) to be partially or fully
resolved, which could impact the astrometry derived from those images. We
therefore looked for indications of bias in our astrometric solutions for these
binaries.

The parallax fits shown in Figure~\ref{fig.curves} for the binaries appear to be
consistent with the fits for single objects. The median \rchi\ for final
IRLS+MCMC solutions for binaries is \varmedrchibin, almost exactly the same as
the median of \varmedrchi\ for all parallax solutions
(Section~\ref{plx.relative.irls}). The \rchi\ value of the 1:1 fit for UKIRT
with Gaia for just the 11 binaries in common (Figure~\ref{fig.comp.gaia}) is
1.28, larger than the $\rchi=1.09$ for all common objects but not enough to
imply a significantly poorer fit. The \gaiat\ parallaxes for these 11 binaries
are systematically $0.28\pm0.31$~standard deviations larger than our UKIRT
measurements (quoting the standard error on the mean), consistent with the mean
offset of single object parallaxes ($0.15\pm0.11$~standard deviations). We
therefore do not see evidence that our parallaxes for binaries are significantly
influenced by non-point-source appearances in our images.

There is one possible exception: the $0.9''$ L3.5+L4 binary
2MASS~J15500845+1455180 \citep{Cruz:2007kb,Burgasser:2009gs} is clearly resolved
in our UKIRT images. Our pipeline nevertheless identified and treated
2MASS~J15500845+1455180 as a single object at the position of the brighter
primary in each frame. The astrometric solution appears to be reasonable, with
$\rchi=1.12$ and no trends or unusual scatter in the residuals.  Our UKIRT
parallax of $24.5\pm3.0$~mas differs by $1.8\sigma$ from the \gaiat\ parallax of
$30.12\pm0.92$~mas.  We note that \gaia\ has also resolved the system and
identifies both objects in DR2, and again, only the primary component has a
parallax.  However, as with our result, there are no indications that this DR2
solution is suspect.  We include our solution in our results but advise using it
with caution.

\section{Individual Objects of Interest}
\label{results.interest}

\subsection{Young Moving Group Candidates}
\label{results.interest.ymg}

Young moving groups (YMGs) are loose associations of young
($\approx$10--200~Myr) stars and brown dwarfs that share common motion through
space \citep[e.g.,][]{Zuckerman:2004ex}. The members of a YMG are thought to
have formed in a single star-forming region and are now gravitationally unbound
and spatially extended, but they retain similar kinematics. The ages of YMGs are
typically estimated from theoretical stellar isochrones
\citep[e.g.,][]{Bell:2015gw}. Associating a brown dwarf with a YMG, usually via
distance, kinematics, and spectroscopic indications of youth, provides a
valuable means of determining the brown dwarf's age that would otherwise be
difficult to constrain \citep[e.g.,][]{Gagne:2015dc,Aller:2016kg}.
Table~\ref{tbl.interest.ymg} lists four objects that were previously identified
as candidate members of YMGs based on sky position, proper motion, photometric
distance, and, in one case, a radial velocity. Our UKIRT parallaxes provide new
data for determining kinematic consistency with the YMGs in question, and we
update their candidacies here.

\floattable
\begin{deluxetable}{lcCccCc}
\tablecaption{Updates to Young Moving Group Membership Status \label{tbl.interest.ymg}}
\tablecolumns{7}
\tabletypesize{\small}
\tablewidth{0pt}
\tablehead{
\colhead{} &
\colhead{} &
\colhead{} &
\colhead{Previous} &
\colhead{} &
\colhead{} &
\colhead{Updated} \\
\colhead{Object} &
\colhead{SpT} &
\colhead{\dphot} &
\colhead{Membership} &
\colhead{References} &
\colhead{$d_\plx$} &
\colhead{Membership} \\
\colhead{} &
\colhead{} &
\colhead{(pc)} &
\colhead{} &
\colhead{} &
\colhead{(pc)} &
\colhead{}
}
\startdata
WISEA~J114724.10$-$204021.3 & L7 (red) & 31.2\pm1.5 & TWA (96\%)     & 1, 2    & 37.3\pm2.8 & Field (97.5\%)\tablenotemark{a} \\
PSO~J071.8769$-$12.2713     & T2       & 25.9\pm3.1 & BPMG (86.4\%)  & 3       & 43.3\pm7.1 & Field (99.9\%) \\
WISE~J180952.53$-$044812.5  & T1       & 13.1\pm1.3 & ABDMG (93.8\%) & 3, 4, 5 & 20.3\pm1.4 & Field (99.9\%) \\
PSO~J319.3102$-$29.6682     & T0       & 15.6\pm1.6 & BPMG (97.1\%)  & 3       & 19.2\pm2.5 & BPMG (99.0\%) \\
\enddata
\tablecomments{Probabilities for previous memberships were calculated using the
  online \banyanii\ tool \citep{Malo:2013gn,Gagne:2014gp}. Probabilities for
  updated memberships were calculated using the online \banyans\ tool
  \citep{Gagne:2018tx}.  Memberships: TWA = TW Hydrae Association, BPMG =
  $\beta$ Pictoris Moving Group, ABDMG = AB Doradus Moving Group.}
\tablenotetext{a}{The change in status from TWA to field is a result of
  differences in the models of TWA and/or the field population between
  \banyanii\ and \banyans.  WISEA~J1147$-$2040 lies at the extreme edge of the
  TWA population, so determining its true membership will require more careful
  analysis.}
\tablerefs{(1) \citet{Schneider:2016iq}, (2) \citet{Gagne:2017gy}, (3)
  \citet{Best:2015em}, (4) \citet{Mace:2013jh}, (5) \citet{Best:2013bp}.}
\end{deluxetable}

\subsubsection{WISEA~J114724.10$-$204021.3}
This extremely red L7 dwarf was discovered and identified by
\citet{Schneider:2016iq} and \citet{Gagne:2017gy} as a likely member of the TW
Hydrae Association \citep[TWA;][]{Kastner:1997fk}, a nearby $10\pm3$~Myr old
\citep{Bell:2015gw} moving group. At this young age, WISEA~J1147$-$2040 would
have a mass of only $6.6\pm1.9$~\mjup\ \citep{Faherty:2016fx}, making it a rare
free-floating planet. We used the \banyans\ tool \citep{Gagne:2018jj} to
calculate an updated probability of membership in TWA for WISEA~J1147$-$2040.
\banyans\ calculates Bayesian estimates for an object's membership in moving
groups using sky position and proper motion, as well as radial velocity and/or
parallax if available. Using our parallax, proper motion, and position from
UKIRT, along with the radial velocity $7\pm3$~\kms\ from \citet{Gagne:2017gy},
we calculate for WISEA~J1147$-$2040 a 0.0\% probability of TWA membership and a
97.5\% probability that it is a field object, casting doubt on its expected
membership in TWA. \banyans\ is a recent update of the \banyanii\ tool
\citep{Malo:2013gn,Gagne:2014gp} that \citet{Schneider:2016iq} originally used
to evaluate WISEA~J1147$-$2040. (\banyanii\ is no longer available online.) We
note that \banyans\ also calculates a 1.5\% probability of TWA membership
(93.2\% for field membership) from the data used by \citet{Schneider:2016iq}.
The drastic decrease in likelihood of TWA membership predicted by \banyans\ must
therefore be a consequence of updates to the models of TWA and/or the field
population used by \banyans\ rather than our parallax measurement.
WISEA~J1147$-$2040 lies at the near and northern extremes of the TWA population,
so estimates of its membership probability are sensitive to the boundaries of
the models used, and determining its true membership will require more careful
analysis.

\citet{Best:2017br} noted that WISEA~J1147$-$2040 may be a binary based on the
similarity of its NIR spectrum and luminosity to the resolved
$\approx$equal-mass binary 2MASS~J1119$-$1137AB, which is also a strong
candidate TWA member \citep{Kellogg:2016fo,Best:2017br}. The distance measured
by our parallax for WISEA~J1147$-$2040, $37.3\pm2.8$~pc, is marginally greater
than the photometric distances estimated for a single young L7~dwarf by
\citet[$31.2\pm1.5$~pc]{Schneider:2016iq} and
\citet[$27.3\pm6.9$~pc]{Best:2017br}. This provides tentative evidence that a
companion contributes to WISEA~J1147$-$2040's luminosity, although not
necessarily a companion of near-equal luminosity, as is the case with
2MASS~J1119$-$1137AB. We also note that similarly to WISEA~J1147$-$2040,
\banyans\ calculates that 2MASS~J1119$-$1137AB is most likely a field object
(63.2\% probability) and not a member of TWA (0.1\%).

\subsubsection{PSO~J071.8769$-$12.2713}
This T2~dwarf was discovered and identified by \citet{Best:2015em} as a
candidate member (86.4\% from \banyanii) of the $\beta$ Pictoris Moving Group
\citep[BPMG;][]{Zuckerman:2001go}, based on PSO~J071.8$-$12.2's proper motion,
sky position, and photometric distance of $25.9\pm3.1$~pc. BPMG has an age of
$24\pm3$~Myr \citep{Bell:2015gw}; to date, no free-floating T~dwarf has a
confirmed age this young. However, our new parallax places PSO~J071.8$-$12.2 at
a larger distance of $43.3\pm7.1$~pc, and \banyans\ now finds a 99.9\%
probability that PSO~J071.8$-$12.2 is a field object. We note that \banyans\
finds the same result with and without our parallax measurement, and it
contrasts with the moderate probability (56.5\%) of membership in the AB Doradus
Moving Group \citep[ABDMG;][]{Zuckerman:2004ds} found by \citet{Vos:2019ks}
using \banyans. The source of this discrepancy is our new UKIRT proper motion in
R.A., which differs significantly from previous measurements
\citep{Best:2015em,Best:2018kw}.

The small photometric distance relative to the trigonometric distance suggests
that PSO~J071.8$-$12.2 may be a binary (or multiple) system with unresolved
components. The object does not show indications of being a spectral blend
\citep{Best:2015em}, indicating that its binary components would have similar
spectra and, therefore, approximately equal mass. Laser guide star adaptive
optics imaging of this object using Keck II/NIRC2 has detected no sign of a
companion at an angular resolution of 100~mas (W. Best et al. 2020, in
preparation).

\subsubsection{WISE~J180952.53$-$044812.5} 
This T1~dwarf \citep{Mace:2013jh,Best:2013bp} was identified by
\citet{Best:2015em} as a candidate member (93.8\% probability from \banyanii) of
ABDMG based on its proper motion, sky position, and photometric distance of
$13.1\pm1.3$~pc. ABDMG has an age of $149^{+51}_{-19}$~Myr \citep{Bell:2015gw};
WISE~J1809$-$0448 would be the first L/T transition member of ABDMG if
confirmed. However, our new parallax places WISE~J1809$-$0448 at
$20.3\pm1.4$~pc. At this distance, \banyans\ calculates a 0.0\% probability of
ABDMG membership (99.9\% for the field). We note that \banyans\ gives a 91.5\%
probability for ABDMG without the parallax, so our UKIRT measurement provides
the key evidence that the object is not in ABDMG. This larger parallax distance
was expected because WISE~J1809$-$0448 has been identified as a close binary
(separation 0.3'') with components having $\Delta J=0.44$~mag \citep[W. Best et
al. 2020, in preparation]{Deacon:2017ja}, making the previous photometric
distance smaller than the true distance.

\subsubsection{PSO~J319.3102$-$29.6682}
This T0~dwarf was discovered and identified by \citet{Best:2015em} as a
candidate member of BPMG (97.1\% from \banyanii) based on its proper motion, sky
position, and photometric distance of $15.6\pm1.6$~pc. With our UKIRT
parallactic distance of $19.2\pm2.5$~pc, the BPMG probability increases to
99.0\% (\banyans). A radial velocity measurement would enable a complete
assessment of the kinematic consistency of PSO~J319.3$-$29.6 with BPMG. If
confirmed as a BPMG member, PSO~J319.3$-$29.6 would be the first known early-T
dwarf known at this age, making it an essential benchmark in studies of the L/T
transition for young objects.

\subsection{Common Proper Motion Companions}
\label{results.interest.cpm}

Our UKIRT measurements include new parallaxes for 14 objects previously
identified as wide common proper motion companions to main-sequence stars. The
primaries have parallaxes from \gaiat, and our UKIRT parallaxes provide
confirmation that the pairs have consistent distances as well as proper motions
and are therefore gravitationally bound. Such pairs formed in the same clusters,
so brown dwarfs confirmed as wide companions to main-sequence stars inherit age
and metallicity constraints from their primaries and, therefore, serve as
valuable benchmarks. Table~\ref{tbl.interest.cpm} lists these wide binaries, all
of which were confirmed by our parallaxes (all agreeing with the primary
parallaxes within $\le1.4\sigma$). We discuss three of these systems in more
detail below.

\floattable
\begin{deluxetable}{lcCCCclcCCCChhl}
\tablecaption{Updates to Common Proper Motion Companion Status \label{tbl.interest.cpm}}
\tablecolumns{13}
\tabletypesize{\tiny}
\setlength{\tabcolsep}{0.04in}
\tablewidth{0pt}
\rotate
\tablehead{
\multicolumn{5}{c}{Primary} &
\colhead{} &
\multicolumn{5}{c}{Companion} &
\colhead{Projected} &
\nocolhead{Previous} &
\nocolhead{Updated} &
\colhead{} \\
\cline{1-5}
\cline{7-11}
\colhead{Name} &
\colhead{SpT} &
\colhead{\plx\tablenotemark{a}} &
\colhead{\mua\tablenotemark{a}} &
\colhead{\mud\tablenotemark{a}} &
\colhead{} &
\colhead{Name} &
\colhead{SpT\tablenotemark{b}} &
\colhead{\plx\tablenotemark{c}} &
\colhead{\mua\tablenotemark{c}} &
\colhead{\mud\tablenotemark{c}} &
\colhead{Separation} &
\nocolhead{Status} &
\nocolhead{Status} &
\colhead{Refs} \\
\multicolumn{2}{c}{} &
\colhead{(mas)} &
\colhead{(\my)} &
\colhead{(\my)} &
\multicolumn{3}{c}{} &
\colhead{(mas)} &
\colhead{(\my)} &
\colhead{(\my)} &
\colhead{($''$)} &
\nocolhead{} &
\nocolhead{} &
\colhead{}
}
\startdata
NLTT 730                  & M4V         & 28.61\pm0.07                & 381.54\pm0.10                 & -227.08\pm0.08                & & 2MASS J00150206+2959323    & L7.5 pec (blue)\tablenotemark{d} & 31.8\pm3.3                  & 372.6\pm2.2                   & -231.7\pm2.3                  & 233.6 & Companion & Confirmed & 7,15 \\
HIP 9269                  & G9V         & 40.38\pm0.04                & 243.76\pm0.07                 & -351.59\pm0.06                & & HIP 9269B                  & L6                               & 43.3\pm3.8                  & 245.7\pm1.7                   & -352.7\pm1.9                  & 52.1  & Companion & Confirmed & 7,11 \\
2MASS J02132062+3648506AB & M4.5V+M6.5V & 70.02\pm0.20                & 65.38\pm0.46                  & 64.89\pm0.38                  & & 2MASS J02132062+3648506C   & T3                               & 70.1\pm4.0                  & 31.1\pm3.6                    & 50.1\pm3.7                    & 16.4  & Companion & Confirmed & 8,13 \\
HIP 26653                 & G5V         & 35.61\pm0.05                & -11.18\pm0.06                 & -142.55\pm0.07                & & HIP 26653B                 & L1.5                             & 32.6\pm3.9\tablenotemark{e} & -13.1\pm2.9\tablenotemark{e}  & -147.2\pm2.7\tablenotemark{e} & 27.0  & Companion & Confirmed & 7,24 \\
HIP 38939                 & K4V         & 54.10\pm0.04                & 362.26\pm0.06                 & -245.89\pm0.06                & & HIP 38939B                 & T4.5                             & 55.7\pm3.6                  & 357.6\pm2.9                   & -250.3\pm2.8                  & 88    & Companion & Confirmed & 5,12 \\
Gl 337AB                  & G8V+K1V     & 49.15\pm0.35                & -523.62\pm0.53                & 244.28\pm0.44                 & & Gl 337CD                   & L8+$>$L8\tablenotemark{f}        & 46.2\pm3.8                  & -543.0\pm3.1                  & 250.1\pm2.9                   & 43    & Companion & Confirmed & 1,3,25 \\
Gl 417A                   & G0V         & 44.14\pm0.04                & -249.39\pm0.09                & -151.59\pm0.07                & & Gl 417BC                   & L4.5+L6\tablenotemark{g}         & 40.3\pm3.4\tablenotemark{h} & -244.1\pm3.8\tablenotemark{h} & -148.4\pm3.3\tablenotemark{h} & 90    & Companion & Confirmed & 2,10,14 \\
LP 374-39                 & M5V         & 67.2\pm3.2\tablenotemark{i} & -995.2\pm7.0\tablenotemark{i} & -318.5\pm6.1\tablenotemark{i} & & WISEPC J112254.73+255021.5 & T6                               & 63.0\pm2.8                  & -1011.7\pm3.0                 & -323.8\pm3.5                  & 265   & Companion & Confirmed & 9,16,17,21 \\
NLTT 31450                & M4V         & 25.89\pm0.05                & -24.55\pm0.08                 & -193.68\pm0.05                & & NLTT 31450B                & L6                               & 27.1\pm2.5                  & -29.1\pm3.0                   & -186.5\pm2.8                  & 12.3  & Companion & Confirmed & 7 \\
LHS 2803                  & M4.5V       & 55.00\pm0.08                & -687.60\pm0.14                & -512.98\pm0.12                & & LHS 2803B                  & T5.5                             & 60.1\pm3.6                  & -686.6\pm2.6                  & -513.8\pm4.6                  & 67.6  & Companion & Confirmed & 6,19 \\
HIP 73169                 & M0V         & 22.25\pm0.05                & -271.07\pm0.09                & -82.51\pm0.11                 & & HIP 73169B                 & L2.5                             & 26.6\pm3.5                  & -270.2\pm5.0                  & -83.5\pm4.5                   & 29.1  & Companion & Confirmed & 7,12 \\
HIP 73786                 & K5V         & 52.59\pm0.07                & -607.66\pm0.10                & -506.50\pm0.13                & & HIP 73786B                 & T6.5                             & 57.1\pm4.3                  & -599.6\pm3.6                  & -502.5\pm3.4                  & 63.8  & Companion & Confirmed & 4,23,18 \\
G 203-50                  & M4.5V       & 47.12\pm0.06                & 252.50\pm0.10                 & 82.22\pm0.12                  & & 2MASS J17114559+4028578    & L5                               & 46.5\pm4.3\tablenotemark{j} & 254.8\pm4.5\tablenotemark{j}  & 67.3\pm3.6\tablenotemark{j}   & 6.4   & Companion & Confirmed & 20,22 \\
PM I23492+3458            & M2V         & 32.76\pm0.04                & -3.81\pm0.05                  & -123.41\pm0.04                & & PM I23492+3458B            & L9                               & 34.4\pm3.1                  & -7.3\pm2.2                    & -112.2\pm2.1                  & 34.9  & Companion & Confirmed & 7 \\
\enddata
\tablenotetext{a}{Astrometry from \gaiat\ \citep{GaiaCollaboration:2018io}
  except where indicated.} 
\tablenotetext{b}{Companion spectral types are NIR. Additional
  optical spectral types for specific objects are given in the footnotes.}
\tablenotetext{c}{Astrometry from UKIRT (this work).}
\tablenotetext{d}{Optical spectral type L7 \citep{Kirkpatrick:2010dc}.}
\tablenotetext{e}{Also has \gaiat\ measurements of $\plx=36.09\pm0.73,
  \mua=-12.71\pm0.87, \mud=-151.51\pm0.94$.} 
\tablenotetext{f}{Optical spectral type L8 \citep[unresolved;][]{Wilson:2001db}.}
\tablenotetext{g}{Optical spectral type L4.5 \citep[unresolved;][]{Kirkpatrick:2000gi}.}
\tablenotetext{h}{Also has \gaiat\ measurements of $\plx=42.87\pm1.10,
  \mua=-236.35\pm2.13, \mud=-152.07\pm2.11$.} 
\tablenotetext{i}{Parallax from \citet{Dittmann:2014cr}, proper motion from
  \citet{Kirkpatrick:2016jt}; \gaiat\ reported only a position for this object.} 
\tablenotetext{j}{Also has \gaiat\ measurements of $\plx=47.44\pm0.65,
  \mua=251.25\pm1.31, \mud=79.57\pm1.25$.} 
\tablerefs{
(1) \citet{Barnaby:2000fo}, (2) \citet{Bidelman:1951ir}, (3) \citet{Burgasser:2006cf}, (4) \citet{Cenarro:2007dc}, 
(5) \citet{Deacon:2012eg}, (6) \citet{Deacon:2012gf}, (7) \citet{Deacon:2014ey}, (8) \citet{Deacon:2017kd}, 
(9) \citet{Dittmann:2014cr}, (10) \citet{Dupuy:2012bp}, (11) \citet{Gray:2003fz}, (12) \citet{Gray:2006ca}, 
(13) \citet{Janson:2012dc}, (14) \citet{Kirkpatrick:2001bi}, (15) \citet{Kirkpatrick:2010dc}, (16) \citet{Kirkpatrick:2011ey}, 
(17) \citet{Kirkpatrick:2016jt}, (18) \citet{Murray:2011ey}, (19) \citet{Muzic:2012hc}, (20) \citet{Radigan:2008jd}, 
(21) \citet{Reid:1995kw}, (22) \citet{Reid:2003hw}, (23) \citet{Scholz:2010cy}, (24) \citet{Scholz:2010cy},
(25) \citet{Wilson:2001db}.
}
\end{deluxetable}

\subsubsection{2MASS J02132062+3648506C}
\citet{Deacon:2017kd} discovered this T3~dwarf (hereinafter 2MASS~J0213+3648C)
and identified it as comoving with the $16.4''$ distant M4.5+M6.5 binary
2MASS~J02132062+3648506AB. \citet{Deacon:2017kd} noted a disagreement of the
photometric parallaxes for this system's binary primary ($68\pm20$~mas) and
candidate companion ($63^{+14}_{-10}$~mas) with a preliminary PS1 parallax for
the companion ($45\pm10$~mas). Our UKIRT parallax of $70.1\pm4.0$~mas and the
\gaiat\ parallax of $70.02\pm0.20$~mas for the primary now resolve this
disagreement, implying that 2MASS~J0213+3648C is indeed a wide companion.
However, we also note a significant discrepancy between the new proper motion
measurements for 2MASS~J0213+3648C from UKIRT ($31.1\pm3.6$, $50.1\pm3.7$~\my)
and for 2MASS~J0213+3648AB from \gaiat\ ($65.38\pm0.46$, $64.89\pm0.38$~\my).
Our UKIRT proper motion is consistent with previous measurements for both the
primary \citep[$24\pm8$, $47\pm8$~\my;][]{Lepine:2011gl} and the companion
\citep[$24\pm10$, $65\pm11$~\my;][]{Deacon:2017kd} suggesting that the \gaiat\
measurement is incorrect. 2MASS~J0213+3648AB is a binary with an evolving
$\approx$0.2$''$ separation \citep{Janson:2012dc,Janson:2014fx} that may be
partially resolved by \gaia, which could explain the discrepant proper motion.
The RUWE statistic for 2MASS~J0213+3648AB is 3.0, much greater than the
$\mathrm{RUWE}\le1.4$ criterion for robust solutions identified by
\citet{Lindegren:2018vg}, indicating that the \gaiat\ astrometry
2MASS~J0213+3648AB may not be reliable.

\subsubsection{Gl 417BC}
\citet{Kirkpatrick:2000gi,Kirkpatrick:2001bi} discovered this object and
identified it as an L4.5~dwarf comoving with Gl~417 at a separation of $90''$.
Gl~417BC was later recognized as a $0.07''$~binary \citep{Bouy:2003eg} with
component spectral types L4.5 and L6 (DL12) and masses
$51.5^{+1.7}_{-1.8}$~\mjup\ and $47.7\pm1.9$~\mjup, respectively
\citep{Dupuy:2014iz}. \citet{Kirkpatrick:2010dc} presented preliminary parallax
measurements for Gl~417A and Gl~417BC based on images from only three epochs
spread over one year. Those preliminary measurements indicated consistency with
each other and with the {\it Hipparcos} parallax for Gl~417A, which, along with
consistent proper motions, provided convincing evidence that the pair was indeed
a comoving system. No other direct parallax measurement for Gl~417BC has been
published until \gaiat\ and this work, which both confirm the bound system.

\subsubsection{HIP 73169B}
This L2.5~dwarf was discovered and identified as a comoving companion to the M0V
star HIP~73169 by \citet{Deacon:2014ey}. However, there was a large difference
in the {\it Hipparcos} parallax for the primary
\citep[$36.6\pm9.4$~mas;][]{vanLeeuwen:2007dc} and the estimated photometric
parallax for the L~dwarf companion ($19.7\pm3.2$~mas). We note that the {\it
  Hipparcos} parallax carries an unusually large uncertainty, suggesting a
problem with that measurement. The discrepancy is now resolved by the \gaiat\
parallax for the primary ($22.25\pm0.05$~mas) and our UKIRT measurement for the
L~dwarf ($26.6\pm3.5$~mas), whose consistency at $1.2\sigma$ indicates
companionship.

\section{Summary}
\label{summary}

We have measured parallaxes, proper motions, and \jmko\ photometry for
\varngoodplx\ ultracool dwarfs with spectral types L0--T8 using UKIRT/WFCAM. We
observed targets over two-thirds of the sky ($\delta=-30\degr$ to $+60\degr$)
having photometric distances within $1\sigma$ of 25~pc that lacked precise
parallax measurements, with the goal of establishing a complete volume-limited
sample of L0--T8~dwarfs. These measurements represent the largest set of
infrared parallax measurements for ultracool dwarfs published to date, and they
include \varnfirst\ objects without a previous parallax measurement. We achieved
a median parallax precision of \varmedplxerr~mas, comparable to several previous
infrared parallax programs but with many more targets in our sample.  We
corrected our parallax and proper motions measurements from relative to absolute
astrometry using Gaia DR2 measurements of field stars in our images.

Our target list shares \varngaia\ objects with \gaiat. Our parallaxes
demonstrate good consistency with \gaiat\ measurements.  Most of the shared
objects are L0--L5 dwarfs, a demonstration that \gaia\ is predominantly a survey
of stars; ground-based infrared observations continue to be essential for
studies of brown dwarfs.

Our UKIRT parallaxes will enable population studies of L and T dwarfs using, for
the first time, a volume-limited sample out to 25~pc with several hundred
members defined entirely by parallaxes.

\vspace{20 pt}

We thank the anonymous referee for a thorough review that improved the quality
of this paper.
We thank Zhoujian Zhang for calculating the Gaia DR2 RUWE statistic for our
parallax targets.
We thank Watson Varricatt and Tom Kerr at UKIRT and Mike Irwin at CASU for their
responsive and ongoing support of our observations and data reduction.
UKIRT is owned by the University of Hawaii (UH) and operated by the UH Institute
for Astronomy; operations are enabled through the cooperation of the East Asian
Observatory.  When some of the data reported here were acquired, UKIRT was
supported by NASA and operated under an agreement among the University of
Hawaii, the University of Arizona, and Lockheed Martin Advanced Technology
Center; operations were enabled through the cooperation of the East Asian
Observatory.  When some other of the data reported here were acquired, UKIRT was
operated by the Joint Astronomy Centre on behalf of the Science and Technology
Facilities Council of the U.K.  All UKIRT data reported here were obtained as
part of the UKIRT Service Programme.
This work has made use of data from the European Space Agency (ESA) mission
\gaia\ (\url{http://www.cosmos.esa.int/gaia}), processed by the \gaia\ Data
Processing and Analysis Consortium (DPAC,
\url{http://www.cosmos.esa.int/web/gaia/dpac/consortium}). Funding for the DPAC
has been provided by national institutions, in particular the institutions
participating in the \gaia\ Multilateral Agreement.
The Pan-STARRS1 Surveys (PS1) and the PS1 public science archive have been made
possible through contributions by the Institute for Astronomy, the University of
Hawaii, the Pan-STARRS Project Office, the Max-Planck Society and its
participating institutes, the Max Planck Institute for Astronomy, Heidelberg and
the Max Planck Institute for Extraterrestrial Physics, Garching, The Johns
Hopkins University, Durham University, the University of Edinburgh, the Queen's
University Belfast, the Harvard-Smithsonian Center for Astrophysics, the Las
Cumbres Observatory Global Telescope Network Incorporated, the National Central
University of Taiwan, the Space Telescope Science Institute, the National
Aeronautics and Space Administration under grant No. NNX08AR22G issued through
the Planetary Science Division of the NASA Science Mission Directorate, the
National Science Foundation grant No. AST-1238877, the University of Maryland,
Eotvos Lorand University (ELTE), the Los Alamos National Laboratory, and the
Gordon and Betty Moore Foundation.
This publication makes use of data products from the Two Micron All Sky Survey
(2MASS), which is a joint project of the University of Massachusetts and the
Infrared Processing and Analysis Center/California Institute of Technology,
funded by the National Aeronautics and Space Administration and the National
Science Foundation.
This research has made use of NASA's Astrophysical Data System
and the SIMBAD database, the {\it VizieR} catalog access tool, the
cross-match service provided by CDS, Strasbourg, France, and the Database of
Ultracool Parallaxes, maintained by Trent Dupuy at
\url{https://www.as.utexas.edu/~tdupuy/plx}.
W.M.J.B. received support from NSF grant AST-0909222.
W.M.J.B, M.C.L., and E.A.M. received support from NSF grant AST-1313455.
TJD acknowledges research support from Gemini Observatory.
Finally, the authors wish to recognize and acknowledge the very significant
cultural role and reverence that the summit of Maunakea has always held within
the indigenous Hawaiian community. We are most fortunate to have the opportunity
to conduct observations from this mountain.

\facility{UKIRT}

\software{
\banyans\ \citep{Gagne:2018tx}, 
CASU pipeline \citep{Irwin:2004ej,Lawrence:2007hu}, 
mpfit \citep{Markwardt:2009wq},
PDS IDL Library \url{https://www.astro.umd.edu/~eshaya/PDS}, 
SCAMP \citep[v1.4.4;][]{Bertin:2006vk}, 
Source Extractor \citep[v2.19.5;][]{Bertin:1996hf}, 
TOPCAT \citep{Taylor:2005wq}.
The IDL routines {\tt queryvizier.pro}, {\tt resistant\_mean.pro}, and {\tt
  robust\_sigma.pro} are part of the IDL Astronomy User's Library, available at
\url{https://idlastro.gsfc.nasa.gov}.  }

\appendix

\section{Comparison of our parallax fitting with Martin et al. (2018)}

We applied the same analysis method used for our UKIRT targets to the 22 objects
(late-T and Y~dwarfs) with \spitzer\ astrometry from \citet{Martin:2018hc}.  Our
motivation for this analysis was Section~6.4.2 of \citet{Martin:2018hc}, which
suggests that our astrometric fitting approach gives systematically smaller
parallaxes than their approach, based on four out of five late-T or Y~dwarfs for
which they measured parallaxes 1--2$\sigma$ larger than those found by
\citet[who used the same method as in this paper]{Dupuy:2013ks}.
\citet{Martin:2018hc} describe our approach as being a three-parameter {\tt
  mpfit} model that does not fit for {R.A.} and {decl.} zero-points. As is
discussed here, and in our original papers, DL12 and \citet{Dupuy:2013ks}, our
analysis actually uses an MCMC method that includes a full five-parameter fit
(R.A., decl., $\mu_{\rm R.A.}$, $\mu_{\rm decl.}$, and $\varpi$).
\citet{Martin:2018hc} also suggested that the shorter time baseline of the
\citet{Dupuy:2013ks} data compared to theirs led to systematically smaller
parallaxes, though their analysis (e.g., their Figure~10) does not reveal any
significant effect when extending the time baseline of the \citet{Dupuy:2013ks}
objects.  Finally, \citet{Martin:2018hc} suggest that a small chromatic effect
may be present in the {[3.6]}-band \spitzer/IRAC astrometry of
\citet{Dupuy:2013ks} due to the differing SEDs of brown dwarfs compared to
background stars in this bandpass. Such an effect can be tested in future
datasets.

\begin{figure*}
  \centering
  \begin{minipage}[t]{0.48\textwidth}
    \includegraphics[width=1.00\columnwidth, trim = 16mm 0 4mm 0]{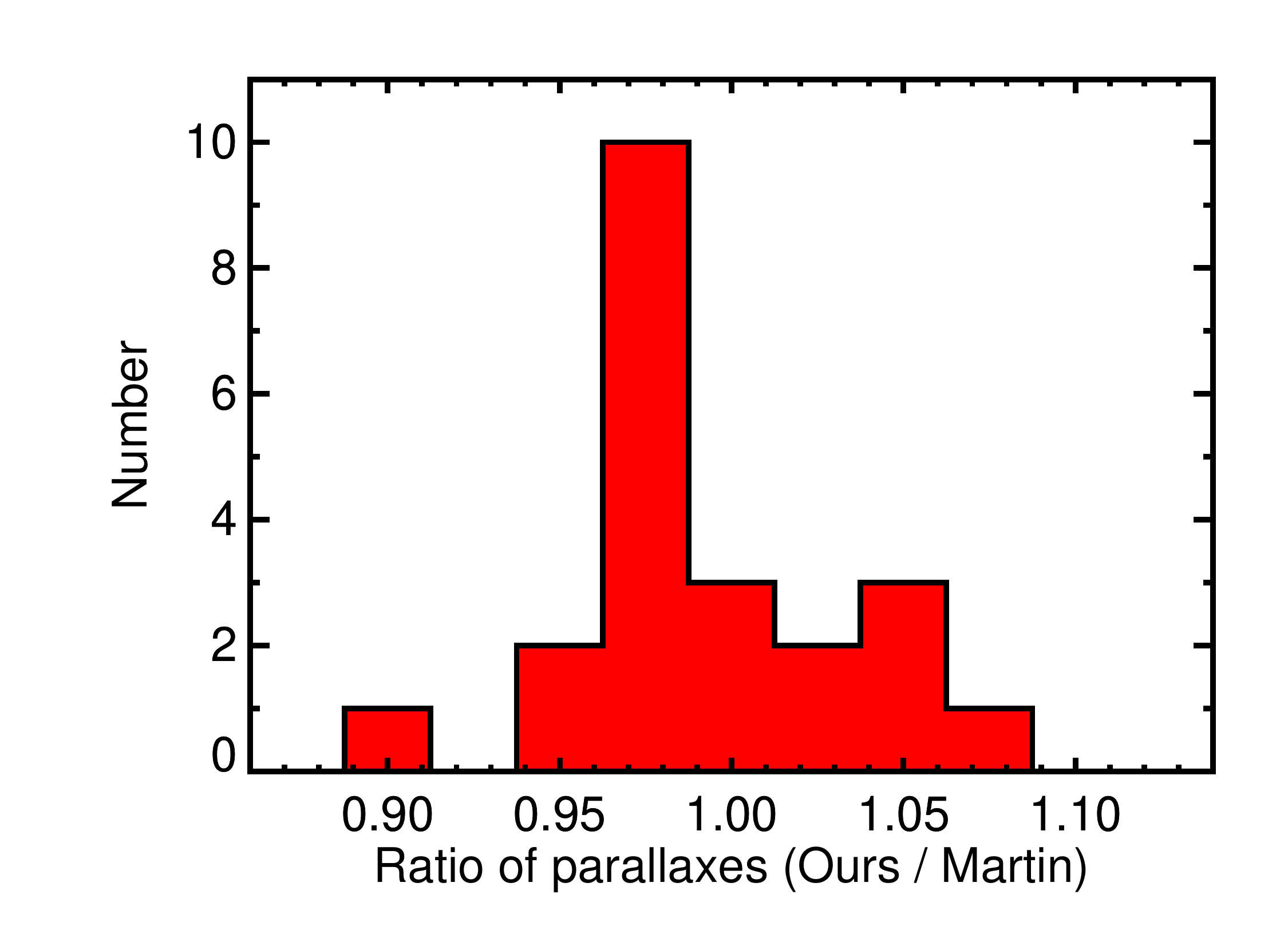}
  \end{minipage}
  \hfill
  \begin{minipage}[t]{0.48\textwidth}
    \includegraphics[width=1.00\columnwidth, trim = 16mm 0 4mm 0]{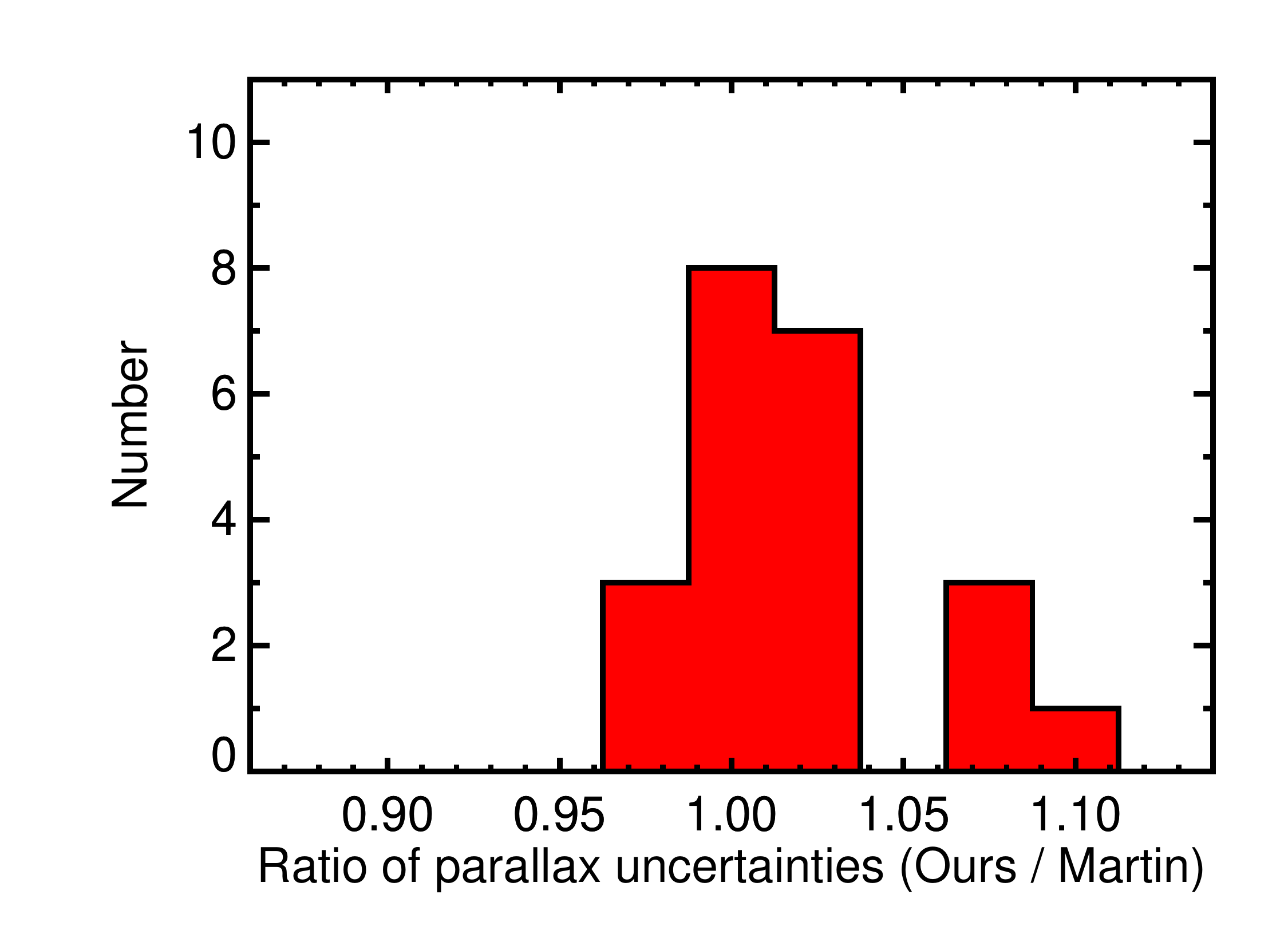}
  \end{minipage}
  \caption{{\it Left panel}: distribution of the ratios of parallax measurements from
    our analysis to those of \citet{Martin:2018hc} for the 22~\spitzer-observed
    objects from \citet{Martin:2018hc}. The mean ratio is 0.993 with an rms of
    0.040. {\it Right panel}: distribution of the ratios of parallax uncertainties.
    The mean ratio is 1.019 with an rms of 0.033, with most of the difference
    coming from the three objects for which our uncertainties are 7\%--11\%
    larger. These results indicate that our fitting method and that of
    \citet{Martin:2018hc} produce parallaxes that are fully consistent with each
    other when applied to the same set of R.A. and decl. measurements.}
  \label{fig.martin.ratios}
\end{figure*}

For the moment, we examine the aforementioned suggestion of
\citet{Martin:2018hc} that our parallax fitting produces systematic offsets
relative to theirs, by analyzing the individual-epoch \spitzer\ astrometry from
\citet{Martin:2018hc} using our same parallax fitting method as this UKIRT work
and past papers (except for the IRLS procedure described in
Section~\ref{plx.relative.irls}, which was not used by DL12 or
\citealt{Dupuy:2013ks}).  The mean ratio of our resulting parallaxes compared to
those of \citet{Martin:2018hc} was 0.993, with an rms of 0.040
(Figure~\ref{fig.martin.ratios}), so, on average, our parallaxes were 0.7\%
smaller but the offset is not significant. Slightly more significant is that our
MCMC parallax uncertainties are on average 1.9\% larger than the uncertainties
reported by \citet{Martin:2018hc}, although, this is entirely driven by three
objects where our uncertainties are 7\%--11\% larger (J0350$-$56, J0359$-$54,
and J0647$-$62), and otherwise, the two sets of uncertainties are consistent
within $0.9\%\pm2.1\%$. We also find excellent agreement between our proper
motion uncertainties and R.A.\ proper motions. The decl.\ proper motions
measured by us and Martin et al. do differ significantly (1.0$\sigma$
dispersion), when we compute the difference in proper motion and divide by the
\citet{Martin:2018hc} uncertainty. In contrast, the R.A.\ proper motions have a
dispersion of only 0.2$\sigma$. This seems somewhat puzzling given that both
fits are based on the same input data.

Overall, we confirm that our parallax fitting method and that of
\citet{Martin:2018hc} produce the same results given the same input data, with
no significant systematic differences in the parallaxes and with our fitted
parallax uncertainties being slightly more conservative. Finally, we note that
the discrepancy identified by \citet{Martin:2018hc} may be a random occurrence,
as having four out of five parallaxes from one analysis being lower than another
analysis can occur by chance with a probability of 0.125.

\end{document}